\documentclass[graybox]{svmult}
\usepackage{mathptmx}       
\usepackage{helvet}         
\usepackage{courier}        
\usepackage{type1cm}        
\usepackage{makeidx}         
\usepackage{graphicx}        
\usepackage{multicol}        
\usepackage[bottom]{footmisc}
\usepackage{hyperref}
\hypersetup{backref=true, pagebackref=true, hyperindex=true, breaklinks=true,colorlinks=true,urlcolor=blue, linkcolor=black,  citecolor=blue,pagecolor=red, bookmarks=true, bookmarksopen=true}

\makeindex             

\newcommand{\apjl}{ApJL,\,}
\newcommand{\apj}{ApJ,\,}
\newcommand{\apjs}{ApJS,\,}
\newcommand{\aj}{AJ,\,}
\newcommand{\aap}{A\&A,\,}
\newcommand{\mnras}{MNRAS,\,}
\newcommand{\araa}{ARA\&A, \,}
\newcommand{\icarus}{Icarus, \,}
\newcommand{\actaa}{Acta Astronomica, \,}
\newcommand{\nar}{New Astronomy Reviews, \,}
\newcommand{\apss}{Astrophysics and Space Science, \,}
\newcommand{\nat}{Nature, \,}
\newcommand{\pasa}{Publications of the Astronomical Society of Australia, \,}

\begin{document}

\title*{Physical Processes in Protoplanetary Disks}
\author{Philip J. Armitage}
\institute{Philip J. Armitage \at JILA, University of Colorado \& NIST, Boulder, Colorado, CO~80309-0440, USA, \email{pja@jilau1.colorado.edu}}

\maketitle

\abstract{This review, based on lectures given at the 45th Saas-Fee Advanced Course 
``From Protoplanetary Disks to Planet Formation", 
introduces physical processes in protoplanetary disks relevant to accretion and the initial stages of planet formation. 
After a brief overview of the observational context, I introduce the elementary theory of disk structure and evolution, 
review the gas-phase physics of angular momentum transport through turbulence and disk winds, and discuss possible 
origins for the episodic accretion observed in 
Young Stellar Objects. Turning to solids, I review the evolution of single particles under aerodynamic forces, and 
describe the conditions necessary for the development of collective gas-particle instabilities. Observations show that 
disks can exhibit pronounced large-scale structure, and I discuss the types of structures that may form from gas and 
particle interactions at ice lines, vortices and zonal flows, prior to the formation of large planetary bodies. I conclude with disk dispersal.}

\newpage

\section{Preamble}
The objective of this review is to introduce the physical processes in protoplanetary disks 
that are relevant to protostellar accretion and the initial stages of planet formation. Protoplanetary disks, as well as being 
interesting objects of study in their own right, are also simultaneously an outcome of star formation and initial conditions  
for planet formation. As such, we need to understand the evolution of both the dominant gaseous component and the 
trace of solid material that is critical for planet formation. Much interesting complexity occurs due to interactions  
between the two.

The review is organized around three motivating questions; how do protoplanetary gas disks evolve over time, 
how are solids transported and concentrated within the gas, and how do gas-phase and solid processes interact 
to form structure within disks? I begin in \S\ref{sec_observations} with a brief summary of the observational context. 
The review proper starts in \S\ref{sec_structure} 
by outlining the equilibrium structure of disks. Disk evolution is described in \S\ref{sec_evolution} and \S\ref{sec_turbulence}, 
first following the classical approach in which the origin of angular momentum transport is unspecified, and subsequently 
in a more modern presentation where angular momentum transport and loss processes are ascribed to specific 
fluid instabilities. \S\ref{sec_episodic} discusses candidate theoretical explanations, some of them directly related 
to angular momentum transport processes, for episodic accretion outbursts. In \S\ref{sec_single} the focus switches to
solids, and I review how single solid particles settle, drift, diffuse and concentrate relative to the gas. \S\ref{sec_structure_form} then 
describes how these processes can give rise to structure within the disk on various scales, either at transient 
``particle traps" or at persistent locations such as ice lines where the disk structure varies rapidly. Finally, \S\ref{sec_dispersal} 
reviews what is known about the processes, including photoevaporation, that can disperse 
protoplanetary disks.

The lectures, even expanded as in this review, could not touch upon all of the physics that an aspiring 
researcher in the field might require. In an effort to cover as much as possible --- and to accommodate the diverse 
backgrounds of participants (and readers) --- the material is presented with varying degrees of detail and rigor. 
For most topics I begin with a self-contained discussion of essential material 
that would often be assumed as background in papers or talks. I then discuss the underlying physics of 
more recent results, and give entry points to the relevant literature. 

One caution is in order. Quantities are generally defined and labeled following 
conventions in the literature, to make it easier for the reader who needs to fill in missing details 
or to explore further. Given the broad scope of the review there is considerable overloading of notation, 
with the same symbols being used to represent unrelated quantities in different sections. Take care!

\newpage

\section{Observational context}
\label{sec_observations}

The challenges of observing protoplanetary disks are formidable. Around Solar mass stars the lifetime of 
primordial disks of gas and dust is a small fraction --- $10^{-3} - 10^{-4}$ --- of the main sequence lifetime. 
Disks around stars are hence rare, and even the nearest examples are relatively distant and hard to 
resolve. A small number of disks, including particularly well-studied examples such as TW Hya, are 
as close as $d \approx 50 \, {\rm pc}$, but for larger samples we need to look out at least as far as low-mass star 
forming regions such as Taurus, at $d \approx 140 \, {\rm pc}$ \cite{torres12}. Regions where massive stars 
are forming alongside low mass stars, such as Orion, are yet further away at $d \approx 400 \, {\rm pc}$ \cite{kounkel17}. 
Spatially resolved imaging of disks with scales comparable to that of the Solar System, $\sim 10^2 \ {\rm AU}$, requires 
sub-arcsecond resolution that has only become available in relatively recent times from the 
{\em Hubble Space Telescope} \cite{odell93}, from the {\em Atacama Large Millimeter / submillimeter Array (ALMA)} 
and earlier mm / sub-mm interferometers \cite{alma15}, and from 
high-contrast imaging instruments on large ground-based telescopes \cite{garufi16,follette17,momose15}. Today 
we have exquisite imaging and spatially resolved spectroscopy of a moderate number of protoplanetary disks, 
along with a larger amount of data that derives from a time when unresolved observations 
were the norm.

Williams \& Cieza (2011) \cite{williams11} review observations of protoplanetary disks up to a time just 
prior to the advent of {\em ALMA}. The focus of this review is on theory, but to set the stage and introduce 
some ubiquitous terminology we begin with a brief discussion of the observational context. What do we know 
about disks that might constrain theoretical models, and how do we know it?

\subsection{The classification of Young Stellar Objects}
\label{subsec_classification}

\begin{figure}[t]
\includegraphics[width=\columnwidth]{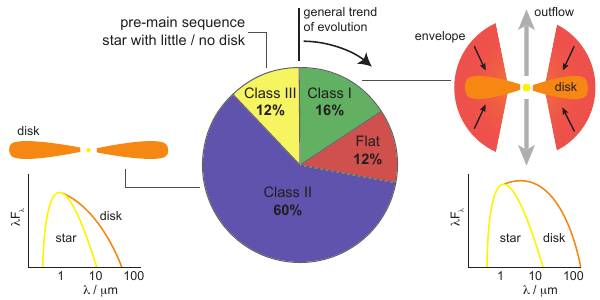}
\caption{The physical picture underlying the classification of Young Stellar Objects \cite{lada84,adams87} is shown together with 
statistics from the {\em Spitzer} c2d Legacy survey for the fraction of sources falling into each class \cite{evans09}. The classification 
scheme is based upon the slope of the spectral energy distribution between the near- and mid-infra-red. This maps to a structural 
(and essentially evolutionary) sequence in which Class~0 and Class~I YSOs are accreting from disks, which themselves are being 
fed by gas falling in from envelopes. Class~II YSOs (also called Classical T Tauri stars) are pre-main-sequence stars with 
surrounding disks, while Class~III YSOs (or Weak-lined T Tauri stars) are pre-main-sequence stars with little or no primordial 
gas remaining in orbit around them.}
\label{fig_YSOs}
\end{figure}

Young Stellar Objects (YSOs) frequently show more emission in the infra-red than would be expected from a 
pre-main-sequence star's photosphere. The IR excess is attributed to the presence of dust in the vicinity of the 
star, and its strength forms the basis for an empirical 
classification scheme for YSOs that dates back to work by Lada \& Wilking \cite{lada84} in the early 1980s. 
We define the slope of the spectral energy distribution (SED) between near-IR and mid-IR wavelengths,
\begin{equation}
 \alpha_{\rm IR} \equiv \frac{ {\rm d} \log \nu F_\nu}{ {\rm d} \log \nu} \equiv \frac{ {\rm d} \log \lambda F_\lambda}{ {\rm d} \log \lambda}.
\end{equation}
The anchor points in the near- and mid-IR for the determination of $\alpha_{\rm IR}$ vary from study to study, but 
are typically something like $2 \ \mu {\rm m}$ and $25 \ \mu {\rm m}$. Based on $\alpha_{\rm IR}$ four or five classes of YSO 
are recognized \cite{williams11},
\begin{itemize}
\item 
{\bf Class 0}: heavily obscured sources with no optical or near-IR emission ($\alpha_{\rm IR}$ is therefore 
undefined).
\item
{\bf Class I}: $\alpha_{\rm IR} > 0.3$.
\item
{\bf Flat spectrum sources}: $-0.3 < \alpha_{\rm IR} < 0.3$.
\item
{\bf Class II}: $-1.6 < \alpha_{\rm IR} < -0.3$.
\item
{\bf Class III}: $\alpha_{\rm IR} < -1.6$. These sources have at most weak IR-excess emission, and have 
SEDs that resemble isolated pre-main-sequence stellar photospheres.
\end{itemize}
Figure~\ref{fig_YSOs} shows how these empirically derived classes match up against the expected 
evolution of circumstellar material over time \cite{adams87}. Stars form from the collapse of molecular 
cloud cores which have vastly more angular momentum than can be accommodated in a star \cite{goodman93}. Some 
of that angular momentum can be lost via magnetic braking \cite{li14}, or subsumed into the orbital angular momentum 
of binaries \cite{reipurth14}, but it is an observational fact that enough is commonly left over to form a rotationally 
supported disk. The free-fall time scale of molecular cloud cores is shorter than the lifetime of disks, so the youngest 
(Class 0 and Class I) YSOs are surrounded by disks which themselves are fed by infall from envelopes. Older YSOs 
(Class II) have lost their envelopes but retain relatively massive and often actively accreting gas disks. Finally 
the gas disk dissipates, leaving behind a Class III YSO.

For the subset of YSOs that are optically visible some of the terminology derives from an even older 
classification scheme that is based on the equivalent width of the H$\alpha$ line \cite{joy45}. H$\alpha$ is a 
diagnostic of gas accreting on to the star, so Class II sources with disks are largely equivalent to 
``Classical T Tauri stars" with high  H$\alpha$ equivalent widths. ``Weak-lined T Tauri stars" are likewise 
essentially the same objects as Class~III sources.

Determining the absolute ages of young stars is a difficult exercise that leads to uncertainty in estimates of the 
mean duration of the different phases. An easier task is to estimate the {\em relative} durations. Provided that 
we survey enough regions of recent and ongoing star formation, we may reasonably assume that the proportions 
of YSOs in the different classes reflect the amount of time that a typical YSO spends in each class. Figure~\ref{fig_YSOs} 
shows the distribution of YSOs among the classes, as determined by the ``Cores to Disks" legacy project that used 
the {\em Spitzer} space telescope \cite{evans09}. One sees that Class I sources with envelopes are much less 
numerous than Flat Spectrum and Class II sources that have disks but little or no envelope. Most of the 
circumstellar disk lifetime therefore consists of relatively isolated star-disk systems that have completed the 
primary phase of accretion from the envelope\footnote{Accretion from lower density gas within the star 
forming region could persist to later times \cite{throop08}.}. 

There is no uniform definition of the term ``protoplanetary disk". In this review, our main focus is on isolated 
disks where envelope accretion has largely ceased. Observationally this would correspond to Flat Spectrum 
and Class~II YSOs. Most of the action in planet formation, reviewed in Willy Kley's contribution to this 
volume, is also commonly assumed to occur in an isolated disk environment. One should be aware, 
however, that scant empirical evidence underlies this assumption, and that the time scales of 
planet formation processes through to (at least) planetesimal scales can be rapid. Important aspects   
of planet formation could be already well-advanced during the embedded Class~0 and Class~I phases 
of YSO evolution.

\subsection{Accretion rates and lifetimes}
The accretion rate $\dot{M}$ of gas through the disk is one of the most important quantities that 
we would like to determine. It can only be measured with any confidence at small radii, where the 
gas from the disk is flowing on to the star. As we will discuss in \S\ref{sec_magnetospheric}, for 
typical Classical T Tauri stars there is abundant evidence that the stellar magnetic field disrupts 
the inner disk at a magnetospheric radius $R_m$. Gas interior to $R_m$ follows trajectories 
that are tied to magnetic field lines, before crashing on to the stellar surface at close to the 
free fall speed \cite{bouvier07}. For a star of mass $M_*$ and radius $R_*$ the accretion 
luminosity will be,
\begin{eqnarray}
 L_{\rm acc} & \simeq & GM_* \dot{M} \left( \frac{1}{R_*} - \frac{1}{R_m} \right) \nonumber \\
 & \sim & 0.2 \left( \frac{M_*}{M_\odot} \right) \left( \frac{R_*}{1.5 \ R_\odot} \right)^{-1} 
 \left( \frac{\dot{M}}{10^{-8} \ M_\odot \ {\rm yr}^{-1}} \right) L_\odot,
\end{eqnarray}
where for the numerical estimate we have taken $R_m \gg R_*$ and adopted 
typical numbers for T Tauri stars. The accretion luminosity will be radiated from 
shocks or hotspots on the stellar surface. We will be able to estimate $\dot{M}$ 
provided that we know the basic stellar parameters and can distinguish the 
emission arising from accretion from the stellar photospheric emission.

In practice the determination of $\dot{M}$ from a stellar spectrum can be attempted 
in several ways, which differ in observational expense and fidelity. The most direct 
measurements require access to the entire ultraviolet (UV) spectrum, including short 
wavelengths that are not accessible from the ground. An example of early work of this type, 
using data from the {\em International Ultraviolet Explorer}, are the accretion rate measurements 
by Gullbring et al. \cite{gullbring00}. Next best is observations from the ground, which can 
cover enough of the UV spectrum (down to about 300~nm) to allow a direct determination 
of $L_{\rm acc}$ \cite{gullbring98,herczeg08}. Recent studies, such as that by Manara et al. 
\cite{manara16} take advantage of instruments with a wide spectral coverage to 
simultaneously determine the accretion rate and the stellar parameters. Finally, 
the luminosity in a number of different spectral lines has been shown to correlate well with the 
total accretion luminosity \cite{rigliaco12}, and measurements of line luminosities can 
be converted to accretion rates at the cost of some additional uncertainty.

The inferred accretion rates in Class~II YSOs are found to vary strongly with the 
stellar mass $M_*$ \cite{muzerolle03}. In the Chamaeleon~I star forming 
region, for example, Manara et al. determine a stellar mass / accretion rate 
relation of the form \cite{manara16},
\begin{equation}
 \log \left( \frac{\dot{M}}{M_\odot \ {\rm yr}^{-1}} \right) = (1.83 \pm 0.35) \log \left( \frac{M_*}{M_\odot} \right) 
 - (8.13 \pm 0.16).
\end{equation} 
This is a fairly representative result. Mean accretion rates through protoplanetary disks 
around Solar mass stars are of the order of $10^{-8} \ M_\odot \ {\rm yr}^{-1}$, with a stellar 
mass scaling that is significantly steeper than linear.

By estimating ages from the position of pre-main-sequence stars in the H-R diagram  
Hartmann et al. \cite{hartmann98} showed that the median accretion rate for T Tauri 
stars in Taurus and Chamaeleon~I declined with time. Consistent with this result, and 
with common sense expectations, the {\em fraction} of stars that host primordial disks 
declines with stellar age. This is commonly expressed not on a star-by-star basis, 
but rather by measuring the fraction of stars in a cluster with near-IR excess, and 
assigning a representative age to that cluster \cite{haisch01}. Using this method 
Hern{\'a}ndez et al. \cite{hernandez08} find that the disk fraction drops to 50\% 
at about 3~Myr. Caution is needed, however, because the ages of young clusters 
remain hard to determine, with some assessments suggesting that uncertainties 
may be as large as a factor of two \cite{soderblom08}.

The astronomically inferred disk lifetime is compatible with the chronology derived 
independently from the radioactive dating of primitive Solar System materials. The oldest 
Solar System samples are calcium-aluminum-rich inclusions (CAIs) found within chondritic 
meteorites. CAIs are dated to $4567.30 \pm 0.16$ million years ago. Chondritic meteorites 
also contain (and are named after) {\em chondrules} --- 0.1-1~mm spheres of rock that were heated to approximately 
their melting temperature in short-lived events in the early Solar System (see e.g. the short 
review by Connolly \& Jones \cite{connolly16}). How chondrules formed remains unclear, but it is known that their production 
(and possibly repeated subsequent heating events) began at the same time as the CAIs and 
continued for about 4~Myr \cite{bollard17}. If chondrule heating occurred predominantly in a 
primordial gas disk environment, one would conclude from this that the lifetime of the Solar System's 
gas disk was fairly typical. At the least, there is no evidence for any inconsistency between astronomical 
and meteoritic dating results.

\subsection{Inferences from the dust continuum}
At disk temperatures $T < 1500~{\rm K}$ the opacity in protoplanetary disks is dominated 
by the contribution from rocky or icy grains (generically ``dust"). Observations of 
continuum radiation from thermal dust emission provide information about the radial 
temperature distribution in the disk, the disk mass, and the size of dust particles. To 
illustrate how this works, consider a toy model of a thin disk in which dust emits thermal 
radiation with a single temperature at each radius (we defer discussion of realistic 
complications associated with the vertical structure to \S\ref{subsec_thermal}). We 
assume that the gas surface density $\Sigma$ and dust temperature $T_d$ are 
power laws in radius,
\begin{eqnarray}
 \Sigma & \propto & r^{-p}, \nonumber \\
 T_d & \propto & r^{-q},
\end{eqnarray} 
and adopt a frequency dependent opacity (defined per unit mass of gas),
\begin{equation}
 \kappa_\nu = \kappa_0 \nu^{\beta}.
\end{equation} 
The vertical optical depth through the disk is then $\tau_\nu = \Sigma \kappa_\nu$. 
For a face-on disk the formal solution of the radiative transfer equation (e.g. \cite{rybicki79}) 
gives the flux density $F_\nu$ as,
\begin{equation}
 F_\nu = \frac{1}{D^2} \int_{r_{\rm in}}^{r_{\rm out}} B_\nu (T_d) (1 - e^{-\tau_\nu}) 2 \pi r {\rm d} r,
\label{eq_flux_density} 
\end{equation}
where $D$ is the distance to the source, $r_{\rm in}$ and $r_{\rm out}$ are the inner and outer 
radii of the disk, and $B_\nu$ is the Planck function,
\begin{equation}
 B_\nu = \frac{2 h \nu^3}{c^2} \frac{1}{\exp [h \nu / k_B T_d] - 1}.
\end{equation} 
Let us consider what we can learn from $F_\nu$ in the limit where the disk is either 
optically thick ($\tau_\nu > 1$) or optically thin.

Taking the $\tau_\nu \gg 1$ limit first, let us assume that we observe the disk at wavelengths 
where the entire disk is optically thick. This will be true in the near- and mid-IR. Setting 
$e^{-\tau_\nu} = 0$ in equation~(\ref{eq_flux_density}) we obtain\footnote{To see this, 
substitute $x \equiv h \nu / k_B T$ and approximate the limits as $r_{\rm in} = 0$ and 
$r_{\rm out} = \infty$.},
\begin{equation}
 \nu F_\nu \propto \nu^{4-2/q}.
\end{equation} 
The slope of the infra-red spectral energy distribution thus provides a constraint on the 
{\bf radial variation of the dust temperature} within the disk.

In the mm / sub-mm region of the spectrum (i.e. $\lambda \sim 1 \ {\rm mm}$) the bulk 
of the emission comes instead from optically thin regions of the disk. Taking the limit 
where the entire disk is optically thin at the frequencies of interest, we have that 
$(1 - e^{-\tau_\nu}) \approx \tau_\nu = \Sigma \kappa_\nu$. Equation~(\ref{eq_flux_density}) 
then takes the form,
\begin{equation}
 F_\nu = \frac{B_\nu (\bar{T_d}) \kappa_\nu}{D^2} \int_{r_{\rm in}}^{r_{\rm out}} 2 \pi r \Sigma {\rm d} r,
\end{equation} 
where $\bar{T}$ is an weighted average of the temperature of the emitting material that we 
can determine from the infra-red part of the SED. From this we can infer two disk 
properties. First, we note that the integral in the above expression is just the disk mass, 
$M_{\rm disk}$. If we know the distance to the source, the opacity, and the disk temperature, 
a measurement of the flux density at optically thin wavelengths {\bf determines the disk 
mass}. Second, at sufficiently long wavelengths we will be on the Rayleigh-Jeans tail of the 
Planck function. In this limit $B_\nu \propto \nu^2$ and we find,
\begin{equation}
 \nu F_\nu \propto \nu^{\beta + 3}.
\end{equation} 
A measurement of the spectral slope at optically thin wavelengths in the mm / sub-mm thus 
determines the {\bf frequency dependence of the opacity}. The opacity, in turn, is determined 
by the size distribution, structure, and composition of the solid particles within the disk \cite{draine06}. 
For many disks, the integrated emission implies a value for $\beta \approx 1 \pm 0.5$ that is 
significantly smaller than the value ($\beta \approx 2$) found for dust in either the interstellar medium or in 
molecular clouds \cite{beckwith91}. This result provides evidence for the growth of solid particles up to sizes of at 
least mm within protoplanetary disks \cite{rodmann06,ricci10}.

It is noteworthy that we can extract constraints on the mass, temperature distribution, and particle properties 
from the SED of even an unresolved disk. Continuum observations become even more powerful if the disk 
can be spatially resolved at multiple wavelengths. Tazzari et al. \cite{tazzari16}, for example,  perform a 
multi-wavelength analysis of three disks (DR~Tau, AS~209 and FT~Tau) using mm and radio data 
covering the range between 0.88~mm and 1~cm. They find evidence for a radial dependence of the 
maximum particle size, that is both broadly consistent with and a critical test of theoretical models 
for particle evolution in disks \cite{birnstiel12}.

With great power comes great responsibility, and several caveats should be borne in mind when 
interpreting dust continuum observations. There are well-known sources 
of uncertainty in disk mass estimates derived from mm / sub-mm data:
\begin{itemize}
\item
The most directly inferred quantity is the mass of {\em optically thin dust} with particle sizes 
roughly the same as the observing wavelengths. Additional mass could be hidden, even 
in similar sized particles, at radii where the emission is optically thick. Moreover, the 
conversion between the observed flux and the mass depends on the opacity, which 
(in detail) is a function of the unknown structural and compositional properties of the 
particles. (Going beyond the simplest continuum observations, some mineralogical information 
is available by modeling the broad silicate feature seen in disk spectra near 10~$\mu$m 
\cite{olofsson10}.)
\item
The total mass of {\em solid material} could be much larger than the mass inferred 
from the optically thin emission if (a) particles have grown to radii $s \gg \lambda_{\max}$ 
(where $\lambda_{\rm max}$, the longest wavelength for which thermal emission 
is securely measurable, is usually a few~cm) and, (b) the size distribution places most 
of the mass in the largest particles. Objects of meter size and above are essentially 
``dark matter" and unobservable in protoplanetary disks.
\item
Conversion to a {\em gas mass} involves educated guesswork, because several   
physical processes affect the solid and gas components in different ways. For example, radial 
drift (see \S\ref{subsec_drift}) is expected to change the gas to dust ratio as a function of 
both radius and time \cite{youdin04}, while photoevaporation (see \S\ref{sec_photoevaporation}) 
preferentially removes gas from the disk leaving all but the smallest solid 
particles behind.
\end{itemize}
Notwithstanding all these uncertainties, the results of sub-mm disk surveys provide 
highly suggestive hints of the typical environment for planet formation. In Taurus-Auriga, 
Andrews \& Williams \cite{andrews05} deduce a lognormal distribution of disk masses, 
with a mean $M_{\rm disk} \approx 5 \times 10^{-3} \ M_\odot$ and a median 
mass ratio between disk and star of $5 \times 10^{-3}$. Larger masses (for the disk 
{\em plus} envelope) are inferred from modeling of Class~I sources in the same 
region \cite{eisner05}. In Ophiuchus, resolved observations of a sample of relatively 
more massive disks suggest a disk surface density profile $\Sigma \propto r^{-1}$ 
\cite{andrews09}.

\subsection{Molecular line observations}
In addition to the continuum emission in the IR and mm / sub-mm, a number of 
molecules and molecular ions have been detected in line radiation. In the mm / sub-mm the 
workhorse molecule is CO (and its isotopologues such as $^{13} {\rm CO}$ 
and ${\rm C}^{18}{\rm O}$), but several other fairly simple and abundant species including 
CS, HCN, ${\rm N}_2 {\rm H}^+$ and ${\rm DCO}^+$ are observable \cite{dutrey14}. 
Some of the same molecules (albeit at different disk radii) are also observable in 
the near-IR, including CO. The far-IR provides access to additional molecules, 
including water \cite{hogerheijde11,podio13} and ammonia \cite{salinas16}.

Given the difficulty in reliably determining disk masses from dust continuum data, one 
might hope to do better using molecular emission. This is also hard. H$_2$ is a 
homonuclear diatomic molecule with no electric dipole moment, and as a result 
cannot produce a rotational or vibrational spectrum in the dipole approximation \cite{rybicki79}. 
The bulk of the mass in cold H$_2$ within protoplanetary disks is thus observationally 
inaccessible. There are some loopholes. Hot H$_2$ in the atmosphere of the disk (with 
$T > 2000 \ {\rm K}$) can be excited by stellar Ly$\alpha$ radiation, and detected via 
electronic transitions that lie in the far-UV \cite{france12}. These observations, however, 
do not furnish any simple path toward measuring disk masses. More promisingly, emission from HD (specifically 
the $J = 1 \rightarrow 0$ rotational transition at $\lambda \approx 112 \ \mu {\rm m}$) was 
detected using {\em Herschel} data for the TW~Hya system \cite{bergin13}. The original 
analysis of this data yielded a disk mass estimate, $M_{\rm disk} \geq 0.05 \ M_\odot$, that 
was surprisingly large given the 3-10~Myr age typically assumed for the star. More recent 
analyses, using different disk models, confirm that the disk mass is high but give somewhat 
lower numbers (Kama et al. find $2.3 \times 10^{-2} \ M_\odot$ \cite{kama16};  Trapman et al. find 
$7.7 \times 10^{-3} \ M_\odot \leq M_{\rm disk} \leq 2.3 \times 10^{-2} \ M_\odot$ \cite{trapman17}). 
Two other disks, GM~Aur and DM~Tau, have HD disk mass estimates \cite{mcclure16}. The 
{\em Herschel} observatory stopped taking data in 2013, leaving us with no ongoing capability for 
further measurements. Determining disk masses using HD is, however, a strong science 
driver for proposed missions sensitive to these far-IR wavelengths.

Other molecules can be used to estimate gas masses. CO itself is not useful as the 
emission is optically thick, but a combination of CO isotopologues can be used instead \cite{williams14}. 
Gas phase CO in disks can be photo-dissociated, freeze-out in cold dense regions, and be 
processed into other species. A subset of these processes can be observationally constrained. 
Qi et al. \cite{qi13}, for example, used imaging in N$_2$H$^+$ to infer that the CO snow line 
in TW~Hya lies at a radius of about 30~AU. In general, however, 
a chemical model is needed to estimate the fraction 
of carbon (typically of the order of 10\%) that resides in observable gas-phase CO \cite{molyarova17}, 
followed by a final step of converting the CO mass to a total gas mass.

Molecular line observations also provide a wealth of kinematic information. The instrumental 
resolution of spectra in the mm / sub-mm is typically better than the expected thermal width 
of molecular lines, so for bright disks that are well-resolved spatially an observation yields a 
``data cube" expressing the line emission as a function of sky position $(x,y)$ and line of 
sight velocity $v_{\rm los}$. At the lowest order, such data can be used to measure the 
rotation profile of the disk gas and hence the mass of the central star. Beyond that  
we may try to measure the contribution that thermal and turbulent broadening make 
to the line profile. The strength of disk turbulence, in particular, is a key 
quantity for both disk evolution (see \S\ref{sec_turbulence}) and planet formation 
(see Willy Kley's contribution in this volume), that desperately needs empirical 
constraint. If the turbulence can be represented as a small-scale fluid process\footnote{Representing 
fluid motions as microturbulence is a standard approximation for stellar atmospheres, but whether 
it is generally valid for disk turbulence is not obvious. Simon et al. \cite{simon15b} showed that it 
works reasonably well in the case where turbulence is driven by the magnetohydrodynamic instabilities discussed 
in \S\ref{sec_transport}.} the summed contribution to the line width is,
\begin{equation}
 \Delta v = \frac{\nu}{c} \sqrt{ \frac{2 k_B T}{\mu m_{\rm H}} + v_{\rm turb}^2},
\end{equation}
where $\mu$ is the molecular weight of the observed species in units of the mass of 
a hydrogen atom $m_H$, $T$ is the temperature, and $v_{\rm turb}$ is a root-mean-square 
estimator of the turbulence. We generally expect turbulence in disks to be subsonic, 
but by observing relatively heavy molecules such as CO or CS it is possible to attain 
sensitivity to $v_{\rm turb}$ values that are significantly below the sound speed. Using 
molecules with differing optical depths (e.g. isotopologues of CO) opens up the possibility 
of mapping $v_{\rm turb}$ as a function of height above the disk mid-plane.

Recent attempts to measure the turbulent velocity have focused on large disks with 
simple kinematics that can be modeled precisely. Flaherty et al. \cite{flaherty15,flaherty17} have analyzed {\em ALMA} 
data for the disk around the nearby A star HD~163296. Using a combination of molecular 
species and transitions (CO, $^{13}$CO, C$^{18}$O and DCO$^+$), the observed data cubes 
were found to be consistent with models that included only orbital motion and thermal 
broadening. No evidence was found for a turbulent contribution. The derived upper limits 
are below a tenth of the sound speed throughout the vertical extent of the disk.  For TW~Hya, 
Teague et al. \cite{teague16} analyzed {\em ALMA} data that included transitions of CO, CN 
and CS. Turbulent velocities in the range of 0.2-0.4~$c_s$ were inferred for this disk on scales 
of around 50-100~AU. These measurements, while important and (for HD~163296) 
provocative to theorists, remain in their infancy. It will be valuable to obtain a larger sample of 
disks, to acquire data in additional molecular lines, and to compare different analysis techniques.

\subsection{Large-scale-structure in disks}
The reliable determination of quantities such as disk lifetimes, masses, and accretion rates has been 
the focus of observational effort for several decades. Although there remain uncertainties the 
strengths and weaknesses of the different methods have been exhaustively litigated, and we think we 
have a decent physical picture for what is going on. The same cannot be said of more recent observations 
that show a variety of large-scale-structure in disks. Very basic questions --- including whether the observed 
structure is an intrinsic property of the fluid dynamics of disks or rather a sign of planet-disk interactions --- 
remain open. In advance of the theoretical discussion in \S\ref{sec_structure_form}, we summarize here 
the main families of observed structures.

\subsubsection{Transition disks}
The term transition disk \cite{espaillat14} is an umbrella for a subset of disks that do not fit neatly into the SED-based 
classification scheme described in \S\ref{subsec_classification}. Strom et al. \cite{strom89} identified 
a number of YSOs that had little-to-no near IR excess (resembling Class~III sources) but robust 
mid- and far-IR emission (resembling Class~II sources). Observations with the {\em Spitzer} space 
telescope showed that this type of SED is by no means uncommon. The exact numbers depend upon the 
adopted definitions but in Taurus, for example, one study found that the 
fraction of disks classified as ``transitional" or ``evolved" (meaning that they are becoming optically thin 
at both near- and mid-IR wavelengths) is about 15\% \cite{luhman10}. The transitional disk class includes 
several well-known systems, including TW~Hya, GM~Aur, IRS~48 and LkCa15, for which a wealth of 
observational data is available.

The geometric interpretation of transition disk SEDs is straightforward. Near-IR excesses originate in the 
inner disk, which in a normal Class~II source is expected to be the region with the highest optical depth. 
Seeing little or no near-IR emission, while mid-IR emission persists, implies a disk with a hole or cavity 
in the dust distribution. This inference is supported by imaging in the sub-mm, which directly reveals 
the presence of dust cavities within transition disks. Andrews et al. \cite{andrews11} used the Submillimeter 
Array (SMA) to image a sample of 12 transition disks in nearby star-forming regions at 880~$\mu$m. 
The found cavities with radii between 15~AU and 73~AU, and estimated that 20-25\% of the brightest 
disks (in mm emission) exhibited such inner clearings.

The cavities in transition disks are for the most part not empty. Most transitional sources are 
accreting, at rates which can be as high as normal Classical T Tauri stars \cite{manara14} 
(though the median accretion rate is probably significantly reduced --- Kim et al. \cite{kim16} 
find a suppression of approximately an order of magnitude for Class~II YSOs in Orion A). 
Moreover, although the defining feature of transitional disks is a cavity in the dust distribution,  
some transition disks' spectra indicate the presence of very low levels of dust close to the 
star (for example GM~Aur \cite{calvet05}). The overall picture is thus one in which a relatively 
unexceptional outer disk is truncated at some cavity radius $r_{\rm cavity}$. Inside the cavity 
the column density of (observable) dust is severely depleted, but gas persists and continues 
to accrete on to the star.

\subsubsection{Rings}
Long baseline observations with {\em ALMA} have enabled imaging of a number of 
protoplanetary disks with an angular resolution as good as 0.025 arcseconds. A 
major surprise has been the discovery of multiple rings of emission in several 
sources. In HL~Tau, continuum imaging at wavelengths between 0.87~mm and 
2.9~mm detects seven pairs of bright / dark ring-like structures at orbital radii 
between 20~AU and 100~AU \cite{alma15}. Qualitatively similar rings, though 
with lesser contrast, are seen between 2.4~AU and 40~AU in TW~Hya \cite{andrews16}. 
TW~Hya's rings are also visible at near-IR wavelengths that probe sub-micron-sized 
dust particles near the disk surface \cite{vanBoekel17}. Other examples of systems 
with rings of dust emission imaged by {\em ALMA} are HD~163296 \cite{isella16}, HD~169142 \cite{fedele17} 
and AA~Tau \cite{loomis17}.

\subsubsection{Azimuthal asymmetries}
Non-axisymmetric structure is also evident in high resolution disk images. One class of non-axisymmetric 
structure is horseshoe-shaped emission in the mm / sub-mm. A prototypical example is
the transition disk system IRS~48. Van der Marel et al. \cite{vdm13} found that the mm-sized particles 
surrounding the cavity at $r \approx 60 \ {\rm AU}$, traced by 0.44~mm continuum emission, were 
strongly concentrated in a crescent-shaped feature on one side of the star. No such asymmetry was 
evident in either the gas or in micron-sized dust traced by mid-IR imaging. A broadly similar 
morphology is observed in HD~142527 \cite{casassus13}, in LkH$\alpha$~330 \cite{isella13}, and 
in a number of other transition disk sources.

Spiral structure represents a separate class of non-axisymmetric features. Spirals are seen 
in the very young protostellar source L1448~IRS3B \cite{tobin16}, in the disk around the young 
star Elias 2-27 \cite{perez16}, and in the transitional disk source MWC~758 
(in high resolution near-IR imaging \cite{grady13}). 

\subsubsection{Interpretation}

The physical origin of the large-scale-structure seen in protoplanetary disks remains unclear. 
Planets orbiting within the gas disk can produce cavities and spirals, and can form local pressure 
maxima and vortices that would trap dust in ring-like or horseshoe-shaped configurations. There is 
no doubt that planets can form or migrate to the large radii where most disk structure is seen --- the 
HR~8799 system has four super-Jovian mass planets with projected orbital separations between 15~AU 
and 70~AU \cite{marois08,marois10}. Some observable disk structure is thus surely of planetary 
origin. Whether all or even most of the currently observed structure is caused by planets is less clear. 
Theoretically, as we will discuss in \S\ref{sec_structure_form}, there are a number of processes that 
may be able to form structures in disks without fully-formed planets. Observationally, direct imaging 
surveys suggest that wide separation planetary systems with planets of roughly Jupiter mass and 
above are not common \cite{galicher16,chauvin15,uyama17}. Empirical constraints on the abundance 
of lower mass planets, however, remain weak.

\begin{svgraybox}
\begin{itemize}
\item The lifetime of disks around Young Stellar Objects is a few Myr. The bulk of this time is spent in the 
relatively isolated Class~II (or Classical T Tauri star) phase, though critical events for subsequent disk 
evolution (such as magnetic field loss) and planet formation (such as planetesimal formation) may start 
earlier while the disk itself is still accreting from an envelope.
\item Key observational diagnostics include the optical / UV spectrum (used to measure the accretion 
rate on to the star), thermal emission from dust in the IR out to mm wavelengths (used to measure 
disk masses and particle properties, albeit with significant uncertainties), and molecular line 
emission (used to measure kinematics, constrain chemical models, and infer the strength of disk 
turbulence).
\item High resolution imaging of protoplanetary disks in both thermal emission and scattered light 
shows that disks can sustain a variety of large-scale-structure, which may be related to intrinsic 
disk processes or to the interaction of planets with the gas and dust within the disk.
\end{itemize}
\end{svgraybox}

\newpage

\section{Disk structure}
\label{sec_structure}

The few Myr lifetime of protoplanetary disks equates to millions of dynamical times in the inner disk and thousands of dynamical times in the outer disk at 100~AU. 
To a first approximation we can treat the disk as evolving slowly through a sequence of axisymmetric static structures 
as mass accretes on to the star and is lost through disk winds, and our first task is to discuss the physics that determines 
those structures. Quantities that we are interested in include the density $\rho(r,z)$, the gas and dust temperatures 
$T(r,z)$ and $T_d (r,z)$, the chemical composition, and the ionization fraction. The density of solid particles 
(``dust") $\rho_d$ is also important, but we will defer saying much about that until we have discussed 
turbulence, radial drift, and the aerodynamic coupling of solids and gas.

\subsection{Vertical and radial structure}

\subsubsection{Vertical structure}
\label{sec_vertical_structure}

\begin{figure}[t]
\includegraphics[width=\columnwidth]{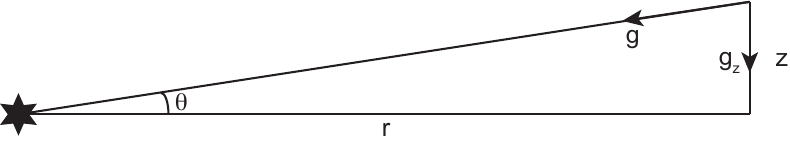}
\caption{The geometry for calculating the vertical hydrostatic equilibrium of a non-self-gravitating protoplanetary 
disk. The balancing forces are the vertical component of stellar gravity and the vertical pressure gradient.}
\label{fig_vertical_geometry}
\end{figure}

The vertical profile of gas density in protoplanetary disks is determined by the condition of hydrostatic equilibrium. 
The simplest case to consider is an optically thick disk that is heated by stellar irradiation, has negligible mass 
compared to the mass of the star, and is supported by gas pressure. We can then approximate the optically 
thick interior of the disk as isothermal, with constant sound speed $c_s$ and pressure $P= \rho c_s^2$. The sound speed is related to the temperature via $c_s^2 = k_{B} T / \mu m_H$, where $k_B$ is Boltzmann's constant, $m_H$ is the mass of a 
hydrogen atom, and where under normal disk conditions the mean molecular weight $\mu \simeq 2.3$. In 
cylindrical co-ordinates, the condition for vertical hydrostatic equilibrium (Figure~\ref{fig_vertical_geometry}) is,
\begin{equation}
 \frac{{\rm d}P}{{\rm d} z} = - \rho g_z = - \frac{GM_*}{r^2 + z^2} \sin \theta \rho,
\label{eq_hydrostatic} 
\end{equation} 
where $M_*$ is the stellar mass. For $z \ll r$, 
\begin{equation}
 g_z = \frac{GM_*}{(r^2 + z^2)^{3/2}} z \simeq \Omega_{\rm K}^2 z,
\end{equation} 
where $\Omega_{\rm K} \equiv \sqrt{GM_*/r^3}$ is the Keplerian orbital velocity (here defined at the mid-plane, later in 
\S\ref{sec_radial_structure} we will need to distinguish between the mid-plane and other locations). Equation~(\ref{eq_hydrostatic}) then 
becomes,
\begin{equation}
 c_s^2 \frac{{\rm d}\rho}{{\rm d} z} = - \Omega_{\rm K}^2 \rho z,
\end{equation}
which integrates to give,
\begin{equation}
 \rho(z) = \rho_0 \exp \left[ -z^2 / 2 h^2 \right],
\label{eq_gaussian} 
\end{equation}
where $\rho_0$ is the mid-plane density and we have defined the {\em vertical scale height} $h \equiv c_s / \Omega_{\rm K}$.  
Because the effective gravity increases with height (and vanishes at the mid-plane) this standard disk profile is 
gaussian, rather than exponential as in a thin isothermal planetary atmosphere. A consequence is that the scale 
over which the density drops by a factor of $e$ gets smaller with $z$; loosely speaking disks become 
more ``two dimensional" away from the mid-plane. Defining the surface density $\Sigma = \int \rho {\rm d}z$, the 
central density is,
\begin{equation} 
 \rho_0 = \frac{1}{\sqrt{2 \pi}} \frac{\Sigma}{h}.
\label{eq_rho0} 
\end{equation} 
Up to straightforward variations due to differing conventions (e.g. some authors define $h = \sqrt{2} c_s / \Omega_{\rm K}$) 
these formulae define the vertical structure of the most basic disk model (isothermal, with a gaussian density profile). 
For many purposes it is an adequate description, especially if one is mostly worried about conditions within a few $h$ of the mid-plane, .

\begin{figure}[t]
\includegraphics[width=\columnwidth]{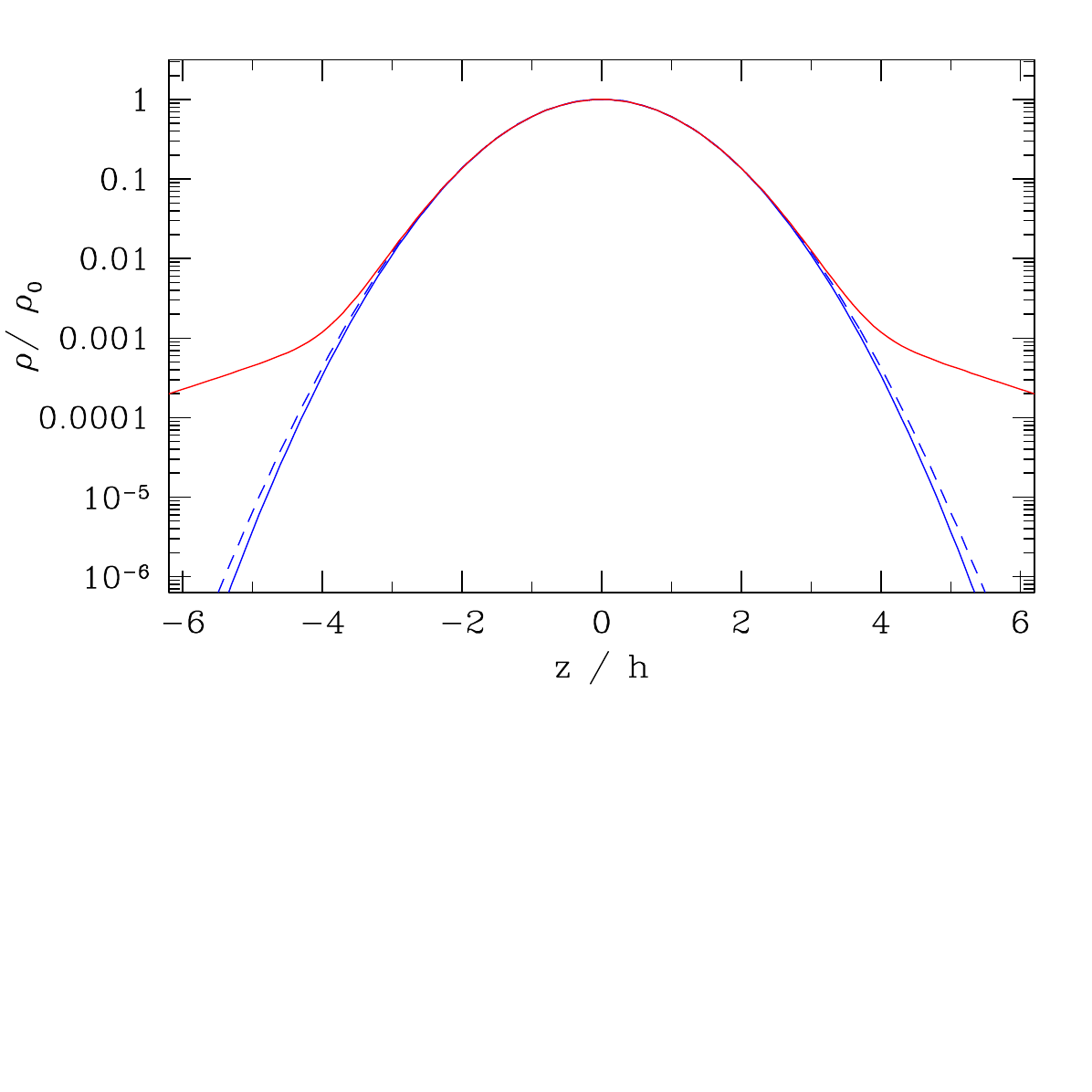}
\vspace{-4.5cm}
\caption{Simple models for the vertical density profile of an isothermal disk, in units of the disk scale height $h$. 
The solid blue line shows the gaussian density profile valid for $z \ll r$, the dashed blue line shows the exact 
solution relaxing this assumption (the two are essentially identical for this disk, with $h/r = 0.05$). The red dashed curve 
shows a fit to numerical simulations that include a magnetic pressure component \cite{hirose11}.}
\label{fig_vertical_density}
\end{figure}

The most obvious cause of gross departures from a gaussian density profile is a non-isothermal temperature 
profile. If the disk is accreting, gravitational potential energy that is thermalized in the optically thick interior will 
need a vertical temperature gradient ${\rm d}T / {\rm d} z < 0$ in order to be transported to the disk photosphere 
and radiated. We will return to this effect in \S\ref{subsec_thermal} and \S\ref{sec_evolution}, after assessing  
the other assumptions inherent in equation~(\ref{eq_gaussian}). We first note that the simplification to $z \ll r$ is 
convenient but not necessary, and that we can integrate equation~(\ref{eq_hydrostatic}) without this assumption to 
give,
\begin{equation}
 \rho = \rho_0 \exp \left[ \frac{r^2}{h^2} \left( \left( 1 + z^2 / r^2 \right)^{-1/2} -1 \right) \right].
\end{equation} 
Protoplanetary disks {\em are} geometrically thin, however, with $h/r \approx 0.05$ being fairly typical. 
In this regime, as is shown in Figure~\ref{fig_vertical_density}, departures from a 
gaussian are negligibly small. We only really need to worry about the full expression for vertical 
gravity when considering disk winds, which flow beyond $z \sim r$.

What about the contribution of the disk itself to the vertical component of gravity? Approximating the disk 
as an infinite sheet with (constant) surface density $\Sigma$, Gauss' theorem tells us that the gravitational 
acceleration above the sheet is independent of height,
\begin{equation}
 g_z = 2 \pi G \Sigma.
\end{equation}
Comparing this acceleration with the vertical component of the star's gravity at $z=h$, we find that the disk 
dominates if,
\begin{equation} 
 \Sigma > \frac{M}{2 \pi} \frac{h}{r^3}.
\end{equation}  
Very roughly we can write the disk mass $M_{\rm disk} \sim \pi r^2 \Sigma$, which allows us to write the 
condition for the disk's own gravity to matter as,
\begin{equation}
 \frac{M_{\rm disk}}{M_*} > \frac{1}{2} \left( \frac{h}{r} \right).
\end{equation} 
For $h/r = 0.05$, a disk mass of a few percent of the stellar mass makes a non-negligible 
change to the vertical structure, and such masses are not unreasonably large. As we will see in \S\ref{sec_turbulence}, 
however, when disk masses $M_{\rm disk} / M_* \sim h/r$ are encountered we tend to have bigger 
fish to fry, as this is also the approximate condition for the onset of disk self-gravity, the formation of 
spiral arms, and substantial departures from axisymmetry.

Magnetic pressure $P_B = B^2 / 8 \pi$ is likely to impact the vertical density profile, at least for $z \gg h$ (here and 
subsequently, we use units such that $B$ is measured in Gauss). No simple principle for predicting the strength 
or vertical variation of the magnetic field is known, so we turn to numerical simulations for guidance. Hirose \& 
Turner \cite{hirose11} completed radiation magnetohydrodynamic (MHD) simulations of the protoplanetary disk 
at 1~AU, adopting fairly typical numbers for the key disk and stellar parameters (a disk surface density $\Sigma = 10^3 \ {\rm g \ cm}^{-2}$, 
stellar mass $M_* = 0.5 \ M_\odot$, stellar effective temperature $T_{\rm eff} = 4000 \ {\rm K}$, and stellar radius $R_* = 2 \ R_\odot$). 
Their simulations included Ohmic diffusion but ignored both ambipolar diffusion and the Hall effect (see \S\ref{sec_turbulence} 
for further discussion of these processes). They found that the model 
disk was gas pressure dominated near the mid-plane, but that the atmosphere (or corona) was magnetically 
dominated. An empirical fit to their density profile is \cite{turner14},
\begin{equation}
 \rho = \frac{\rho_0}{1 + \epsilon} \left[ \exp \left( - z^2 / 2 h^2 \right) + \epsilon \exp \left( - |z| / k h \right) \right],
\end{equation}
with $\epsilon \simeq 1.25 \times 10^{-2}$ and $k \simeq 1.5$. This fit is shown in Figure~\ref{fig_vertical_density}. 
Magnetic pressure beats out gas pressure for $|z| > 4 h$, leading to a low density exponential atmosphere that is much 
more extended than a standard isothermal disk. Even if the atmosphere itself gives way to a disk wind at still higher 
altitudes (as suggested by other simulations), these results suggest that observational probes of conditions near the 
disk surface may be sampling regions where the magnetic field dominates. Moreover, 
simulations of the inner disk ($r \sim 1 \ {\rm AU}$) that include the Hall effect \cite{lesur14} show that it may be possible to 
generate strong azimuthal magnetic fields whose magnetic pressure may exceed that of the gas within one 
scale height, or even at the mid-plane. Such fields would lead to larger departures from the standard purely 
thermal definition of the scale height.

\subsubsection{Radial structure}
\label{sec_radial_structure}

The radial run of the surface density cannot be predicted from considerations of static disk structure, because two dimensional 
equilibrium density distributions $\rho(r,z)$ can be constructed for a broad class of surface density profiles. Not all such 
distributions would be stable against rapid hydrodynamic instabilities, but even if we enforce stability as an additional 
requirement we have no way to discriminate between commonly adopted surface density profiles (e.g. $\Sigma \propto r^{-1}$ 
versus $\Sigma \propto r^{-3/2}$). Instead, the surface density profile must either be measured observationally \cite{andrews09} or 
studied using time-dependent models (\S\ref{sec_evolution})\footnote{The {\em minimum mass Solar Nebula} (MMSN) \cite{weidenschilling77b,hayashi81}, an approximate lower bound for the amount of disk gas needed to form the planets in the Solar System, can be useful as a reference model despite its tenuous connection to actual conditions in the disk at the time of planet formation. The MMSN has a gas surface density profile $\Sigma(r) = 1.7 \times 10^3 (r / AU)^{-3/2} \ {\rm g \ cm^{-2}}$.}. 
Given an assumed surface density profile, however, we can derive useful results for 
the azimuthal velocity $v_\phi$. If the disk is static (and even if it is slowly evolving) the azimuthal component of the 
momentum equation,
\begin{equation}
 \frac{\partial {\bf v}}{\partial t} + \left( {\bf v} \cdot \nabla \right) {\bf v} = -\frac{1}{\rho} \nabla P - \nabla \Phi
\label{eq_S1_momentum} 
\end{equation} 
can be written in the mid-plane as,
\begin{equation}
 \frac{v_\phi^2}{r} = \frac{GM_*}{r^2} + \frac{1}{\rho} \frac{{\rm d}P}{{\rm d}r}.
\label{eq_force_balance} 
\end{equation}
Here $P$ is the pressure and all quantities are mid-plane values. Let's start with an explicit example of the 
consequences of this force balance in protoplanetary disks. Consider a disk with $\Sigma \propto r^{-1}$ and 
central temperature $T_c \propto r^{-1/2}$. We then have $c_s \propto r^{-1/4}$, $\rho \propto r^{-9/4}$ and 
$P \propto r^{-11/4}$. Substituting into equation~(\ref{eq_force_balance}) yields, 
\begin{equation}
 v_\phi = v_{\rm K} \left[ 1 - \frac{11}{4} \left( \frac{h}{r} \right)^2 \right]^{1/2}.
\label{eq_kepler_deviation} 
\end{equation} 
From this we deduce,
\begin{itemize}
\item
The deviation from strict Keplerian rotation, $v_{\rm K} = \sqrt{GM_*/r}$, is of the order of $(h/r)^2$.
\item
Its magnitude is {\em small}. For a disk with $h/r = 0.03$ at 1~AU, the difference between the disk 
azimuthal velocity and the Keplerian value is about 0.25\%, or, in absolute terms $| v_\phi - v_{\rm K} | \simeq 
70 \ {\rm m \ s}^{-1}$.
\end{itemize}
When we come to discuss the evolution of particles within disks (\S\ref{sec_single}), it will turn out that 
this seemingly small effect is of paramount importance. Particles do not experience the radial pressure 
gradient that is the cause of the mismatch in speeds, and as a result develop a differential velocity 
with respect to the gas that leads to aerodynamic drag and (usually) inspiral. Because this process is 
so important it is worth studying not just 
the magnitude of the effect but also its vertical dependence. To do so, we follow Takeuchi \& Lin \cite{takeuchi02} 
and consider an axisymmetric vertically isothermal disk supported against gravity by gas pressure. The vertical 
density profile is then gaussian (equation~\ref{eq_gaussian}) and in equilibrium (equation~\ref{eq_S1_momentum}) 
we have,
\begin{equation}
 r \Omega_{\rm g}^2 = \frac{GM_*}{\left( r^2 + z^2 \right)^{3/2}} r + \frac{1}{\rho} \frac{\partial P}{\partial r}.
\end{equation}
We distinguish between the gas angular velocity, $\Omega_{\rm g} (r,z)$, the Keplerian angular velocity 
$\Omega_{\rm K} (r,z) = GM_* / (r^2 + z^2)^{3/2}$, and its mid-plane value $\Omega_{\rm K, mid}$. The disk 
is fully specified by the local power-law profiles of surface density and temperature,
\begin{eqnarray} 
 \Sigma & \propto & r^{-\gamma} \\
 T_c & \propto & r^{-\beta},
\end{eqnarray}
with $\gamma = 1$ and $\beta = 1/2$ being typically assumed values. Evaluating $\partial P / \partial r$ 
using equation~(\ref{eq_gaussian}) with $h = h(r)$ allows us to determine the equilibrium gas angular velocity 
in terms of the mid-plane Keplerian value,
\begin{equation}
 \Omega_{\rm g} \simeq \Omega_{\rm K,mid} 
 \left[ 1 - \frac{1}{4} \left( \frac{h}{r} \right)^2 \left( \beta + 2 \gamma + 3 + \beta \frac{z^2}{h^2} \right) \right].
\label{eq_vertical_shear}
\end{equation} 
Provided that the temperature is a locally decreasing function of radius ($\beta > 0$), the sense of the 
vertical shear is that the gas rotates slower at higher $z$. Like the sub-Keplerian mid-plane velocities, 
the magnitude of the shear is only of the order of $(h/r)^2$, but this small effect may nevertheless be detectable 
with {\sc ALMA} data \cite{rosenfeld13}. 

\begin{figure}[t]
\includegraphics[width=\columnwidth]{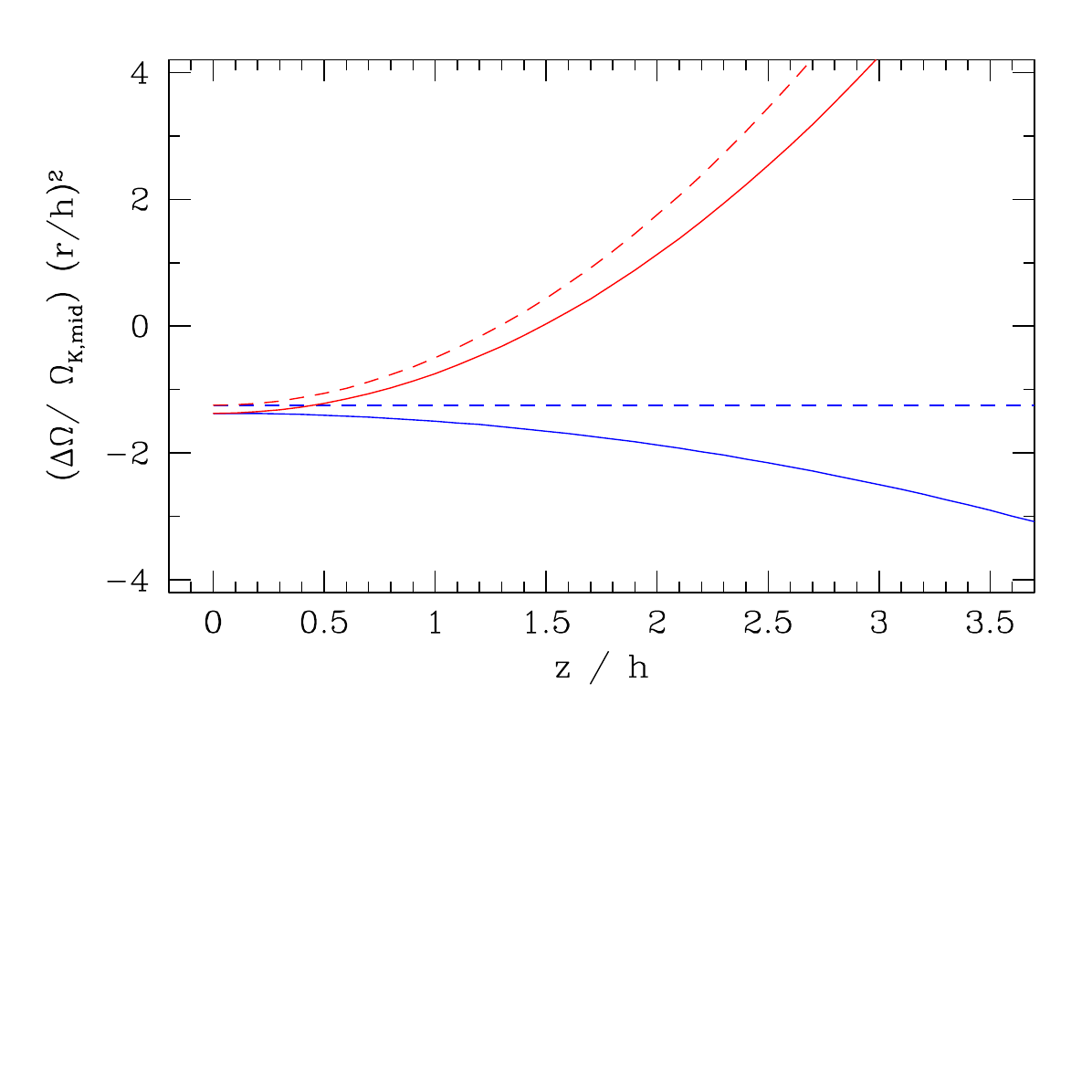}
\vspace{-4.5cm}
\caption{The variation in angular velocity with height in the disk. In blue, the angular velocity of gas 
relative to the mid-plane Keplerian value, $(\Omega_{\rm g} - \Omega_{\rm K, mid}) / \Omega_{\rm K, mid} \times (r/h)^2$.  In red, 
the difference between the Keplerian angular velocity and the {\em local} gas angular velocity, 
$(\Omega_g - \Omega_{\rm K}) / \Omega_{\rm K, mid} \times (r/h)^2$. The assumed disk 
has $\Sigma \propto r^{-1}$ and $T_c \propto r^{-1/2}$ (solid curves) or radially constant $T_c$ (dashed curves).}
\label{fig_vertical_omega}
\end{figure}

For particle dynamics, what matters is the difference between the gas velocity and the local Keplerian speed. 
To order $z^2 / r^2$, the vertical dependence of the Keplerian velocity is,
\begin{equation}
 \Omega_{\rm K} \simeq \Omega_{\rm K, mid} \left( 1 - \frac{3}{4} \frac{z^2}{r^2} \right).
\end{equation}
This function decreases faster with height than $\Omega_{\rm g}$, so the difference between them, plotted in 
Figure~\ref{fig_vertical_omega}, switches 
sign for sufficiently large $z$,
\begin{equation}
 \Omega_{\rm g} - \Omega_{\rm K} \simeq -\frac{1}{4} \left( \frac{h}{r} \right)^2 
 \left[ \beta + 2 \gamma + 3 + \left( \beta - 3 \right) \frac{z^2}{h^2} \right] \Omega_{\rm K, mid}.
\end{equation}  
Particles that orbit the star at the local Keplerian speed move slower than the gas near the mid-plane (and thus 
experience a ``headwind"), but faster at high altitude. For typical parameters, the changeover occurs at about 
$z \approx 1.5 h$.

The orbital velocity will also deviate from the point mass Keplerian form if the disk mass is sufficiently high. 
The gravitational potential of a disk is {\em not} that of a point mass (and does not have a simple form 
for realistic disk surface density profiles), but for an approximation we assume that it is. Then the 
modified Keplerian velocity depends only upon the enclosed disk mass,
\begin{equation}
 v_{\rm K}^\prime \simeq v_{\rm K} \left( 1 + \frac{M_{\rm disk}}{M_*} \right)^{1/2}.
\end{equation} 
For disk masses $M_{\rm disk} \sim 10^{-2} \, M_*$ the effect on the rotation 
curve is of comparable magnitude (but opposite sign) to the effect of the radial pressure gradient.

The deviations from Keplerian rotation due to pressure gradients in a planar axisymmetric disk are 
relatively subtle effects. Larger kinematic departures are possible if the disk is either eccentric 
or warped. Observationally, there are known examples of strongly warped disks (such as HD~142527 
\cite{casassus15}) that can be traced using molecular line observations. Theoretically, we might 
expect disks formed from the collapse of turbulent molecular cloud cores to start out kinematically 
disturbed, and the rate of 
decay such disturbances remains a subject of active research \cite{barker14}. 
It's thus worth remembering, especially when interpreting precise kinematic observations, that un-modeled 
warps or eccentricities as small as $e \sim (h/r)^2$ could be significant. 

\subsection{Thermal physics}
\label{subsec_thermal}

We seek to determine the temperature of gas and dust as $f(r,z)$. Our first task is to calculate the 
interior temperature of a disk heated solely by starlight. This is straightforward. At most radii of interest 
the dust opacity is high enough for the disk to be optically thick to both stellar radiation and to its own 
re-emitted radiation, which hence has a thermal spectrum. It is then a geometric problem to work out 
how much stellar radiation each annulus of the disk intercepts, and what equilibrium $T$ results. We will then 
consider the temperatures of gas and dust in the surface layers of disks. These problems are trickier. 
The surface layers are both optically thin and of low density, 
so we have to account explicitly for the heating and cooling processes and allow for the possibility that 
the dust and gas are too weakly coupled to maintain the same temperature. We defer until \S\ref{sec_evolution} 
the question of how accretion heating modifies these solutions.

\begin{figure}[t]
\includegraphics[width=\columnwidth]{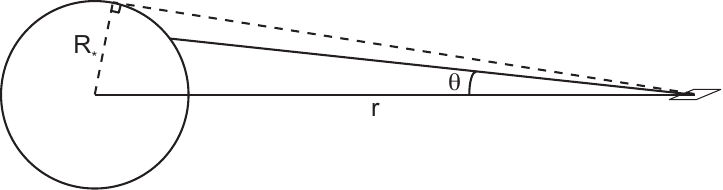}
\caption{The setup for calculating the radial temperature distribution of an optically thick, razor-thin disk. We 
consider a ray that makes an angle $\theta$ to the line joining the area element to the center of the star.
Different rays with the same $\theta$ are labeled with the azimuthal angle $\phi$; $\phi=0$ corresponds to the 
``twelve o'clock" position on the stellar surface.}
\label{fig_sed_razor}
\end{figure}

A disk whose temperature is set by stellar irradiation is described as ``passive". The model problem is a 
flat razor-thin disk that absorbs all incoming 
stellar radiation and re-emits it locally as a blackbody. We seek the temperature of the 
blackbody disk emission as $f(r)$. Modeling the star as a sphere of radius $R_*$, and 
constant  brightness $I_*$, we define
spherical polar coordinates such that the axis of the coordinate 
system points to the center of the star (figure \ref{fig_sed_razor}).  
The stellar flux passing through a surface at distance $r$ is,
\begin{equation} 
 F = \int I_* \sin \theta \cos \phi d \Omega,
\label{eq_C2_fluxintegral} 
\end{equation}
where $d \Omega$ represents the element of solid angle. 
We count the flux coming from the top half of the star only (and 
later equate that to radiation from only the top surface 
of the disk), so the integral has limits,
\begin{eqnarray}
 -\pi / 2 < & \phi & \leq \pi / 2 \nonumber \\
 0 < & \theta & < \sin^{-1} \left( \frac{R_*}{r} \right).
\end{eqnarray} 
Substituting $d\Omega = \sin \theta d\theta d\phi$, the integral 
evaluates to,
\begin{equation}
 F = I_* \left[ \sin^{-1} \left( \frac{R_*}{r} \right) - 
 \left( \frac{R_*}{r} \right) \sqrt{1 - \left( \frac{R_*}{r} \right)^2} \right].
\end{equation} 
A star with effective temperature $T_*$ has brightness $I_* = (1 / \pi) 
\sigma T_*^4$, with $\sigma$ the Stefan-Boltzmann constant.   
Equating $F$ to the one-sided disk emission $\sigma T_{\rm disk}^4$ the  
temperature profile is,
\begin{equation}
 \left( \frac{T_{\rm disk}}{T_*} \right)^4 = \frac{1}{\pi} 
 \left[ \sin^{-1} \left( \frac{R_*}{r} \right) - 
 \left( \frac{R_*}{r} \right) \sqrt{1 - \left( \frac{R_*}{r} \right)^2} \right].
\label{equation_C2_tpassive} 
\end{equation}
The exact result is unnecessarily complicated. To simplify, we expand the right hand side in a Taylor series 
for $(R_* / r) \ll 1$ (i.e. far from the stellar surface) to obtain,
\begin{equation}
 T_{\rm disk} \propto r^{-3/4},
\label{eq_tpassive2} 
\end{equation} 
as the power-law temperature profile of a thin, flat, passive disk. This implies a sound speed profile,
$c_s \propto r^{-3/8}$, and a disk thickness $(h/r) \propto r^{1/8}$. We therefore predict that the disk 
becomes geometrically thicker (``flares") at larger radii. We can also integrate equation~(\ref{equation_C2_tpassive}) 
exactly over $r$, with the result that the luminosity from both side of the disk sums to $L_{\rm disk} / L_* = 1/4$. 

A more detailed calculation of the dust emission from passive disks requires consideration of two additional 
physical effects \cite{kenyon87}. First, as we just noted, the disk thickness as measured by the gas scale 
height flares to larger radii. If dust is well-mixed with the gas --- which may or may not be a reasonable 
assumption --- a flared disk intercepts more stellar radiation in its outer regions than a flat one, which 
will tend to make it flare even more strongly. We therefore need to solve for the self-consistent {\em shape} 
of the disk that simultaneously satisfies hydrostatic and thermal equilibrium at every radius. This is 
conceptually easy, and the slightly messy geometry required to generalize the flat disk calculation is 
clearly described by Kenyon \& Hartmann \cite{kenyon87}. Second, small dust grains that are directly 
exposed to stellar irradiation (i.e. those where the optical depth to stellar radiation along a line toward the star 
$\tau < 1$) emit as a dilute blackbody with a temperature higher than if they were true blackbody emitters \cite{kenyon87}. 
The reason for this is that small dust grains, of radius $s$, have an emissivity $\epsilon = 1$ only for 
wavelengths $\lambda \leq 2 \pi s$. At longer wavelengths, their emissivity declines. The details depend 
upon the composition and structure of the dust grains, but roughly the emissivity (and opacity $\kappa$) 
scale inversely with the wavelength. In terms of temperature,
\begin{equation}
 \epsilon = \left( \frac{T}{T_*} \right)^\beta
\end{equation}
with $\beta = 1$. A dust particle exposed to the stellar radiation field is then in radiative equilibrium at 
temperature $T_{\rm s}$ when absorption and emission are in balance,
\begin{equation}
 \frac{L_*}{4 \pi r^2} \pi s^2 = \sigma T_{\rm s}^4 \epsilon \left( T_{\rm s} \right) 4 \pi s^2.
\end{equation}
The resulting temperature,
\begin{equation}
 T_{\rm s} = \frac{1}{\epsilon^{1/4}} \left( \frac{R_*}{2 r} \right)^{1/2} T_*,
\end{equation}
exceeds the expected blackbody temperature by a substantial factor if $\epsilon \ll 1$.   

An illustrative analytic model that incorporates these effects was developed by Chiang \& Goldreich \cite{chiang97}. 
They considered a disk with a surface density profile $\Sigma = 10^3 (r / {\rm 1 \, AU})^{-3/2} \ {\rm g \ cm}^{-2}$ around a star 
with $M_* = 0.5 \ M_\odot$, $T_* = 4000 \ {\rm K}$ and $R_* = 2.5 \ R_\odot$. Within about 100~AU, their solution has half 
of the bolometric luminosity of the disk emitted as a blackbody at the interior temperature,
\begin{equation}
 T_{\rm i} \approx 150 \left( \frac{r}{\rm 1 \ AU} \right)^{-3/7} \ {\rm K},
\end{equation}
with equal luminosity at each radius emerging from a hot surface dust layer at,
\begin{equation}
 T_{\rm s} \approx 550 \left( \frac{r}{\rm 1 \ AU} \right)^{-2/5} \ {\rm K}.
\end{equation} 
The Chiang \& Goldreich solution is a two-layer approximation to dust continuum radiative transfer for a 
passive, hydrostatic disk. Approximations in the same spirit have been developed that incorporate 
heating due to accretion \cite{garaud07}, but the 
full problem requires numerical treatment. Several codes are 
available for the efficient solution of the full radiative transfer problem \cite{bjorkman01,ercolano05,robitaille11,steinacker13}.

Under most circumstances dust dominates both the absorption of starlight and the thermal emission of 
reprocessed stellar radiation and accretion heating. If the density is high enough, collisions between 
dust particles and gas molecules will establish a common temperature for both, and there is no 
need to explicitly consider the thermal physics of the gas. Kamp \& Dullemond (2004), for example, find that 
$T_g$ and $T_d$ are within about 10\% of each other for an optical extinction 
$A_V > 0.1$ \cite{kamp04}. This criterion will be met in the disk mid-plane within the normal planet forming region 
(i.e. excluding very large orbital distances where the disk is becoming optically thin). 
The gas temperature near the surface of the disk is, however, of critical importance for a 
number of applications,
\begin{itemize}
\item
Interpretation of sub-mm data, where the observable emission is rotational transitions of molecules such 
as CO and HCO$^+$. These observations frequently probe the outer regions of disks, at depths where the 
molecules are not photo-dissociated but where the gas is warm and not in equilibrium with the 
dust.
\item
Interpretation of near-IR and far-IR data, often from the inner disk, where we are seeing ro-vibrational 
transitions of molecules along with fine-structure cooling lines such as [CII] and [OI].
\item
Chemistry. It's cold at the disk mid-plane, and chemical reactions are sluggish. Although the densities are 
much lower in the disk atmosphere, the increased temperatures and exposure to higher energy stellar 
photons make the upper regions of the disk important for chemistry \cite{henning13}.
\end{itemize}
The properties of gas near the surface of disks are very closely tied to the incident flux of ultraviolet 
radiation from the star. Stellar UV radiation ionizes and dissociates atoms and molecules, and heats 
the gas by ejecting electrons from dust grains (grain photoelectric heating). Depending upon the temperature 
and density, the heating is balanced by cooling from rotational transitions of molecules (especially CO) and 
atomic fine structure lines. Also important is energy exchange due to inelastic collisions between gas 
molecules and dust particles (thermal accommodation) --- if this process is too efficient the gas temperature 
will revert to match the dust which is absorbing and emitting the bulk of the star's bolometric luminosity.

Photoelectric heating \cite{draine78} is typically the dominant process for dusty gas exposed to an ultraviolet radiation 
field. The work function of graphite grains (the minimum energy required to free an electron from them) 
is around 5~eV, so 10~eV FUV photons can eject electrons from uncharged grains with 5~eV of kinetic 
energy that ultimately heats the gas. Ejection occurs with a probability of the order of 0.1, so the overall 
efficiency (the fraction of the incident FUV energy that goes into heating the gas rather than the dust) can be 
rather high, around 5\%. 

A detailed evaluation of the photoelectric heating rate is involved, and resistant to a fully first-principles 
calculation. Weingartner \& Draine \cite{weingartner01} give a detailed description. Here, we sketch 
the main principles following Kamp \& van Zadelhoff \cite{kamp01}, who developed models for the 
gas temperature in A star disks. We consider a stellar radiation spectrum $F_\nu$ impinging on 
grains of graphite (work function $w = 4.4 \ {\rm eV}$ \cite{weingartner01}) and silicate ($w = 8.0 \ {\rm ev}$). 
For micron-sized grains the work function, which is a property of the bulk material, is equivalent to the 
ionization potential --- the energy difference between infinity and the highest occupied energy level in the 
solid. Additionally, the probability for absorption when a photon strikes a grain is $Q_{\rm abs} \approx 1$. 
Most absorbed photons, however, do {\em not} eject electrons, rather their energy goes entirely into 
heating the dust grain. The yield of emitted electrons is some function of photon energy $Y(h \nu)$, 
and they have some spectrum of kinetic energy $E$, roughly described by \cite{draine78} 
$f (E, h \nu) \propto ( h \nu - w )^{-1}$. If the grains are charged (e.g. by prior emission of 
photoelectrons) then the kinetic energy $(E - e U)$ available to heat the gas is that left over once the 
electron has escaped the electrostatic potential $eU$ of the grain. The heating rate is then \cite{kamp01},
\begin{equation}
 \Gamma_{pe} = 4 n_H \sigma \int_{E_{\rm min}}^{E_{\rm max}} \left( 
 \int_{\nu_{\rm th}}^{\nu_{\rm max}} Q_{\rm abs} Y (h \nu) f (E, h \nu) F_\nu {\rm d}\nu \right) 
 \left( E - eU \right) {\rm d}E, 
\end{equation} 
where $n_H$ is the number density of hydrogen atoms, $\sigma$ is the geometric cross-section per 
hydrogen nucleus, and the lower limits express the minimum frequency $\nu_{\rm th}$ of a photon 
that can overcome the work function and the minimum energy of a photoelectron that can escape 
from a charged grain. An assessment of the photoelectric heating rate then requires knowledge of the 
functions $Y$ and $f$, specification of the radiation field $F_\nu$, and calculation of the typical 
charge on grains of different sizes \cite{draine87,weingartner01}. The physics is conceptually identical 
but quantitatively distinct when the grains in question are {\em extremely small} (e.g. Polycyclic Aromatic 
Hydrocarbons, PAHs) \cite{bakes94,weingartner01}.

The rate of energy exchange from inelastic gas-grain collisions can be calculated with a collision rate 
argument. Consider grains with geometric cross-section $\sigma_d = \pi \langle s^2 \rangle$ and number 
density $n_d$, colliding with hydrogen atoms with number density $n_H$. The thermal speed of the 
hydrogen atoms is $v_{\rm th} = ( 8 k_B T_g / \pi m_H)^{1/2}$ and the average kinetic energy of the 
molecules on striking the surface is $2 k_B T_g$. The cooling rate per unit volume due to gas-grain 
collisions can then be written in the form \cite{burke83},
\begin{equation}
 \Lambda_{g-d} = n_d n_g \sigma_d \left( \frac{8 k_B T_g}{\pi m_H} \right)^{1/2} 
 \alpha_T \left( 2 k_B T_g - 2 k_B T_d \right),
\end{equation}
where $T_g$ and $T_d$ are the temperatures of the gas and dust respectively. The subtleties 
of the calculation are reflected in the ``accommodation co-efficient" $\alpha_T$, which is 
typically $\alpha_T \approx 0.3$ for silicate and carbon grains. For a specified volumetric 
heating rate (and assumptions as to the gas to dust ratio and properties of the grains), this 
expression can be used to estimate the density below which the thermal properties of gas 
and dust decouple.

\begin{figure}[t]
\includegraphics[width=\columnwidth]{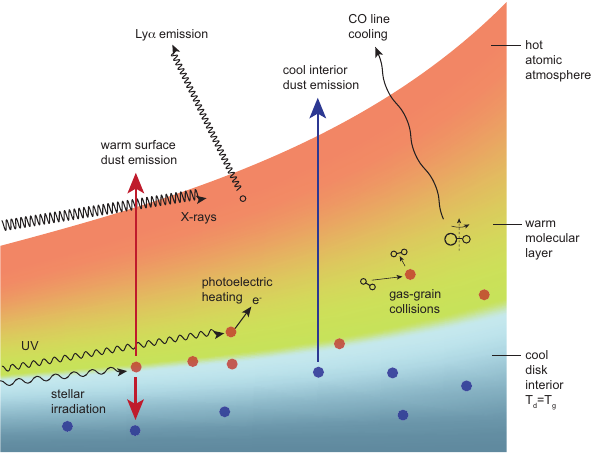}
\caption{Illustration of some of the physical processes determining the temperature and emission properties of 
irradiated protoplanetary disks.}
\label{fig_heating}
\end{figure}

In addition to cooling that occurs indirectly, as a consequence of gas-grain collisions, gas in the 
upper layers of disks also cools radiatively. In the molecular layer of the disk, 
the dominant coolant is typically CO, as this is the most abundant molecule that is not homonuclear 
(diatomic molecules, such as H$_2$, have no permanent electric dipole moment and hence radiate 
inefficiently). At higher temperatures --- only attained in the very rarefied uppermost regions of the 
disk atmosphere --- cooling by Ly$\alpha$ emission becomes important. Qualitatively, there are 
then three distinct layers in the disk:
\begin{itemize}
\item
A cool mid-plane region, where dust and gas have the same temperature and dust cooling is 
dominant.
\item
A warm surface layer in which both dust and gas have temperatures that exceed the mid-plane 
value. The gas in the warm layer can be substantially hotter than the dust ($T \sim 10^3 \ {\rm K}$  
at 1~AU), and cools both by dust-gas collisions and by CO rotational-vibrational transitions.
\item
A hot, low-density atmosphere, where Ly$\alpha$ radiation and other atomic lines (e.g. O[I]) 
cool the gas.
\end{itemize}
The disk structure that results from these heating and cooling processes is illustrated in Figure~\ref{fig_heating}.

\subsection{Ionization structure}
The degree of ionization of the gas in protoplanetary disks is important because it is key to 
understanding how gas couples to magnetic fields, and thence to understanding the role of 
magnetic fields in the formation of disks, in the sustenance of turbulence within them, and in 
the generation of jets and magnetohydrodynamic (MHD) winds. At the most basic level, we 
care about the ratio of the number density of free electrons $n_e$ to the number density of 
neutrals,
\begin{equation}
 x_e \equiv \frac{n_e}{n_n},
\end{equation}
though we should remember that dust grains can also bear charges and carry currents. We 
will consider separately the thermodynamic equilibrium process of {\em thermal} (or collisional) 
ionization, which typically dominates above $T \sim 10^3 \ {\rm K}$, and {\em non-thermal} ionization 
due to photons or particles that have an energy well in excess of the typical thermal energy in the gas.

In anticipation of results that will be derived in \S\ref{sec_turbulence}, we note that very low and 
seemingly negligible levels of ionization --- $x_e \ll 10^{-10}$ -- often suffice to couple magnetic 
fields to the fluid. We need to worry about small effects when considering ionization.

\subsubsection{Thermal ionization}
\label{sec_thermal_ionization}
Thermal ionization of the alkali metals is important in the 
innermost regions of the disk, usually well inside 1~AU. In thermal equilibrium 
the ionization state of a single species 
with ionization potential $\chi$ is obeys the Saha equation 
\cite{rybicki79},
\begin{equation} 
 \frac{n^{\rm ion} n_e}{n} = \frac{2 U^{\rm ion}}{U} 
 \left( \frac{2 \pi m_e k_B T}{h^2} \right)^{3/2} 
 \exp[ - \chi / k_BT ].
\end{equation}
Here, $n^{\rm ion}$ and $n$ are the number densities of the 
ionized and neutral species, and $n_e$ ($=n^{\rm ion}$) is the electron 
number density. The partition functions for the ions and neutrals are 
$U^{\rm ion}$ and $U$, and the electron mass is $m_e$. The 
temperature dependence is not quite just  
the normal exponential Boltzmann factor, because the ionized state is 
favored on entropy grounds over the 
neutral state. 

\begin{figure}[t]
\includegraphics[width=\columnwidth]{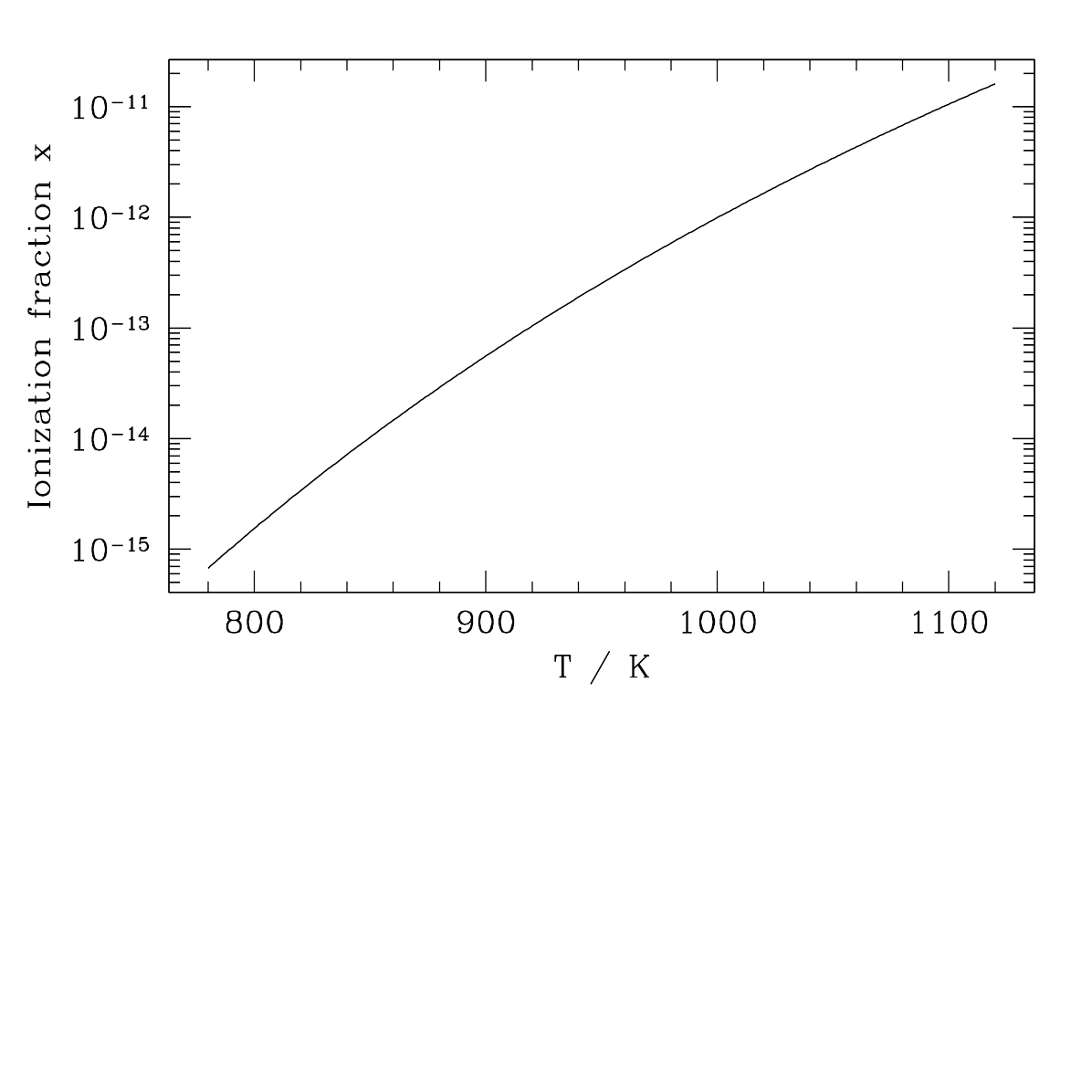}
\vspace{-4.5cm}
\caption{The thermal ionization fraction as a function of temperature predicted 
by the Saha equation for the inner disk. Here we assume that potassium, 
with ionization potential $\chi = 4.34 \ {\rm eV}$ and fractional abundance 
$f = 10^{-7}$, is the only element of interest for the ionization. 
The number density of neutrals is taken to be $n_n = 10^{15} \ {\rm cm}^{-3}$.}
\label{fig_saha}
\end{figure}

In protoplanetary disks thermal ionization 
becomes significant when the temperature becomes high enough to start ionizing alkali 
metals. For potassium, the ionization potential $\chi = 4.34 \ {\rm eV}$. 
We write the abundance of potassium relative to all other neutral 
species as $f = n_K / n_n$, and define the ionization fraction $x$, 
\begin{equation}
 x \equiv \frac{n_e}{n_n}.
\label{eq_C2_efraction} 
\end{equation} 
While potassium remains weakly ionized, the Saha equation gives,
\begin{equation}
 x \simeq 10^{-12} \left( \frac{f}{10^{-7}} \right)^{1/2} 
 \left( \frac{n_n}{10^{15} \ {\rm cm}^{-3}} \right)^{-1/2} 
 \left( \frac{T}{10^3 \ {\rm K}} \right)^{3/4} 
 \frac{ \exp[-2.52 \times 10^4 / T]}{1.14 \times 10^{-11}}
\end{equation}
where the final numerical factor in the denominator is the 
value of the exponent at $10^3$~K. 
The ionization fraction at different temperatures is shown in 
Figure~\ref{fig_saha}. Ionization fractions 
large enough to be interesting for studies of magnetic 
field coupling are reached at temperatures of $T \sim 10^3 \ {\rm K}$ 
although the numbers remain extremely small -- of the order 
of $x \sim 10^{-12}$ for these parameters.  

\subsubsection{Non-thermal ionization}
\label{sec_non_thermal}

Outside the region close to the star where thermal ionization is possible, any remnant levels 
of ionization are controlled by non-thermal processes. Considerations of thermodynamic 
equilibrium are not relevant, and we need to explicitly balance the rate of ionization by 
high-energy particles or photons against the rate of recombination within the disk gas.

There are several potentially important sources of ionization. Ordering them roughly in 
order of their penetrating power, ideas that have been suggested include,
\begin{itemize}
\item
Ultraviolet photons (from the star, or from other stars in a cluster)
\item
Stellar X-rays
\item
Cosmic rays
\item
Energetic protons from a stellar corona \cite{turner09}
\item
Particles produced from radioactive decay of nuclides within the disk \cite{stepinski92}
\item
Electric discharges. \cite{muranushi12}
\end{itemize}
We will limit our discussion to the first three of these processes. Ionization due to radioactive decay, 
although undoubtedly present, leads to a very small electron fraction, while the level of ionization due to 
energetic protons and electric discharges is quite uncertain\footnote{Note that the amount of 
{\em power} involved in any of these non-thermal ionization processes is rather small when compared to that liberated by 
accretion \cite{inutsuka05}. Any additional processes that could convert even a small fraction of the accretion energy 
into non-thermal particles would likely matter for the ionization state.}.

The coronae of T~Tauri stars are powerful sources of keV X-rays \cite{preibisch05}. 
Typical luminosities are $L_X \simeq 10^{28} - 10^{31} \ {\rm erg \ s}^{-1}$, in X-rays with 
temperatures $k_B T_X$ of a few keV. The physics of the interaction of these X-rays 
with the disk gas involves Compton scattering and absorption by photo-ionization, 
which has a cross-section $\sigma \sim 10^{-22} \ {\rm cm}^2$ for keV energies, 
decreasing with photon energy roughly as $E_{\rm phot}^{-3}$. Given an input 
stellar spectrum and assumptions as to where the X-rays originate, the scattering and 
absorption physics can be calculated using radiative transfer codes to deduce the 
ionization rate within the disk \cite{ercolano13}.

Depending upon the level of detail needed for a particular application, the results of 
numerical radiative transfer calculations can be approximated analytically. For a 
relatively hard stellar spectrum ($k_B T_X = 5 \ {\rm keV}$), the ionization rate 
fairly deep within the disk scales with radius $r$ and vertical column from the 
disk surface $\Sigma$ as $r^{-2} \exp[-\Sigma / {8 \ {\rm g \ cm}^{-2}}]$ \cite{turner08}. 
A more detailed fit is given by Bai \& Goodman (2009) \cite{bai09}. For an X-ray 
luminosity scaled to $L_{X,29} = L_X / 10^{29} \ {\rm erg \ s}^{-1}$ they represent 
the numerical results with two components,
\begin{equation}
 \frac{\zeta_X}{L_{X,29}} \left( \frac{r}{1 \ {\rm AU}} \right)^{-2.2} = 
 \zeta_1 \exp[ - (\Sigma / \Sigma_1)^\alpha ] + 
 \zeta_2 \exp[ - (\Sigma / \Sigma_2)^\beta ] + ...
\label{eq_bai_ionization} 
\end{equation}
where $\Sigma$ is the vertical column density from the top of the disk and symmetric terms 
in the column density from the bottom of the disk are implied. For $k_B T_X = 3 \ {\rm keV}$ and Solar 
composition gas the 
fit parameters are $\zeta_1 = 6 \times 10^{-12} \ {\rm s}^{-1}$, $\zeta_2 = 10^{-15} \ {\rm s}^{-1}$, 
$\Sigma_1 = 3.4 \times 10^{-3} \ {\rm g \ cm}^{-2}$, $\Sigma_2 = 1.59 \ {\rm g \ cm}^{-2}$, 
$\alpha = 0.4$ and $\beta = 0.65$. For $k_B T_X = 5 \ {\rm keV}$ the 
fit parameters are $\zeta_1 = 4 \times 10^{-12} \ {\rm s}^{-1}$, $\zeta_2 = 2 \times 10^{-15} \ {\rm s}^{-1}$, 
$\Sigma_1 = 6.8 \times 10^{-3} \ {\rm g \ cm}^{-2}$, $\Sigma_2 = 2.27 \ {\rm g \ cm}^{-2}$, 
$\alpha = 0.5$ and $\beta = 0.7$.

\begin{figure}[t]
\includegraphics[width=\columnwidth]{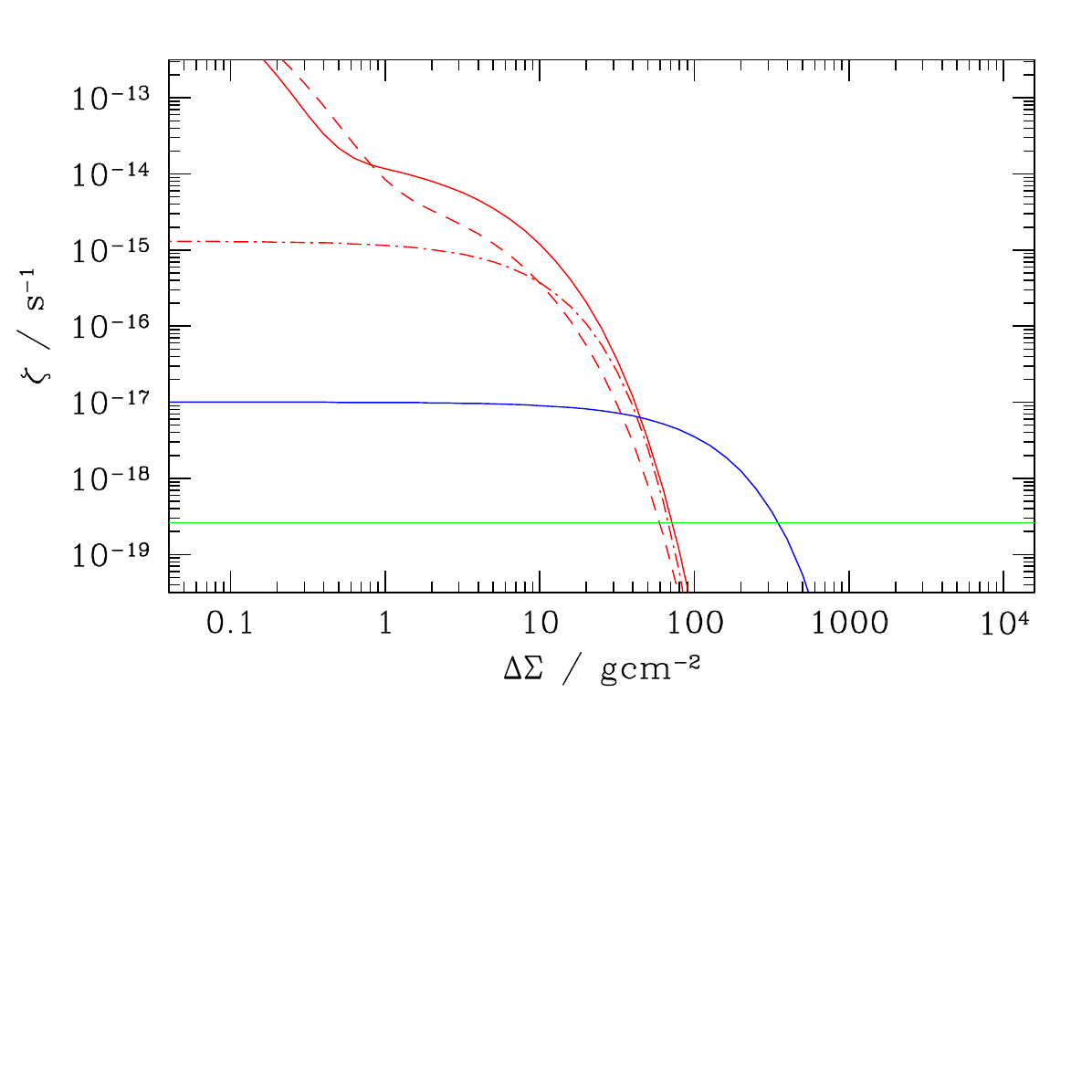}
\vspace{-4.5cm}
\caption{Estimates of the non-thermal ionization rate due to X-rays (red curves), unshielded cosmic 
rays (blue) and radioactive decay of short-lived nuclides (green), plotted as a function of the vertical 
column density from the disk surface. The solid red curve shows the Bai \& Goodman (2009) result 
for an X-ray temperature $k_B T_X = 5 \ {\rm keV}$, the dashed curve their result for $k_B T_X = 3 \ {\rm keV}$. 
The dot-dashed red curve shows a simpler formula proposed by Turner \& Sano (2008). All of the X-ray 
results have been normalized to a flux of $L_X = 10^{30} \ {\rm erg \ s}^{-1}$ and a radius of 1~AU.}
\label{fig_ionization}
\end{figure}

Figure~\ref{fig_ionization} shows the estimates for $\zeta(\Sigma)$ for a stellar X-ray luminosity of 
$L_X = 10^{30} \ {\rm erg \ s}^{-1}$ incident on the disk at 1~AU. If one is mainly interested in 
regions of the disk more than $\approx 10 \ {\rm g \ cm}^{-2}$ away from the surfaces the single 
exponential fit given by Turner \& Sano (2008) \cite{turner08} may suffice. The more complex 
fitting function given by equation~(\ref{eq_bai_ionization}) captures the much higher rates of 
ionization due to X-rays higher up in the disk atmosphere.

Cosmic rays are another potential source of disk ionization. A standard description of the 
interstellar cosmic ray flux gives them an unattenuated ionization rate of $\zeta_{CR} \sim 10^{-17} - 
10^{-16} \ {\rm s}^{-1}$ and an exponential stopping length of $96 \ {\rm g \ cm}^{-2}$ (substantially 
greater than even high energy stellar X-rays)\footnote{Umebayashi \& Nakano noted that if 
cosmic rays have an approximately isotropic angular distribution at the disk surface, geometric 
effects lead to a faster than exponential attenuation deep in the disk \cite{umebayashi09}.}. With 
these parameters, X-rays would remain the 
primary source of ionization in the upper $\approx 50 \ {\rm g \ cm}^{-2}$ of the disk, but cosmic 
rays would dominate in the region between about 50 and $500 \ {\rm g \ cm}^{-2}$. It is unclear, 
however, whether the unattenuated interstellar medium flux of cosmic rays typically reaches the 
surfaces of protoplanatary disks. The magnetic fields embedded in the {\em Solar} wind form a 
partial barrier to incoming cosmic ray particles, whose effect is seen in a modulation of the 
observed flux with the Solar cycle. T~Tauri stars could have much stronger stellar winds 
that exclude cosmic rays efficiently. Indeed, chemical modeling of molecular line data suggests that 
cosmic rays are substantially excluded (to a level $\zeta_{CR} \sim 10^{-19}$) from the disk around the 
nearby star TW~Hya \cite{cleeves15}, though 
how pervasive this phenomenon is remains unknown. If cosmic rays are not present, 
the only guaranteed source of ionization at columns more than $\approx 100 \ {\rm g \ cm}^{-2}$ 
away from the disk surfaces is radioactive decay.

If our main interest is in conditions at $r \sim 1 \ {\rm AU}$ the surface density in gas is 
typically $\Sigma \sim 10^3 \ {\rm g cm}^{-2}$ and X-rays, which are our main concern, will 
not reach the mid-plane. The situation is different further out. At 100~AU 
typical surface densities are much lower --- $1 \ {\rm g \ cm}^{-2}$ might be reasonable --- 
and X-rays will sustain a non-zero rate of ionizations throughout that column. On these scales 
ultraviolet photons can also be important. Stellar FUV radiation will ionize carbon and sulphur 
atoms near the disk surface, yielding a relatively high electron fraction $x_e \sim 10^{-5}$. 
The ionized skin that results is shallow, penetrating to a vertical column of just 
$0.01 - 0.1 \ {\rm g \ cm}^{-2}$, but enough to be significant in the tenuous outer disk \cite{perezbecker11}.

As with ionization, the rate of recombination within the disk can be calculated from complex 
numerical models that track reactions (often numbering in the thousands) between dozens 
of different species. The following discussion, which borrows heavily from the description given by Ilgner \& 
Nelson (2006) \cite{ilgner06} and Fromang (2013) \cite{fromang13}, is intended only to outline 
some of the important principles. At the broadest level of discussion we need to consider 
gas-phase recombination reactions (involving molecular and gas-phase metal ions) along 
with recombination on the surface of dust grains.

The principles of gas-phase recombination can be illustrated by considering the possible 
reactions between electrons and generic molecules ${\rm m}$ and metal atoms ${\rm M}$ \cite{oppenheimer74,ilgner06}. 
The basic reactions are then,
\begin{itemize}
\item
Ionization,
\begin{equation}
 {\rm m} \rightarrow {\rm m}^+ + {\rm e}^-,
\end{equation}
with rate $\zeta$. A specific example is ${\rm H}_2 \rightarrow {\rm H}_2^+ + {\rm e}^-$. 
\item
Recombination with molecular ions,
\begin{equation}
 {\rm m}^+ + {\rm e}^- \rightarrow {\rm m},
\end{equation}
with rate $\alpha = 3 \times 10^{-6} T^{-1/2} \ {\rm cm^3 \ s^{-1}}$. An example is the 
dissociative recombination reaction ${\rm HCO}^+ + {\rm e}^- \rightarrow {\rm CO} + {\rm H}$. 
\item
Recombination with gas-phase metal ions,
\begin{equation}
 {\rm M}^+ + {\rm e}^- \rightarrow {\rm M} + h \nu,
\end{equation}
with rate $\gamma = 3 \times 10^{-11} T^{-1/2}  \ {\rm cm^3 \ s^{-1}}$. An example is 
${\rm Mg}^+ + {\rm e}^- \rightarrow {\rm Mg} + h \nu$.
\item
Charge exchange reactions,
\begin{equation}
 {\rm m}^+ + {\rm M} \rightarrow {\rm m} + {\rm M}^+,
\end{equation}
with rate $\beta = 3 \times 10^{-9}  \ {\rm cm^3 \ s^{-1}}$. An example is 
${\rm HCO}^+ + {\rm Mg} \rightarrow {\rm Mg}^+ + {\rm HCO}$.
\end{itemize}
From such a set of reactions we form differential equations describing the time 
evolution of the number density of species involved. For the molecular abundance 
$n_{\rm m}$, for example, we have,
\begin{equation}
 \frac{ {\rm d}n_{\rm m} }{ {\rm d}t } = 
 -\zeta n_{\rm m} + 
 \alpha n_e n_{\rm m^+} + 
 \beta n_{\rm M} n_{\rm m^+} = 0,
\label{eq_recombination_ode} 
\end{equation}
where the second equality follows from assuming that the system has reached equilibrium. 
The resulting system of algebraic equations has simple limiting solutions. For example, if 
there are no significant reactions involving metals then the above equation, together with the 
condition of charge neutrality ($n_{\rm m^+} = n_e$), gives an electron fraction 
$x_e = n_e / n_{\rm m}$,
\begin{equation}
 x_e = \sqrt{ \frac{\zeta}{\alpha n_{\rm m} } }.
\end{equation}
In the more general case, the network yields a cubic equation which can be solved for the 
electron fraction as a function of the gas-phase metal abundance \cite{ilgner06}. Typically 
the presence of metal atoms and ions is important for the ionization level.

Recombination can also occur on the surfaces of dusty or icy grains. The simplest 
reactions we might consider are,
\begin{eqnarray}
 {\rm e}^- + {\rm gr} & \rightarrow & {\rm gr}^- \nonumber \\
 {\rm m}^+ + {\rm gr}^- & \rightarrow & {\rm gr} + {\rm m}.
\end{eqnarray} 
If the first of these reactions is rate-limiting, then we can write a modified version 
of the ordinary differential equation (equation~\ref{eq_recombination_ode}) that 
includes grain processes. Ignoring metals for simplicity,
\begin{equation}
  \frac{ {\rm d}n_{\rm m} }{ {\rm d}t } = 
 -\zeta n_{\rm m} + 
 \alpha n_e n_{\rm m^+} + 
 \sigma v_e n_{\rm gr} n_{\rm e^-},
\end{equation}
where $\sigma$ is the cross-section of grains to adhesive collisions with free 
electrons and $v_e$ is the electron thermal velocity. In the limit where {\em only} 
grains contribute to recombination we then find,
\begin{equation}
 x_e = \frac{\zeta}{\sigma v_e n_{\rm gr}}.
\end{equation}  
If the grains are mono-disperse with radius $s$, then $x_e \propto 1 / \sigma n_{\rm gr} \propto s$, 
and recombination on grain surfaces will be more important for small grain sizes. We also note 
that the dependence on the ionization rate $\zeta$ is linear, rather than the square root dependence 
found in the gas-phase case. As for metals, grain populations with commonly assumed size distributions are 
found to matter for the ionization level.

The above discussion of recombination leaves a great deal unsaid. For grains, an important 
additional consideration is related to the typical charge state, which needs to be calculated \cite{draine87}. 
A good comparison of different networks for the calculation of the ionization state is given by 
Ilgner \& Nelson (2006) \cite{ilgner06}, while Bai \& Goodman (2009) \cite{bai09} provide a clear discussion 
of the important processes. In wading into this literature the reader who encounters a problem involving 
the ionization level is advised to first evaluate whether a simple 
analytic approximation is adequate for their application, or whether solution of a full chemical network is required. 
The accuracy needed from calculations of ionization equilibrium is strongly problem-dependent, and in some 
cases, such as if we don't know if cosmic rays are present for a particular system, high accuracy may be illusory. 

\begin{svgraybox}
\begin{itemize}
\item The vertical density profile of disks is determined by hydrostatic equilibrium. The simplest (vertically 
isothermal) model yields a gaussian density profile. More complex models need to incorporate vertical thermal 
gradients and the possible contribution of magnetic pressure. Generally, the optically thin atmospheres of 
disks have higher temperatures than at the mid-plane, and distinct temperatures for the gas and dust components.
\item The radial temperature profile is set by the balance between stellar irradiation and local cooling, supplemented 
by dissipation of accretion energy if $\dot{M}$ is high enough. These processes yield a mid-plane temperature 
scaling roughly in the range between $r^{-3/4}$ and $r^{-1/2}$. The radial profile of the surface density cannot 
be determined by similarly simple physical considerations. However, on average we expect the mid-plane 
pressure to decline with radius, and this causes the gas to orbit at slightly less than the Keplerian speed.
\item The disk close to the star is thermally ionized at levels sufficient to couple magnetic fields to the gas 
where the temperature exceeds about $10^3 \ {\rm K}$. At larger radii the ionization is non-thermal, due to 
stellar X-rays, FUV photons, radioactive decays, and cosmic rays if they are not screened from the disk. The ionization 
state varies radially and vertically depending on the balance of ionization and recombination reactions, which may 
occur in the gas phase or on dust grain surfaces.
\end{itemize}
\end{svgraybox}

\newpage

\section{Disk evolution} 
\label{sec_evolution}
The population of protoplanetary disks is observed to evolve, but the dominant 
physical processes responsible for this evolution remain unclear. 
For a geometrically thin, low-mass disk, 
the deviation from a point-mass Keplerian rotation curve is small 
(c.f. equation~\ref{eq_kepler_deviation}) and the specific angular momentum,
\begin{equation}
 l(r) = r^2 \Omega_{\rm K} = \sqrt{GM_* r} \propto r^{1/2},
\end{equation}
is an increasing function of orbital radius. To accrete, gas in the disk must 
lose angular momentum, and the central theoretical problem in disk evolution 
is to understand this process.

Within any shearing fluid momentum is transported in the cross-stream direction 
because the random motion of molecules leads to collisions between particles 
that have different velocities. The classical approach to disk evolution \cite{lyndenbell74,pringle81} treats 
the disk as a vertically thin axisymmetric sheet of viscous fluid, and leads to 
a fairly simple equation for the time evolution of the disk surface density $\Sigma(r,t)$. 
There appears to be a fatal flaw to this approach, because the molecular viscosity 
of the gas is much too small to lead to any significant rate of disk evolution. But it's not 
as bad as it seems. The classical disk evolution equation involves few assumptions 
beyond the immutable laws of mass and angular momentum conservation, and as 
we shall see is therefore approximately valid if the ``viscosity" is re-interpreted as 
the outcome of a turbulent process. We will have (much) more to say about the possible origin 
of disk turbulence in \S\ref{sec_turbulence}.

Redistribution of angular momentum within the gas disk is not the only route to 
evolution. An almost equally long-studied suggestion \cite{blandford82} is that gas accretes 
because a magnetohydrodynamic (MHD) wind {\em removes} angular momentum 
entirely from the disk. Winds and viscosity have frequently between 
seen as orthogonal and competing hypotheses for disk evolution, but there is 
evidence suggesting that both processes are simultaneously 
important in regions of protoplanetary disks.

\subsection{The classical equations}
The evolution of a flat, circular and geometrically thin ($(h/r) \ll 1$) viscous disk follows from 
the equations of mass and angular momentum conservation \cite{pringle81}. Given a surface density 
$\Sigma(r,t)$, radial velocity $v_r (r,t)$ and angular velocity $\Omega (r)$, the continuity 
equation in cylindrical co-ordinates yields,
\begin{equation}
 r \frac{\partial \Sigma}{\partial t} + \frac{\partial}{\partial r} \left( r \Sigma v_r \right) = 0.
\label{eq_disk_continuity} 
\end{equation}  
Angular momentum conservation gives,
\begin{equation}
 r \frac{\partial}{\partial t} \left( r^2 \Omega \Sigma \right) + 
 \frac{\partial}{\partial r} \left( r^2 \Omega \cdot r \Sigma v_r \right) = 
 \frac{1}{2 \pi} \frac{\partial G}{\partial r},
\label{eq_disk_angular} 
\end{equation} 
where $\Omega(r)$ is time-independent but need not be the point mass Keplerian 
angular velocity. The rate of change of disk angular momentum  
is given by the change in surface density due to radial flows 
and by the {\em difference} 
in the torque exerted on the annulus by stresses at the inner and 
outer edges. For a viscous fluid the torque $G$ has the form,
\begin{equation}
 G = 2 \pi r \cdot \nu \Sigma r \frac{{\rm d}{\Omega}}{{\rm d}r} \cdot r,
\end{equation}
where $\nu$ is the kinematic viscosity. The torque is 
the product of the circumference, the viscous force per unit length, and 
the lever arm $r$, and scales with the gradient of the angular 
velocity.

To obtain the surface density evolution equation in its usual form we first 
eliminate $v_r$ by substituting for $\partial \Sigma / \partial t$ in equation~(\ref{eq_disk_angular}) 
from equation~(\ref{eq_disk_continuity}). This gives an expression for $r \Sigma v_r$, which 
we substitute back into equation~(\ref{eq_disk_continuity}) to yield,
\begin{equation}
 \frac{\partial \Sigma}{\partial t} = -\frac{1}{r} 
 \frac{\partial}{\partial r} \left[ \frac{1}{\left( r^2 \Omega \right)^\prime} 
  \frac{\partial}{\partial r}
  \left( \nu \Sigma r^3 \Omega^\prime \right) \right],
 \label{eq_disk_evolve_general} 
 \end{equation} 
where the primes denote differentiation with respect to radius.  
Specializing to a point mass Keplerian potential 
($\Omega \propto r^{-3/2}$) we then find that viscous redistribution of angular momentum 
within a thin disk obeys an equation,
\begin{equation}
 \frac{\partial \Sigma}{\partial t} = \frac{3}{r} 
 \frac{\partial}{\partial r} \left[ r^{1/2} 
 \frac{\partial}{\partial r} \left( \nu \Sigma r^{1/2} \right) \right].
\label{eq_disk_evolve}
\end{equation}
This equation is a diffusive partial differential equation for the evolution of the 
gas, which has a radial velocity,
\begin{equation}
 v_r = - \frac{3}{\Sigma r^{1/2}}  \frac{\partial}{\partial r} \left( \nu \Sigma r^{1/2} \right).
\label{eq_vr}
\end{equation} 
The equation is linear if the viscosity $\nu$ is independent of $\Sigma$.

Some useful rules of thumb for the rate of evolution implied by 
equation~(\ref{eq_disk_evolve}) can be deduced with a change of variables. Defining,
\begin{eqnarray}
 X &\equiv& 2 r^{1/2},  \\
 f &\equiv& \frac{3}{2} \Sigma X, 
\end{eqnarray}
and taking the viscosity $\nu$ to be constant, we get a simpler 
looking diffusion equation,
\begin{equation}
  \frac{\partial f}{\partial t} = D \frac{ \partial^2 f }{\partial X^2},
\label{eq_diffusion_prototype}  
\end{equation}
with a diffusion coefficient $D$,
\begin{equation}
 D = \frac{12 \nu}{X^2}.
\end{equation} 
The diffusion time scale across a scale $\Delta X$ for an equation of the form of 
equation~(\ref{eq_diffusion_prototype}) 
is just $(\Delta X)^2 / D$. Converting back to the physical variables, the 
time scale on which viscosity will smooth surface density gradients on a 
scale $\Delta r$ is $\tau_\nu \sim {(\Delta r)^2}/{\rm \nu}$.
For a disk with characteristic size $r$, the surface density at all radii will 
evolve on a time scale,
\begin{equation}
 \tau_\nu \approx \frac{r^2}{\nu}.
\label{eq_viscous_time} 
\end{equation}
This is the {\em viscous time scale} of the disk.   

\begin{figure}[t]
\includegraphics[width=\columnwidth]{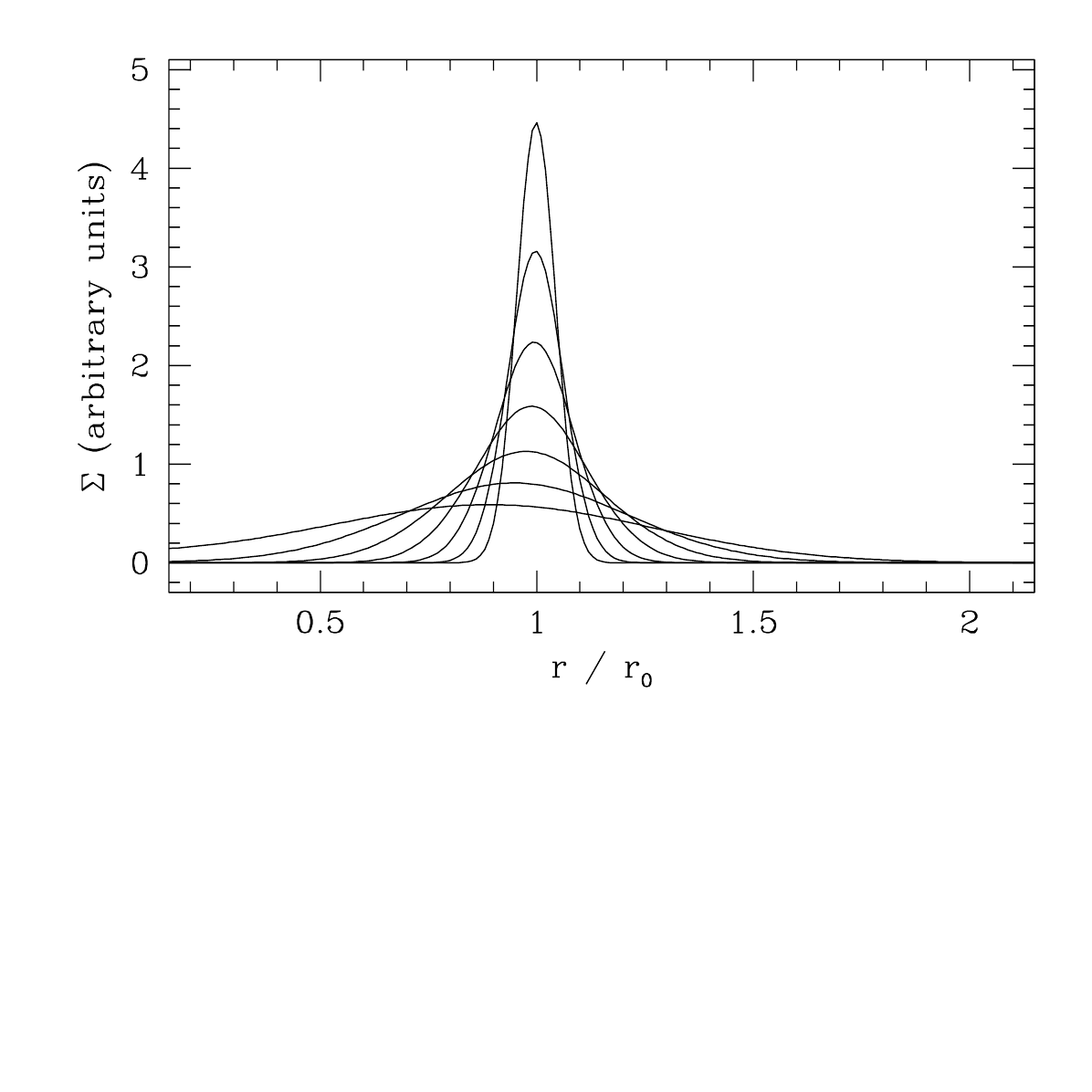}
\vspace{-4.5cm}
\caption{The time-dependent solution to the disk evolution equation with 
$\nu = {\rm constant}$, showing the spreading of a ring of gas  
initially orbiting at $r = r_0$. From top down the curves show the 
surface density as a function of the scaled time variable $\tau=12 \nu r_0^{-2} t$, 
for $\tau = 0.004$, $\tau=0.008$, $\tau=0.016$, $\tau=0.032$, $\tau=0.064$, 
$\tau=0.128$, and $\tau=0.256$.}
\label{fig_diffusion_analytic}
\end{figure}

We can gain some intuition into how equation~(\ref{eq_disk_evolve}) works 
by inspecting time-dependent analytic solutions that can be derived for 
special forms of the viscosity $\nu(r)$. For  
$\nu = {\rm constant}$ a Green's function solution is possible.
Suppose that at $t=0$ the gas 
lies in a thin ring of mass $m$ at radius $r_0$,
\begin{equation}
 \Sigma(r,t=0) = \frac{m}{2 \pi r_0} \delta (r-r_0),
\end{equation}
where $\delta (r-r_0)$ is a Dirac delta function. 
With boundary conditions that impose zero-torque at $r=0$ and allow 
for free expansion toward $r = \infty$ the solution is \cite{lyndenbell74},
\begin{equation}
 \Sigma (x,\tau) = \frac{m}{\pi r_0^2} \frac{1}{\tau} 
 x^{-1/4} \exp \left[ -\frac{(1+x^2)}{\tau} \right] I_{1/4} \left( \frac{2x}{\tau} \right), 
\label{eq_C3_analytic} 
\end{equation}
in terms of dimensionless variables 
$x \equiv r / r_0$, $\tau \equiv 12 \nu r_0^{-2} t$, and where $I_{1/4}$ is a 
modified Bessel function of the first kind. This solution is plotted in 
Figure~\ref{fig_diffusion_analytic} and illustrates generic features 
of viscous disk evolution. As $t$ increases the ring spreads 
diffusively, with the mass flowing toward $r = 0$ while  
the angular momentum is carried by a negligible fraction of the mass 
toward $r = \infty$. This segregation of mass and angular momentum 
is generic to the evolution of a viscous disk, and must occur if accretion 
is to proceed without overall angular momentum loss (for example in a 
magnetized disk wind).

\subsubsection{Limits of validity}
Protoplanetary disks are not viscous fluids in the same way that honey is 
a viscous fluid (or, for that matter, in the same way as the mantle of the Earth 
is viscous). To order of magnitude precision, the viscosity of a gas 
$\nu \sim v_{\rm th} \lambda$, where $v_{\rm th}$ is the thermal speed 
of the molecules and the mean-free path $\lambda$ is,
\begin{equation}
 \lambda \sim \frac{1}{n \sigma}.
\end{equation}
Here $n$ is the number density of molecules with collision cross-section $\sigma$. 
Taking $\sigma$ to be roughly the physical size of a hydrogen molecule, $\sigma 
\sim \pi (10^{-8} \ {\rm cm})^2$, and conditions appropriate to 1~AU ($n \sim 10^{15} \ 
{\rm cm}^{-3}$, $v_{\rm th} \sim 10^5 \ {\rm cm \ s}^{-1}$) we estimate,
\begin{eqnarray}
 \lambda & \sim & 3 \ {\rm cm} \nonumber \\
 \nu & \sim & 3 \times 10^5 \ {\rm cm^2 \ s}^{-1}.
\label{eq_nu_estimate} 
\end{eqnarray} 
This is not a large viscosity. The implied viscous time according to equation~(\ref{eq_viscous_time}) 
is of the order of $10^{13} \ {\rm yr}$, far in excess of observationally inferred time scales of 
protoplanetary disk evolution. If we nevertheless press on and use equation~(\ref{eq_disk_evolve}) 
to model disk evolution, we are implicitly modeling a system that is not an ordinary viscous 
fluid with a viscous equation. We need to understand when this is a valid approximation.

The first possibility, introduced by Shakura \& Sunyaev (1973) in their paper on black hole 
accretion disks \cite{shakura73}, retains the idea that angular momentum is conserved within 
the disk system, but supposes that turbulence rather than molecular processes is the agent 
of angular momentum transport. Looking back at the derivation of the disk evolution 
equation~(\ref{eq_disk_evolve}), we note that the fluid properties of molecular viscosity 
only enter twice, (i) in the specific expression for $G$ (for example in the fact that the torque 
is hardwired to be linear in the rate 
of shear) and (ii) in the more basic assumption that angular momentum transport is 
determined by the {\em local} fluid properties. The rest of the derivation involves 
only conservation laws that hold irrespective of the nature of transport. Plausibly then, 
a disk in which angular momentum is redistributed by the action of turbulence should 
still be describable by a diffusive equation, provided that the turbulence is a local 
process. Proceeding rigorously, Balbus \& Papaloizou (1999) \cite{balbus99} showed 
that MHD turbulence is in principle local in this sense, whereas angular momentum transport 
by self-gravity is in principle not. At the level of the basic axisymmetric evolution equation then --- before 
we turn to questions of what determines $\nu$, or how boundary layers behave where the 
shear is reversed --- we have not committed any cardinal sin in starting from the viscous 
disk equation.

Greater care is needed in situations where the disk flow is no longer axisymmetric. Fluids obey 
the Navier-Stokes equations, but there is no guarantee that a turbulent 
disk with a complex geometry will behave in the same way as a viscous Navier-Stokes flow with 
effective kinematic and bulk viscosities. In eccentric disks, for example, even the most basic 
properties (such as whether the eccentricity grows or decays) depend upon the nature of the 
angular momentum transport \cite{ogilvie99}.

The disk evolution equation will also need modification if there are external sources or sinks 
of mass or angular momentum. If the disk gains or loses mass at a rate $\dot{\Sigma} (r,t)$, 
and {\em if that gas has the same specific angular momentum as the disk}, then the 
modification is trivial,
\begin{equation}
 \frac{\partial \Sigma}{\partial t} = \frac{3}{r} 
 \frac{\partial}{\partial r} \left[ r^{1/2} 
 \frac{\partial}{\partial r} \left( \nu \Sigma r^{1/2} \right) \right] + \dot{\Sigma}.
\label{eq_disk_evolve_pe} 
\end{equation}
Disk evolution in the presence of thermally driven winds (such as photo-evaporative 
flows) can be described with this equation. Alternatively, we may consider a disk 
subject to an external torque that drives a radial flow with velocity $v_{r, ext}$. 
This adds an advective term,
\begin{equation}
 \frac{\partial \Sigma}{\partial t} = \frac{3}{r} 
 \frac{\partial}{\partial r} \left[ r^{1/2} 
 \frac{\partial}{\partial r} \left( \nu \Sigma r^{1/2} \right) \right] - 
 \frac{1}{r} \frac{\partial}{\partial r} \left( r \Sigma v_{r, ext} \right).
\end{equation} 
The qualitative evolution of the disk --- for example the tendency for the outer regions to 
expand to conserve angular momentum, or the steady-state surface density profile at small radii \cite{suzuki16}  --- 
can be changed if there is even a modest external torque on the system.

From an observational point of view the relative simplicity of equation~(\ref{eq_disk_evolve}) means 
that it is often used to model the evolution of disk populations and to fit the surface density profile 
of individual disks. This provides a useful connection to disk theory, but it should be remembered 
that the general validity of the simple diffusion equation is itself an open question. In the outer regions, especially, 
it is possible that the initial surface density distribution is modified more by thermal or magnetic winds than 
by internal redistribution of angular momentum.

\subsubsection{The $\alpha$ prescription}
Molecular viscosity depends in a calculable way upon the density, temperature and composition 
of the fluid. Can anything similar be said about the ``effective viscosity" present in disks? The 
standard approach is to write the viscosity as the product of characteristic velocity and 
spatial scales in the disk,
\begin{equation}
 \nu = \alpha c_s h,
\label{eq_alpha_prescription} 
\end{equation}
where $\alpha$ is a dimensionless parameter. This ansatz (introduced in a related form in 
\cite{shakura73}) is known as the Shakura-Sunyaev $\alpha$ prescription.

We can view the $\alpha$ prescription in two ways. The weak version is to regard it as 
a re-parameterization of the viscosity that describes the leading order scaling 
expected in disks (so that $\alpha$ is a more slowly varying function of temperature, radius 
etc than $\nu$). This is useful, and along with convention is the reason why numerical 
simulations of turbulent transport are invariably reported in terms of an effective $\alpha$. 
One can also adopt a strong version of the prescription 
in which $\alpha$ is assumed to be a constant. This is powerful as it allows for the development 
of a predictive theory of disk structure that is based on only one free parameter (for a textbook 
discussion see Frank, King \& Raine \cite{frank02}, or for an application to protoplanetary disks 
see, e.g. \cite{bell97}). However, its use must be justified on a case by case basis, as there is 
no reason why $\alpha$ should be a constant. Constant $\alpha$ models probably work better 
in highly ionized disks around black holes and neutron stars, where angular momentum transport across a broad range 
of radii occurs via the simplest version of the magnetorotational instability \cite{balbus98}, than 
in protoplanetary disks where the physical origin of angular momentum transport is more 
complex \cite{armitage11}.

\subsection{Boundary conditions}
Solving equation~(\ref{eq_disk_evolve}) requires the imposition of boundary conditions. The most 
common, and simplest, is a zero-torque inner boundary condition, which exactly conserves the 
initial angular momentum content of the disk. If the star has a dynamically significant magnetic field, 
however, or if the disk is part of a binary system, other boundary conditions may be more appropriate.

\subsubsection{Zero-torque boundary conditions}
\label{sec_zero_torque}

\begin{figure}[t]
\center
\includegraphics[width=0.9\columnwidth]{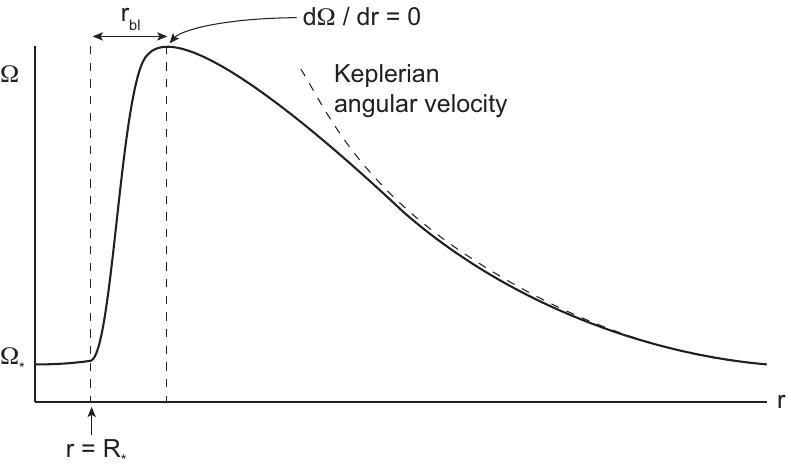}
\caption{A sketch of what the angular velocity profile $\Omega(r)$ must look like if the disk 
extends down to the surface of a slowly rotating star. By continuity there must be a point --- usually 
close to the stellar surface --- where ${\rm d} \Omega / {\rm d} r = 0$ and the viscous stress vanishes.}
\label{figure_boundary}
\end{figure}

A steady-state solution to equation~(\ref{eq_disk_evolve}) with a zero-torque 
inner boundary condition is derived by starting from the angular momentum 
conservation equation (equation~\ref{eq_disk_angular}). Setting the time 
derivative to zero and integrating we have,
\begin{equation}
 2 \pi r \Sigma v_r \cdot r^2 \Omega = 2 \pi r^3 \nu \Sigma 
 \frac{{\rm d}\Omega}{{\rm d}{r}} + {\rm constant}.
\end{equation}
In terms of the mass accretion rate $\dot{M} = - 2 \pi r \Sigma v_r$ we 
can write this in the form,
\begin{equation}
 - \dot{M} \cdot r^2 \Omega = 2 \pi r^3 \nu \Sigma 
 \frac{{\rm d}\Omega}{{\rm d}{r}} + {\rm constant},
\label{eq_disk_blconstant} 
\end{equation}
where the constant of integration, which is an angular momentum 
flux, is as yet undermined. To specify the constant we note that if there is a 
point where ${\rm d}\Omega / {\rm d}r = 0$ the viscous stress vanishes, and the 
constant is just the advective flux of angular momentum,
\begin{equation}
 {\rm constant} = - \dot{M} \cdot r^2 \Omega.
\end{equation} 
The physical situation where ${\rm d}\Omega / {\rm d}r = 0$ is where the 
protoplanetary disk extends all the way 
down to the surface of a slowly rotating star. The disk and the 
star form a single fluid system, and the angular velocity (shown  
in Figure~\ref{figure_boundary}) must be a continuous function  
that connects $\Omega \approx 0$ in the star to $\Omega \propto r^{-3/2}$ 
within the disk. The viscous stress 
must then vanish at some radius $R_* + r_{\rm bl}$, 
where $r_{\rm bl}$ is the width of the {\em boundary layer} that separates the 
star from the Keplerian part of the disk. Within the boundary layer the angular 
velocity increases with radius, and gravity is balanced against a combination 
of rotation and radial pressure support. Elementary arguments \cite{pringle77} 
show that in many cases the boundary layer is narrow, so that $R_* + r_{\rm bl} \simeq R_*$.
We then find that,
\begin{equation}
 {\rm constant} \simeq -\dot{M} R_*^2 \sqrt{\frac{GM_*}{R_*^3}},
\end{equation}
and the steady-state solution for the disk simplifies to,
\begin{equation}
 \nu \Sigma = \frac{\dot{M}}{3 \pi} \left( 
 1 - \sqrt{\frac{R_*}{r}} \right).
\label{eq_disk_steady} 
\end{equation}  
For a specified viscosity this equation gives the steady state surface density 
profile of a disk with a constant accretion rate $\dot{M}$. Away 
from the inner boundary $\Sigma(r) \propto \nu^{-1}$, and the radial 
velocity (equation~\ref{eq_vr}) is $v_r = -3 \nu / 2 r$.

In obtaining equation~(\ref{eq_disk_steady}) we have derived an expression for a 
{\em Keplerian} disk via an argument that relies on the {\em non-Keplerian} form 
of $\Omega(r)$ in a boundary layer. The resulting expression for the surface density 
is valid in the disk at $r \gg R_*$, but would not work well close to the star 
{\em even if there is a boundary layer}. To model the boundary layer properly, 
we would need equations that self-consistently determine the angular velocity 
along with the surface density \cite{popham93}. 

\subsubsection{Magnetospheric accretion}
\label{sec_magnetospheric}
For protoplanetary disks the stellar magnetic field can have a dominant 
influence on the disk close to the star \cite{konigl91}. The simplest magnetic geometry involves  
a dipolar stellar magnetic field that is aligned with the stellar rotation axis and 
perpendicular to the disk plane. The unperturbed field then has a vertical 
component at the disk surface,
\begin{equation}
 B_z = B_* \left( \frac{r}{R_*} \right)^{-3}.
\end{equation}
In the presence of a disk, the vertical field will 
thread the disk gas and be distorted by differential 
rotation between the Keplerian disk and the star. The differential rotation 
twists the field lines that couple the disk to the star, generating an 
azimuthal field component at the disk surface $B_\phi$ and a magnetic 
torque per unit area (counting both upper and lower disk surfaces),
\begin{equation} 
 T = \frac{B_z B_\phi}{2 \pi} r.
\end{equation}
Computing the perturbed field accurately is hard (for simulation results see, e.g. \cite{romanova12}), 
but it is easy to identify 
the qualitative effect that it has on the disk. For a star with rotation period $P$, we define 
the co-rotation radius $r_{\rm co}$ as the radius where the field lines have the same 
angular velocity as that of Keplerian gas in the disk,
\begin{equation}
 r_{\rm co} = \left( \frac{GM_* P^2}{4 \pi^2} \right)^{1/3}. 
\end{equation}
There are then two regions of star-disk magnetic interaction:
\begin{itemize}
\item
Interior to co-rotation ($r < r_{\rm co}$) the disk gas has a greater 
angular velocity than the field lines. Field lines that link the disk 
and the star here are dragged forward by the disk, and exert a braking torque 
that removes angular momentum from the disk gas.
\item
Outside co-rotation ($r > r_{\rm co}$) the disk gas has a smaller 
angular velocity than the field lines. The field lines are dragged 
backward by the disk, and there is a positive torque on the disk gas.
\end{itemize}  
Young stars are typically rapid rotators \cite{bouvier14}, so the co-rotation radius  
lies in the inner disk. For $P = 7$~days, for example, the co-rotation 
radius around a Solar mass star is at $r_{\rm co} \simeq 15 \ R_\odot$ 
or 0.07~AU.

The presence of a stellar magnetic torque violates the assumption of a 
zero-torque boundary condition, though the steady-state solution we 
derived previously (equation~\ref{eq_disk_steady}) will generally still 
apply at sufficiently large radius. The strong radial dependence of the 
stellar magnetic torque means that there is only a narrow window of 
parameters where the torque will be significant yet still allow the disk 
to extend to the stellar surface. More commonly, a dynamically significant 
stellar field will disrupt the inner disk entirely, yielding a magnetospheric 
regime of accretion in which the terminal phase of accretion is along 
stellar magnetic field lines. The disruption (or magnetospheric) radius 
$r_m$ can be estimated in various ways \cite{konigl91}, but all yield 
the same scaling as the spherical Alfv\'en radius that is obtained by 
equating the magnetic pressure of a dipolar field to the ram pressure of 
spherical infall,
\begin{equation}
 r_m \simeq \left( \frac{k B_*^2 R_*^6}{\dot{M} \sqrt{GM_*}} \right)^{2/7}.
\end{equation}
Here $B_*$ is the stellar surface field (defined such that $B_* R_*^3$ is 
the dipole moment) and $k$ a constant of the order of unity. Taking $k=1$ for a 
Solar mass star,
\begin{equation}
 r_m \simeq 14 \left( \frac{B_*}{\rm kG} \right)^{4/7}
 \left( \frac{R_*}{2 \ R_\odot} \right)^{12/7} 
 \left( \frac{\dot{M}}{10^{-8} \ M_\odot \ {\rm yr}^{-1}} \right)^{-2/7} R_\odot.
\label{eq_rm} 
\end{equation} 
Often, the magnetospheric radius is comparable to the co-rotation radius. This is 
to some extent expected, since if $r_m$ is substantially different from $r_{\rm co}$ 
the magnetic torque acting on the star will tend to modify $P$ in the direction of 
reducing the difference. 

\subsubsection{Accretion on to and in binaries}
\label{sec_binary_bc}
Boundary conditions for disk evolution also need modification in binary systems. For a 
coplanar disk orbiting {\em interior} to a prograde stellar binary companion, tidal torques 
from the companion remove angular momentum from the outer disk and prevent it 
from expanding too far \cite{papaloizou77}. The tidal truncation radius roughly corresponds 
to the largest simple periodic orbit in the binary potential \cite{paczynski77}, which is at 
about 40\% of the orbital separation for a binary with mass ratio $M_2 / M_1 = 0.5$\footnote{The size 
of the disk (and even whether it is tidally truncated at all) will be different if the 
disk is substantially misaligned with respect to the orbital plane of the binary \cite{lubow15,miranda15}.}. The 
tidal torque is a function of radius, but to a first approximation one may assume that tides impose 
a rigid no-expansion condition at $r = r_{\rm out}$. From equation~(\ref{eq_vr}),
\begin{equation}
 \left. \frac{\partial}{\partial r} \left( \nu \Sigma r^{1/2} \right) \right|_{r=r_{\rm out}} = 0.
\end{equation} 
This type of boundary condition may also be an appropriate approximation for a 
circumplanetary disk truncated by stellar tides \cite{martin11}.

\begin{figure}[t]
\includegraphics[width=\columnwidth]{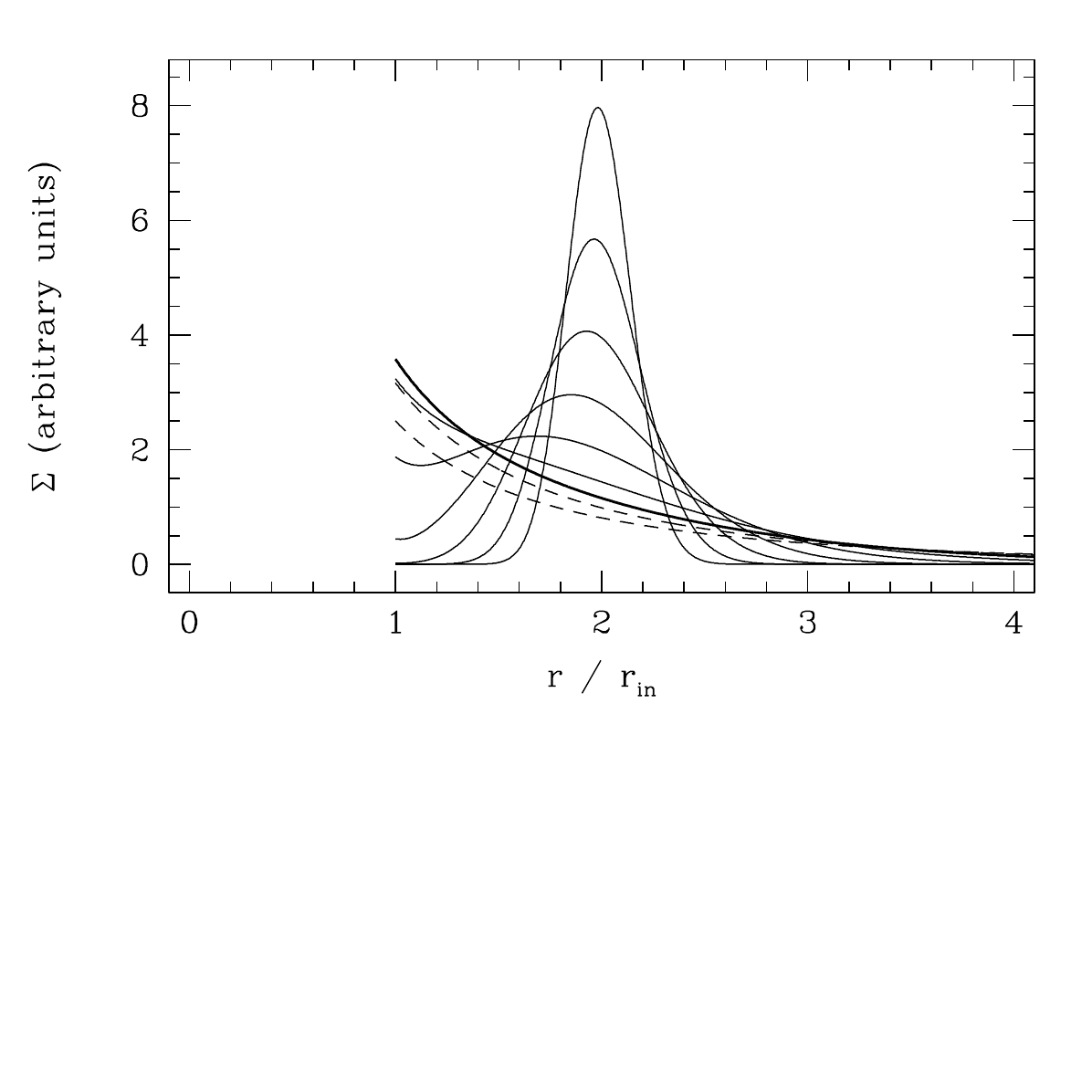}
\vspace{-4.5cm}
\caption{The time-dependent analytic solution (equation~\ref{eq_circumbinary}) to the disk evolution equation with a 
$v_r = 0$ boundary condition at $r=1$ for the case $\nu = r$. The solid curves show the 
evolution of a ring of gas initially at $r=2$, at times $t=0.002$, $t=0.004$, 
$t=0.008$ etc. The bold curve is at $t=0.128$, and dashed curves show 
later times. Gas initially accretes, but eventually {\em decretes} due to the 
torque being applied at the boundary.}
\label{fig_external}
\end{figure}

An exterior {\em circumbinary} disk will also experience stellar gravitational torques, which 
in this case add angular momentum to the disk and slow viscous inflow. How best to model 
these torques is an open question, particularly in the case of extreme mass ratio binaries 
composed of a star and a massive planet. Pringle \cite{pringle91} derived an illuminating analytic 
solution for circumbinary disk evolution under the assumption that tidal 
torques {\em completely prevent} inflow past some radius $r=r_{\rm in}$. With this 
assumption the boundary condition at $r=r_{\rm in}$ is $v_r = 0$, and the task is to 
find a solution to equation~(\ref{eq_disk_evolve}) with this finite radius boundary condition. 
A simple solution is possible for $\nu = k r$, with $k$ a constant. Defining scaled variables,
\begin{eqnarray}
 x & = & r^{1/2} \nonumber \\
 \sigma & = & \Sigma r^{3/2},
\end{eqnarray}
the $t > 0$ solution for an initial mass distribution,
\begin{equation} 
 \sigma(x,t=0) = \sigma_0 \delta (x-x_1),
\end{equation}
is \cite{pringle91},  
\begin{equation}
 \sigma = \frac{\sigma_0 t^{-1/2}}{4 \left( 3 \pi k \right)^{1/2}} 
 \left\{ \exp \left[ -(x-x_1)^2 / 3kt \right] 
 + \exp \left[ -(x+x_1 - 2x_{\rm in})^2 / 3kt \right] \right\}.
\label{eq_circumbinary} 
\end{equation} 
The solution, plotted in Figure~\ref{fig_external}, can be compared to the 
zero-torque solution (Figure~\ref{fig_diffusion_analytic}, though note this 
is for a constant viscosity). The initial evolution is similar, but at late times 
the torque that precludes inflow past $r_{\rm in}$ causes qualitatively 
different behavior. The disk switches from an accretion to a {\em decretion} 
disk, with an outward flow of mass driven by the binary torque.

The classical decretion disk solution was developed as a model for disk evolution 
around binaries. It is not clear, however, whether it is ever realized in the binary 
context. Numerical simulations of the interaction between a binary and a circumbinary disk 
show that angular momentum transfer to the disk co-exists 
with persistent inflow into a low density cavity containing the binary \cite{artymowicz96,dorazio13}. 
How best to represent this complexity in a one-dimensional model is not entirely obvious. 
The decretion disk solution may be a better description of other astrophysical situations, 
such as disks around rapidly rotating and strongly magnetized 
stars. 

\subsection{Viscous heating}
\label{sec_viscous_heating}
Although stellar irradiation is often the dominant source of heat for protoplanetary disks (\S\ref{subsec_thermal}), 
dissipation of gravitational potential energy associated with accretion is also important. Ignoring irradiation 
for the time being, we can derive the effective temperature profile of a steady-state viscous disk. In the 
regime where the classical equations are valid, the fluid dissipation per unit area is \cite{pringle81},
\begin{equation}
 Q_+ = \frac{9}{4} \nu \Sigma \Omega^2.
\label{eq_qplus} 
\end{equation}
Using the steady-state solution for $\nu \Sigma$ (equation~\ref{eq_disk_steady}) we equate $Q_+$ to the 
rate of local energy loss by radiation. If the disk is optically thick, the disk radiates (from both sides) at a 
rate $Q_- = 2 \sigma T_{\rm disk}^4$, with $\sigma$ being the Stefan-Boltzmann constant. This yields an  
effective temperature profile,
\begin{equation}
 T_{\rm disk}^4 = \frac{3 G M_* \dot{M}}{8 \pi \sigma r^3} \left( 1 - \sqrt{\frac{R_*}{r}} \right).
\label{eq_viscous_temp} 
\end{equation}
Away from the inner boundary, the steady-state temperature profile for a viscous disk 
($T_{\rm disk} \propto  r^{-3/4}$) is steeper than for irradiation. For any accretion rate, 
we then expect viscous heating to be most important in the inner disk, whereas irradiation 
always wins out at sufficiently large radii.

The viscous disk temperature profile is {\bf not} what we get from considering just the local 
dissipation of potential energy. The gradient of the potential energy per unit mass $\epsilon$, 
is ${\rm d} \epsilon / {\rm d} r = GM_* / r^2$. For an accretion rate $\dot{M}$, the luminosity 
available to be radiated from an annulus of width $\Delta r$ due to local potential energy 
release would be,
\begin{equation}
 L = \frac{1}{2} \frac{G M_* \dot{M}}{r^2} \Delta r,
\end{equation}
where the factor of a half accounts for the fact that half the energy goes into increased 
kinetic energy, with only the remainder available to be thermalized and radiated. Equating 
this luminosity to the black body emission from the annulus, $2 \sigma T_{\rm disk}^4 \cdot 
2 \pi r \Delta r$, would give a profile that is a factor three different from the asymptotic 
form of equation~(\ref{eq_viscous_temp}). The difference arises because the radial transport 
of angular momentum is accompanied by a radial transport of energy. The local luminosity 
from the disk surface at any radius then has a contribution from potential energy liberated 
closer in.

The optically thick regions of irradiated protoplanetary disks will be vertically isothermal. When 
viscous heating dominates, however, there must be a vertical temperature gradient to allow
energy to be transported from the mid-plane toward the photosphere. What this gradient looks 
like, in detail, depends on the vertical distribution of the heating, which is not well known. 
However, an approximation to $T(z)$ can be 
derived assuming that the energy dissipation due 
to viscosity is strongly concentrated toward the mid-plane. 
We define the optical depth to the disk mid-plane,
\begin{equation} 
 \tau = \frac{1}{2} \kappa_R \Sigma,
\end{equation}
where $\kappa_R$ is the Rosseland mean opacity. The vertical density profile of 
the disk is $\rho(z)$. If the vertical energy transport occurs 
via radiative diffusion then for $\tau \gg 1$ the vertical energy flux $F(z)$ 
is given by the equation of radiative diffusion \cite{rybicki79}
\begin{equation} 
 F_z(z) = - \frac{16 \sigma T^3}{3 \kappa_R \rho} \frac{{\rm d}T}{{\rm d}z}.
\end{equation} 
Now assume for simplicity that {\em all} of the dissipation 
occurs at $z=0$. In that case $F_z(z) = \sigma T_{\rm disk}^4$ 
is constant with height. We integrate from the mid-plane to the 
photosphere at $z_{\rm ph}$ assuming that the 
opacity is also constant,
\begin{eqnarray}
 - \frac{16 \sigma}{3 \kappa_R} \int_{T_{\rm c}}^{T_{\rm disk}} T^3 dT & = & 
 \sigma T_{\rm disk}^4 \int_0^{z_{\rm ph}} \rho ( z^\prime ) dz^\prime  \\
 - \frac{16}{3 \kappa_R} \left[ \frac{T^4}{4} \right]_{T_{\rm c}}^{T_{\rm disk}} 
 & = & T_{\rm disk}^4 \frac{\Sigma}{2},
\label{eq_C3_TcTeff} 
\end{eqnarray} 
where the final equality relies on the fact that for $\tau \gg 1$ 
almost all of the disk gas lies below the photosphere. For large optical depth 
$T_{\rm c}^4 \gg  T_{\rm disk}^4$ and the equation 
simplifies to,
\begin{equation} 
 \frac{T_{\rm c}^4}{T_{\rm disk}^4} \simeq \frac{3}{4} \tau.
\label{eq_C3_Tc} 
\end{equation}
Often {\em both} stellar irradiation and accretional 
heating contribute significantly to the thermal balance of the disk. If we define 
$T_{\rm disk, visc}$ to be the effective temperature that would result 
from viscous heating in the absence of irradiation (i.e. the quantity 
called $T_{\rm disk}$, with no subscript, above) and $T_{\rm irr}$ 
to be the irradiation-only effective temperature, then,
\begin{equation}
 T_c^4 \simeq \frac{3}{4} \tau T_{\rm disk, visc}^4 + T_{\rm irr}^4
\end{equation}
is an approximation for the central temperature, again valid for 
$\tau \gg 1$. 

These formulae can be applied to estimate the location of the snow line. In the 
Solar System meteoritic evidence \cite{morbidelli00} places the transition between 
water vapor and water ice, which occurs at a mid-plane temperature of 
150-180~K, at around 2.7~AU. This is substantially further from the Sun than 
would be expected if the only disk heating source was starlight. Including 
viscous heating, however, an accretion rate of $2 \times 10^{-8} \ M_\odot \ {\rm yr}^{-1}$ 
could sustain a mid-plane temperature of 170~K at 2.7~AU in a disk with $\Sigma = 400 \ {\rm g \ cm}^{-2}$ 
and $\kappa_R = 1 \ {\rm cm^2 \ g^{-1}}$. This estimate (from equation~\ref{eq_C3_Tc}) 
is consistent with more detailed models for protoplanetary disks \cite{bell97}, though 
considerable variation in the location of the snow line is introduced by uncertainties 
in the vertical structure \cite{martin12}.

\subsection{Warped disks}
\label{sec_warp}
The classical equation for surface density evolution needs to be rethought if the disk 
is non-planar. Disks may be warped for several reasons; the direction of the angular 
momentum vector of the gas that forms the disk may not be constant,  the disk may 
be perturbed tidally by a companion \cite{larwood96,nixon10}, or warped due to 
interaction with the stellar magnetosphere \cite{lai99}.

\begin{figure}[t]
\center
\includegraphics[width=\columnwidth]{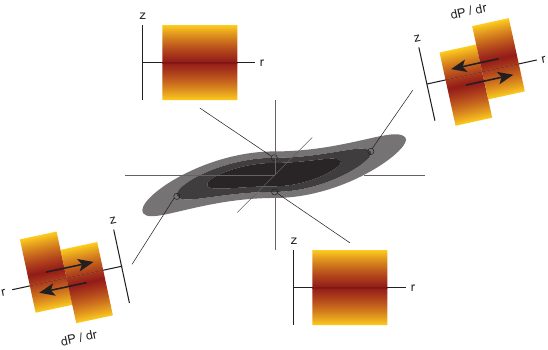}
\caption{Illustration (after \cite{lodato07,nixon15}) of how a warp introduces an oscillating 
{\em radial} pressure gradient within the disk. As fluid orbits in a warped disk, vertical 
shear displaces the mid-planes of neighboring annuli. This leads to a time-dependent radial 
pressure gradient ${\rm d}P / {\rm d}r (z)$. Much of the physics of warped disks is determined 
by how the disk responds to this warp-induced forcing.}
\label{fig_warp_geometry}
\end{figure}

A warp affects disk evolution through physics that is independent of its origin (for a 
brief review, see \cite{nixon15}). In a 
warped disk, neighboring annuli have specific angular momenta that differ in 
direction as well as in magnitude. If we define a unit tilt vector ${\bf l} (r,t)$ that 
is locally normal to the disk plane, the shear then has a vertical as well as a radial 
component \cite{ogilvie13}, 
\begin{equation}
 {\bf S} = r \frac{ {\rm d}\Omega}{{\rm d}r} {\bf l} + r \Omega \frac{\partial {\bf l}}{\partial r}.
\end{equation}
The most important consequence of the vertical shear is that it introduces a periodic 
vertical displacement of radially separated fluid elements. As illustrated in Figure~\ref{fig_warp_geometry} 
this displacement, in turn, 
results in a horizontal pressure gradient that changes sign across the mid-plane and 
is periodic on the orbital frequency. In a Keplerian disk this forcing frequency is 
resonant with the epicyclic frequency.

How the disk responds to the warp-generated horizontal forcing depends on the 
strength of dissipation \cite{papaloizou83}. If the disk is sufficiently viscous, 
specifically if,
\begin{equation}
 \alpha > \frac{h}{r},
\end{equation} 
the additional shear is damped locally. The equation for the surface density 
and tilt evolution (the key aspects of which are derived in \cite{pringle92}, though 
see \cite{ogilvie99b} for a complete treatment) then includes terms which diffusively 
damp the warp at a rate that is related to the radial redistribution of angular momentum. 
Normally, warp damping is substantially faster than the viscous evolution of a planar 
disk. Even for a Navier-Stokes viscosity --- which is fundamentally isotropic --- the 
{\em effective} viscosity which damps the warp is a factor $\approx 1 / 2 \alpha^2$ 
larger than its equivalent in a flat disk. Rapid evolution also occurs in the opposite 
limit of an almost inviscid disk with,
\begin{equation}
 \alpha < \frac{h}{r},
\end{equation}
but in this case the component of angular momentum associated with the warp 
is communicated radially in the form of a wave. For a strictly inviscid disk 
the linearized fluid equations for the evolution of the tilt vector take a 
simple form \cite{lubow00},
\begin{equation}
 \frac{\partial^2 {\bf l}}{\partial t^2} = \frac{1}{\Sigma r^3 \Omega} 
 \frac{\partial}{\partial r} \left( \Sigma r^3 \Omega \frac{c_s^2}{4} 
  \frac{\partial {\bf l}}{\partial r} \right).
\end{equation}  
The speed of the warp wave is $v_w \approx c_s / 2$.

In most cases we expect protoplanetary disks to have $\alpha < h/r$, and warps 
will evolve in the wave-like regime. We expect, however, that the details of warp evolution 
will depend upon the nature of angular momentum transport, and little is known 
about how warps behave for the transport mechanisms (such as non-ideal MHD) 
most relevant to protoplanetary disks. The extent of the differences between warp evolution with 
realistic transport and that with a Navier-Stokes viscosity are undetermined.

\subsection{Disk winds}
Viscous evolution driven by redistribution of angular momentum is a 
consequence of turbulent (or possibly laminar) stresses that are internal to the 
fluid. Evolution can also be driven by external torques, the most important of 
which is the magnetic torque that an MHD wind exerts on the surface of the 
disk. An excellent pedagogical introduction to disk winds is the review by 
Spruit \cite{spruit96}, while K\"onigl \& Salmeron \cite{konigl11} provide a more 
recent account that addresses the peculiarities specific to protoplanetary disks.

\begin{figure}[t]
\center
\includegraphics[width=0.9\columnwidth]{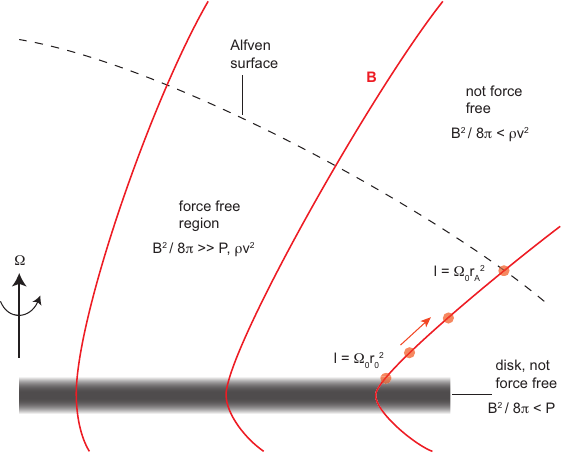}
\caption{Illustration, after Spruit \cite{spruit96}, of the different regions of a disk wind solution.}
\label{fig_spruit}
\end{figure}

Winds that are driven solely by pressure gradients (``thermal winds") are of 
interest as a mechanism for disk dispersal (\S\ref{sec_dispersal}) but do not 
change the qualitative character of disk evolution. Instead, we 
consider here MHD winds, with the simplest case being that of a well-ionized 
disk is threaded by a large-scale ordered poloidal magnetic field. In the ideal MHD 
limit the fluid is tied to magnetic field lines, which can facilitate acceleration 
of the wind while exerting a back reaction on the disk that removes angular momentum. This 
type of MHD wind is known as a Blandford-Payne wind \cite{blandford82}. Other varieties 
of MHD outflow could also occur. Winds could be launched by a gradient of {\em toroidal} magnetic field 
pressure \cite{lyndenbell03}, perhaps during or shortly after the collapse of the cloud that forms 
both the star and the disk. Jets can also originate from the interaction between the 
stellar magnetosphere and the disk \cite{shu94}.

Returning to the specific case of Blandford-Payne winds, their structure, illustrated in 
Figure~\ref{fig_spruit}, generically has 
three regions. Within the disk the energy density in the magnetic field, $B^2 / 8 \pi$, 
is smaller than $\rho c_s^2$, the thermal energy\footnote{We can also consider 
situations where the magnetic pressure in the disk is stronger than the gas pressure, 
though it must always be weaker than $\rho v_\phi^2$.}. Due to flux conservation, 
however, the energy in the vertical field component, $B_z^2 / 8 \pi$, is roughly 
constant with height for $z < r$, while the gas pressure typically decreases at 
least exponentially with a scale height $h \ll r$. This leads to a region above the 
disk surface where magnetic forces dominate. The magnetic force per unit 
volume is,
\begin{equation}
 \frac{ {\bf J} \times {\bf B} }{c} = - \nabla \left( \frac{B^2}{8 \pi} \right) + 
 \frac{ {\bf B} \cdot \nabla {\bf B} }{4 \pi},
\end{equation} 
where the current,
\begin{equation}
 {\bf J} = \frac{c}{4 \pi} \nabla \times {\bf B}.
\end{equation} 
The force, in general, can be written as shown above as the sum of a magnetic pressure 
gradient and a force due to magnetic tension. In the disk wind region where magnetic 
forces dominate, the requirement that they exert a finite acceleration on the low density 
gas implies that the force approximately vanishes, i.e. that,
\begin{equation}
 {\bf J} \times {\bf B} \approx 0.
\end{equation}
The structure of the magnetic field in the magnetically dominated region is then described as 
being ``force-free", and in the disk wind case (where $B$ changes slowly with $z$) the field 
lines must be approximately straight to ensure that the magnetic tension term is also  
small. If the field lines support a wind, the force-free structure persists up to where the 
kinetic energy density in the wind, $\rho v^2$, first exceed the magnetic energy density. This 
is called the Alfv\'en surface. Beyond the Alfv\'en surface, the inertia of the gas in the wind 
is sufficient to bend the field lines, which tend to wrap up into a spiral structure as the disk 
below them rotates.

\begin{figure}[t]
\center
\includegraphics[width=0.5\columnwidth]{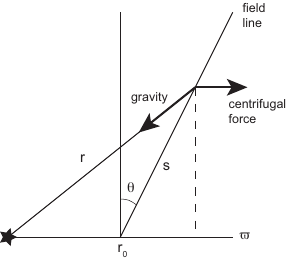}
\caption{Geometry for the calculation of the critical angle for magneto-centrifugal wind launching. A 
magnetic field line $\bf s$, inclined at angle $\theta$ from the disk normal, enforces rigid rotation at the 
angular velocity of the foot point, at cylindrical radius $\varpi = r_0$ in the disk. Working in the 
rotating frame we consider the balance between centrifugal force and gravity.}
\label{fig_wind_geometry}
\end{figure}

Magneto-centrifugal driving can launch a wind from the surface of a cold gas disk if the 
magnetic field lines are sufficiently inclined to the disk normal. The critical inclination angle 
in ideal MHD can be derived via an {\em exact} mechanical analogy. To proceed, we 
note that in the force-free region the magnetic field lines are (i) basically straight lines, and 
(ii) enforce rigid rotation out to the Alfv\'en surface at an angular velocity equal to that 
of the disk at the field line's footpoint. The geometry is shown in Figure~\ref{fig_wind_geometry}. 
We consider a field line that intersects the disk 
at radius $r_0$, where the angular velocity is $\Omega_0 = \sqrt{GM_*/r_0^3}$, and that 
makes an angle $\theta$ to the disk normal. We define the spherical polar radius $r$, 
the cylindrical polar radius $\varpi$, and measure the distance along the field line from its 
intersection with the disk at $z=0$ as $s$. In the frame co-rotating with $\Omega_0$ 
there are no magnetic forces along the field line to affect the acceleration of a wind; 
the sole role of the magnetic field is to constrain the gas to move along a straight line 
at constant angular velocity. Following this line of argument, the acceleration of a 
wind can be fully described in terms of an effective potential,
\begin{equation}
 \Phi_{\rm eff} (s) = -\frac{GM_*}{r(s)} -\frac{1}{2} \Omega_0^2 \varpi^2 (s).
\end{equation}
The first term is the gravitational potential, while the second describes the 
centrifugal potential in the rotating frame. 

\begin{figure}[t]
\includegraphics[width=\columnwidth]{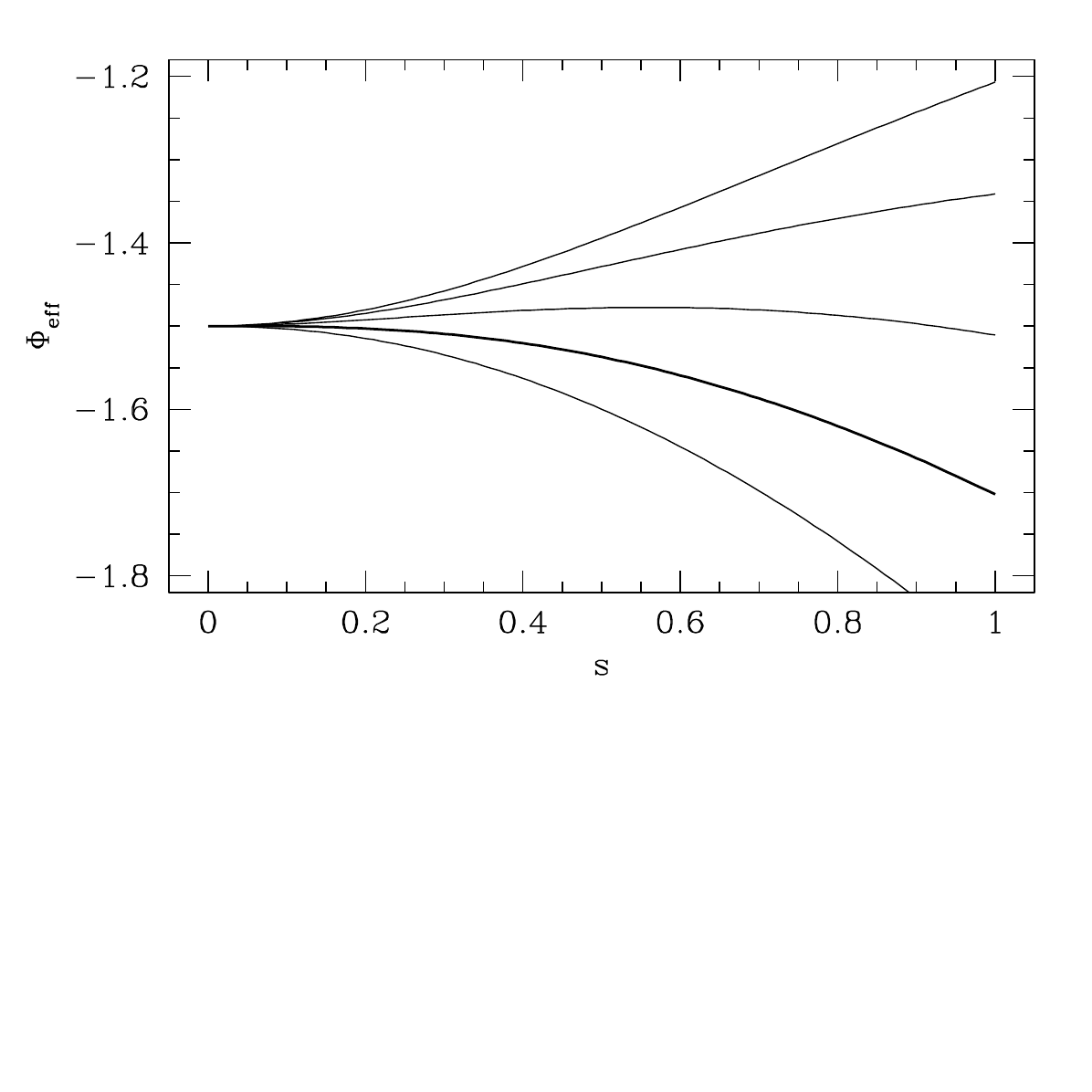}
\vspace{-4.5cm}
\caption{The variation of the disk wind effective potential $\Phi_{\rm eff}$ (in arbitrary units) 
with distance $s$ along a field line. From top downwards, the curves 
show field lines inclined at $0^\circ$, $10^\circ$, $20^\circ$, $30^\circ$ (in bold) and $40^\circ$ 
from the normal to the disk surface. For angles of $30^\circ$ and more from the vertical, there is 
no potential barrier to launching a cold MHD wind directly from the disk surface.}
\label{fig_wind_potential}
\end{figure}

Written out explicitly, the effective potential is,
\begin{equation}
 \Phi_{\rm eff} (s) = -\frac{GM_*}{ (s^2 + 2 s r_0 \sin \theta + r_0^2 )^{1/2}} 
 - \frac{1}{2} \Omega_0^2 \left( r_0 + s \sin \theta \right)^2.
\label{eq_eff_potential} 
\end{equation} 
This function is plotted in Figure~\ref{fig_wind_potential} for various values of the angle $\theta$. 
If we consider first a vertical field line ($\theta = 0$) the effective potential is a monotonically increasing function 
of distance $s$. For modest values of $\theta$ there is a potential barrier defined by a maximum 
at some $s = s_{\rm max}$, while for large enough $\theta$ the potential decreases from $s=0$. 
In this last case purely magneto-centrifugal forces suffice to accelerate a wind off the disk 
surface, even in the absence of any thermal effects. The critical inclination angle of the field 
can be found by computing $\theta_{\rm crit}$, specified though the condition,
\begin{equation}
 \left. \frac{\partial^2 \Phi_{\rm eff}}{\partial s^2} \right\vert_{s=0} = 0.
\end{equation} 
Evaluating this condition, we find,
\begin{eqnarray} 
 1 - 4 \sin^2 \theta_{\rm crit} & = & 0 \nonumber \\
 \Rightarrow \theta_{\rm crit} & = & 30^\circ,
\end{eqnarray}
as the minimum inclination angle from the vertical needed for unimpeded wind launching 
in ideal MHD \cite{blandford82}. Since most of us are more familiar with mechanical rather 
than magnetic forces, this derivation in the rotating frame offers the easiest route to this 
result. But it can, of course, be derived just as well by working in the inertial frame of 
reference \cite{spruit96}.

The rigid rotation of the field lines interior to the Alfv\'en surface means that gas being 
accelerated along them increases its specific angular momentum. The magnetic field, 
in turn, applies a torque to the disk that removes a corresponding amount of angular 
momentum. If a field line, anchored to the disk at radius $r_0$, crosses the Alfv\'en 
surface at (cylindrical) radius $r_A$, it follows that the angular momentum flux is,
\begin{equation}
 \dot{L}_w = \dot{M}_w \Omega_0 r_A^2,
\end{equation}
where $\dot{M}_w$ is the mass loss rate in the wind. Removing angular momentum 
at this rate from the disk results in a local accretion rate $\dot{M} = 2 \dot{L}_w / \Omega_0 r_0^2$. 
The ratio of the disk accretion rate to the wind loss rate is,
\begin{equation}
 \frac{\dot{M}}{\dot{M}_w} = 2 \left( \frac{r_A}{r_0} \right)^2.
\end{equation} 
If $r_A$ substantially exceeds $r_0$ (by a factor of a few, which is reasonable for detailed
disk wind solutions) a relatively weak wind can carry away enough angular momentum 
to support a much larger accretion rate.

The behavior of a disk that evolves under wind angular momentum loss depends on how 
the wind and the poloidal magnetic field respond to the induced accretion. It is not 
immediately obvious that a steady accretion flow is even possible. The form of the 
effective potential (Figure~\ref{fig_wind_potential}) suggests that the rate of mass and 
angular momentum loss in the wind ought to be a strong function of the inclination 
of the field lines --- for $\theta < 30^\circ$ there is a potential barrier to wind launching, 
while for $\theta \ge 30^\circ$ there is no barrier at all. How $\theta$ responds to 
changes in the inflow rate through the disk is of critical importance \cite{lubow94b,cao02,ogilvie01}, 
and there is no simple analog of the diffusive disk evolution equation.
Despite this, viscous and wind-driven disks exhibit some 
qualitative difference that may enable observational tests. The classical test is the 
evolution of the outer disk radius, which expands in viscous models ({\em if} there is 
no mass loss, even in the form of a thermal wind) but contracts if an MHD wind 
dominates. Old and almost forgotten observations \cite{smak84} of disk radius changes in 
dwarf novae (accreting white dwarfs in mass transfer binary systems) provided 
empirical support for viscous disk evolution {\em in those specific systems}. Disk 
winds also remove energy, and so another potential test is to look for evidence of the 
dissipation of accretion energy within the disk that is present in viscous models but 
absent for winds. At fixed accretion rate a wind-driven disk will have a lower effective 
temperature at small radii than its viscous counterpart, and this will alter the predicted 
spectral energy distribution (at large radii the temperature of both types of disk is 
set by irradiation, and no significant differences are expected).

\subsubsection{Magnetic field transport}
\label{sec_field_transport}
The strength and radial profile of the vertical magnetic field threading the disk are 
important quantities for disk winds, and for turbulence driven by MHD processes. 
Disks form from the collapse of magnetized molecular clouds, and it is inevitable 
that they will inherit non-zero flux at the time of formation. The poloidal component 
of that flux can subsequently be advected radially with the disk gas, diffuse relative 
to the gas, or (if the flux has varying sign across the disk) reconnect.

A theory for the radial transport of poloidal flux within geometrically thin accretion 
disks was developed by Lubow, Papaloizou \& Pringle (1994) \cite{lubow94}. They 
considered a disk within which turbulence generates an effective viscosity $\nu$ and 
an effective magnetic diffusivity $\eta$. The disk is threaded by a vertical magnetic 
field $B_z(r,t)$, which is supported by a combination of currents within the disk and 
(potentially) a current external to the disk. Above the disk, as in Figure~\ref{fig_spruit} 
the field is force-free. The field lines bend within the disk, such that the poloidal field 
has a radial component $B_{rs} (r,t)$ at the disk surface.

To proceed (following the notation in \cite{guilet14}), we first write the poloidal field in terms of a magnetic flux function $\psi$, 
such that ${\bf B} = \nabla \psi \times {\bf e}_\phi$, where ${\bf e}_\phi$ is a unit 
vector in the azimuthal directions. The components of the field are,
\begin{eqnarray}
 B_r & = & - \frac{1}{r} \frac{\partial \psi}{\partial z}, \nonumber \\
 B_z & = & \frac{1}{r} \frac{\partial \psi}{\partial r}.
\end{eqnarray}  
From the second of these relations we note that $\psi$ is (up to a factor of $2 \pi$) just the 
vertical magnetic flux interior to radius $r$. We split $\psi$ into two pieces, a piece $\psi_{\rm disk}$ 
due to currents within the disk, and a piece $\psi_\infty$ due to external currents (``at infinity"),
\begin{equation}
 \psi = \psi_{\rm disk} + \psi_{\infty}.
\end{equation}
The external current generates a magnetic field that is uniform across the disk. 

With these definitions, the evolution of the poloidal field in the simplest analysis \cite{lubow94,guilet14} 
obeys,
\begin{equation}
 \frac{\partial \psi}{\partial t} + r \left( v_{\rm adv} B_z + v_{\rm diff} B_{rs} \right) = 0,
\label{eq_flux_evolve} 
\end{equation}
where $v_{\rm adv}$ is the advective velocity of magnetic flux and $v_{\rm diff}$ its 
diffusive velocity due to the turbulent resistivity within the disk. The disk component of the 
flux function is related to the surface radial field via an integral over the disk. Schematically,
\begin{equation}
 \psi_{\rm disk} (r) = \int_{r_{\rm in}}^{r_{\rm out}} F(r,r^\prime) B_{rs} (r^\prime) {\rm d}r^\prime,
\label{eq_flux_integral} 
\end{equation} 
where $F$ is a rather complex function that can be found in Guilet \& Ogilvie (2014) \cite{guilet14}.  
The appearance of this integral reflects the inherently global nature of the problem --- a current at some 
radius within the disk affects the poloidal magnetic field everywhere, not just locally --- and makes 
analytic or numerical solutions for flux evolution more difficult. Nonetheless, equation~(\ref{eq_flux_integral})
can be inverted to find $B_{rs}$ from $\psi$, after which the more familiar equation~(\ref{eq_flux_evolve}) 
can be solved for specified transport velocities to determine the flux evolution.

Equation~(\ref{eq_flux_evolve}) expresses a simple competition, the inflow of gas toward the star will 
tend to drag in poloidal magnetic field, but this will set up a radial gradient and be opposed by diffusion. 
The physical insight of Lubow et al. \cite{lubow94} was to note that although both of these 
are processes involving turbulence (and, very roughly, we might guess that $\nu \sim \eta$), 
the scales are quite distinct. From Figure~\ref{fig_spruit}, we note that because the field lines bend 
{\em within the disk}, a moderately inclined external field (with $B_{rs} \sim B_z$) above the disk 
only has to diffuse across a scale $\sim h$ to reconnect with its oppositely directed counterpart 
below the disk. Dragging in the field with the mean disk flow, however, requires angular momentum 
transport across a larger scale $r$. In terms of transport velocities, in a steady-state we have,
\begin{eqnarray}
 v_{\rm adv} & \sim & \frac{\nu}{r}, \nonumber \\
 v_{\rm diff} & \sim & \frac{\eta}{h} \frac{B_{\rm rs}}{B_z},
\end{eqnarray}
so that for $\nu \sim \eta$ and $B_{rs} \sim B_z$ diffusion beats advection by a factor $\sim (h/r)^{-1} \gg 1$. 
Defining the magnetic Prandtl number $P_m = \nu / \eta$ as the ratio of the turbulent viscosity to the 
turbulent resistivity, we would then expect that in steady-state \cite{lubow94},
\begin{equation}
 \frac{B_{rs}}{B_z} \sim \frac{h}{r} P_m.
\end{equation}  
This argument is the origin of the claim that {\em thin disks do not drag in external magnetic fields}.
It suggests that obtaining enough field line bending to launch a magneto-centrifugal wind ought to be 
hard, and that whatever primordial flux the disk is born with may be able to escape easily.

The physical arguments given above are robust, but a number of authors have emphasized that 
the calculation of the transport velocities that enter into equation~(\ref{eq_flux_evolve}) involves 
some subtleties \cite{ogilvie01,lovelace09,guilet12,guilet14,takeuchi14}. The key point is that 
the viscosity and resistivity that enter into the equation for flux transport should not be computed 
as density-weighted vertical averages, but rather (in the case of the induction equation) as 
conductivity-weighted averages \cite{guilet12}. This makes a large difference for protoplanetary 
disks, where the conductivity is both generally low, and highest near the disk surface where the 
density is small. The derived transport velocities are, moreover, functions of the poloidal field strength, 
in the sense that diffusion becomes relatively less efficient as the field strength decreases. It should 
be noted that none of the flux transport calculations fully includes all of the MHD effects expected 
to be present in protoplanetary disks (see \S\ref{sec_mhd_transport}). It seems possible, though, 
that the variable efficiency of flux diffusion could simultaneously allow,
\begin{itemize}
\item
For rapid flux loss from the relatively strongly magnetized disks formed from star formation \cite{li14}, 
averting overly rapid wind angular momentum loss that would be inconsistent with 
observed disk lifetimes.
\item
For convergence toward a weak but non-zero net poloidal flux (possibly with a ratio of thermal to 
poloidal field magnetic pressure at the mid-plane $\beta \sim 10^4-10^7$) later in the disk lifetime \cite{guilet14}.
\end{itemize}
As we will discuss in the next section, poloidal field strengths in roughly this range are of interest for their 
role in stimulating MHD instabilities within weakly ionized disks, so this is a speculative but interesting scenario.

\begin{svgraybox}
\begin{itemize}
\item Isolated protoplanetary disks are expected to evolve under a combination of (i) internal 
redistribution of angular momentum, often referred to informally as viscosity, (ii) mass loss 
due to accretion and thermal winds, and (iii) mass and angular momentum loss in MHD winds. 
The relative importance of these processes for disk evolution is not clearly established.
\item In the limit where local internal redistribution of angular momentum dominates, the disk 
surface density evolves according to a diffusion equation that matches the one derived for 
a viscous fluid. The viscosity in this equation must be interpreted as an effective viscosity 
resulting from a turbulent process.
\item The use of the $\alpha$ prescription, and the extension of viscous models to complex 
non-axisymmetric geometries, are common and sometimes useful approximations. Neither, 
however, has a clear physical justification.
\item The dissipation of accretion energy within the disk modifies its vertical temperature structure, 
and can lead to much higher mid-plane temperatures within a few AU of the star.
\item Magnetic winds can be launched along sufficiently inclined field lines that thread the 
disk surface. The long-term evolution of such winds depends upon the efficiency of 
magnetic flux transport relative to gas accretion.
\end{itemize}
\end{svgraybox}

\newpage

\section{Turbulence}
\label{sec_turbulence}

Turbulence within protoplanetary disks is important for two independent reasons. 
First, if it is strong enough and has the right properties, it could account for disk 
evolution by redistributing angular momentum much faster than molecular viscosity. 
Second, turbulence has its fingers in a plethora of planet formation processes, 
ranging from the collision velocities of small particles \cite{ormel07} to the 
formation of planetesimals \cite{johansen14} and the migration rate of low-mass 
planets \cite{kley12}. For these reasons we would like to understand disk turbulence, 
even if (as is possible) it is not always responsible for disk evolution.

The first order of business when considering possibly turbulent fluid systems is 
usually to estimate the Reynolds number, which is a dimensionless measure of 
the relative importance of inertial and viscous forces. For a system with 
characteristic size $L$, velocity $U$, and (molecular) viscosity $\nu$, the 
Reynolds number is defined as,
\begin{equation}
 {\rm Re} = \frac{UL}{\nu}.
\label{eq_Re} 
\end{equation} 
There is no unique or ``best" definition of $U$ and $L$ for protoplanetary disks, 
but whatever choice we make gives a very large number. For example, taking 
$L=h$ and $U=c_s$ then our estimate of the viscosity at 1~AU 
(equation~\ref{eq_nu_estimate}) implies ${\rm Re} \sim 10^{11}$. By 
terrestrial standards this is an enormous Reynolds number. Experiments on 
flow through pipes, for example --- including those of Osborne Reynolds himself --- 
show that turbulence is invariably present once ${\rm Re} > 10^4$ \cite{eckhardt07}. 
If turbulence is present within protoplanetary disks there is no doubt that viscous 
forces will be negligible on large scales, and the turbulence will exhibit a broad 
inertial range.

\begin{figure}[t]
\center
\includegraphics[width=0.85\columnwidth]{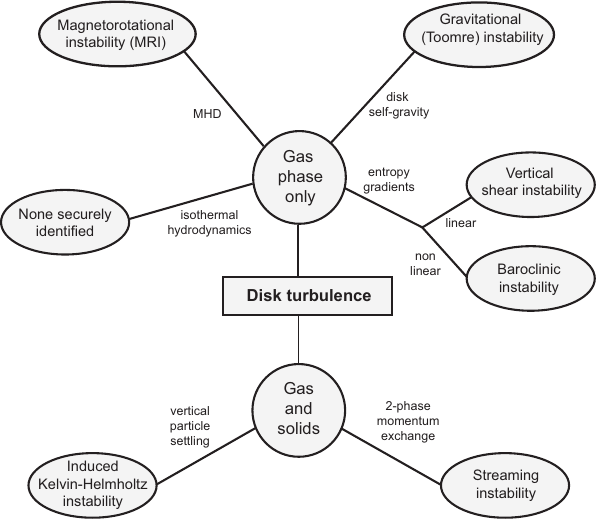}
\caption{A menu of the leading suspects for creating turbulence within protoplanetary disks.}
\label{fig_turbulence}
\end{figure}

\begin{figure}[t]
\center
\includegraphics[width=\columnwidth]{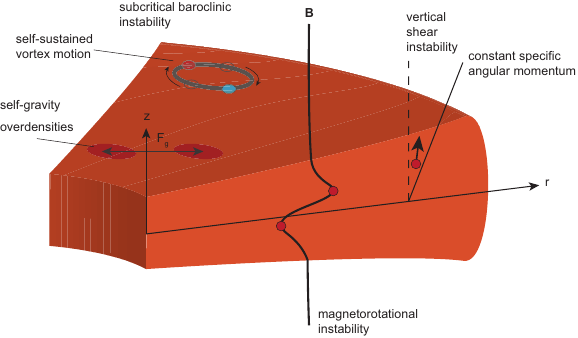}
\caption{A summary of the most important instabilities that can be present in protoplanetary disks. {\em Self-gravity} 
is important for sufficiently massive and cold disks. It leads to spiral arms and gravitational torques between regions 
of over-density. The {\em magnetorotational instability} occurs whenever a weak magnetic field is sufficiently coupled 
to differential rotation. The magnetic field acts to couple fluid elements at different radii, leading to an instability 
that can sustain MHD turbulence and angular momentum transport. The {\em vertical shear instability} feeds off the 
vertical shear that is set up in disks with realistic temperature profiles. It is a linear instability characterized by near-vertical 
growing modes. The {\em subcritical baroclinic instability} is a non-linear instability that operates in the presence of a 
sufficiently steep radial entropy gradient. It resembles radial convection, and leads to self-sustained vortices within the 
disk.}
\label{fig_instabilities}
\end{figure}

Figure~\ref{fig_turbulence} lists some of the possible sources of turbulence 
within protoplanetary disks. It's a long list! We can categorize the candidates 
according to various criteria,
\begin{itemize}
\item
The physics involved in generating the turbulence. The simplest possibility (which 
appears unlikely) is that turbulence develops spontaneously in an isothermal, 
purely hydrodynamical shear flow. More complete physical models invoke 
entropy gradients, disk self-gravity or magnetic fields as necessary elements 
for the origin of turbulence.
\item
The origin of the free energy that sustains the turbulence, which could be the 
radial or vertical shear, heating from the star, or velocity differences between 
gas and solid particles.
\item
The character of the instabilities proposed to initiate turbulence from an initially 
non-turbulent flow. {\em Linear} instabilities grow exponentially from arbitrarily 
small perturbations, while {\em non-linear} instabilities require a finite amplitude 
disturbance. Demonstrating the existence of linear instabilities is relatively 
easy, whereas proving that a fluid system is non-linearly stable is very hard.
\item
The species involved. In this section we concentrate on instabilities present 
in purely gaseous disks; additional instabilities are present once we consider 
how gas interacts aerodynamically with its embedded solid component (\S\ref{sec_streaming}).
\end{itemize}
Figure~\ref{fig_instabilities} illustrates the dominant fluid motions or forces involved in some 
of the most important disk instabilities.

For each candidate instability we would like to know the disk conditions under which 
it would be present, its growth rate, and the strength and nature of the turbulence 
that eventually develops. For disk evolution we are particularly interested in how 
efficiently the turbulence transports angular momentum (normally characterized by 
an effective $\alpha$). In most cases the efficiency of transport can only be determined 
using numerical simulations, whose fluctuating velocity and magnetic fields can be analyzed 
to determine $\alpha$ via the relation \cite{balbus98},
\begin{equation}
 \alpha = \left\langle \frac{ \delta v_r \delta v_\phi }{c_s^2} - \frac{ B_r B_\phi }{4 \pi \rho c_s^2} \right\rangle_{\rho},
\label{eq_def_alpha} 
\end{equation} 
where the angle brackets denote a density weighted average over space (and time, in some instances). 
The first term in this expression is the Reynolds stress from correlated fluctuations in the radial and 
perturbed azimuthal velocity, the second term is the Maxwell stress from MHD turbulence. We speak of the 
stress as being ``turbulent" if the averages in the above relation are dominated by contributions from small 
spatial scales. It is also possible for a disk to sustain large scale stresses --- for example at some radius we 
might have non-zero {\em mean} radial and azimuthal magnetic fields --- which are normally described as being 
``laminar". Note that it is possible for the velocity field to exhibit turbulence on small scales even if the stress is 
dominated by large scale contributions.

\subsection{Hydrodynamic turbulence}
The dominant motion in protoplanetary disks is Keplerian orbital motion about a central 
point mass. Simplifying as much as possible, we first ask whether, in the absence of 
magnetic fields\footnote{Ignoring magnetic fields in astrophysical accretion flows is 
generally a stupid thing to do, and indeed there is broad consensus that the magnetorotational 
instability (MRI) \cite{balbus98} is responsible for turbulence and angular momentum transport 
in most accretion disks. In protoplanetary disks, however, the low ionization fraction means 
that the dominance of MHD instabilities is much less obvious, and purely hydrodynamic 
effects could in principle be important.}, the radial shear present in a low-mass disk would be unstable to the 
development of turbulence. We first consider (rather unrealistically) a radially isothermal 
disk, where according to equation~(\ref{eq_vertical_shear}) there is no {\em vertical} shear. 
We then turn to the more general case where the temperature varies with radius, giving 
rise both to vertical shear and qualitatively distinct possibilities for instability. 

\subsubsection{Linear and non-linear stability}
The linear stability of a shear flow with a smoothly varying $\Omega(r)$ against axisymmetric 
perturbations is given by Rayleigh's criterion (this is derived in most fluids textbooks, see e.g. 
\cite{pringle07}). The flow is stable if the specific angular momentum increases with radius,
\begin{equation}
 \frac{{\rm d}l}{{\rm d} r} = \frac{{\rm d}}{{\rm d} r} \left( r^2 \Omega \right) > 0 \rightarrow {\rm stability}.
\label{eq_rayleigh} 
\end{equation} 
A Keplerian disk has $l \propto \sqrt{r}$ and is linearly stable.

There is no mathematical proof of the non-linear stability of Keplerian shear flow, but nor is there 
any known instability. The apparently analogous cases of pipe flow and Cartesian shear flows --- 
which are linearly stable but undergo non-linear transitions to turbulence --- are in fact sufficiently 
different problems as to offer no guidance \cite{balbus98}. There are analytic and numerical 
arguments against the existence of non-linear instabilities \cite{balbus96}, which although not 
decisive \cite{rebusco09} essentially rule out the hypothesis that a non-linear instability could 
result in {\em astrophysically interesting} levels of turbulence \cite{lesur05}. The same conclusion 
follows from laboratory experiments that have studied the stability of quasi-Keplerian rotation 
profiles in Taylor-Couette experiments \cite{edlund14}. One caveat is that 
laboratory experiments, and most theoretical work, consider the stability of unstratified 
cylindrical shear flows. Marcus and collaborators have identified a new instability (of a distinct character, 
related to the existence of locations in the flow known as critical layers, for a review of this physics see 
\cite{maslowe86}) that can arise when the vertical stratification present in disks is included \cite{marcus13,marcus14}. 
The existence of this instability, which leads to self-replication of vortices, has been reproduced 
independently \cite{lesur16}, and also shown to depend upon the radiative properties of the disk. 
Conditions in most regions of protoplanetary disks do not appear especially propitious for it 
to play a major role, but investigations remain in their infancy at the time of writing.

\subsubsection{Entropy-driven instabilities}
A separate class of purely hydrodynamic instabilities (no self-gravity, no magnetic fields) are 
what might loosely be called ``entropy-driven" instabilities, in that they rely on the existence of 
a non-trivial temperature structure. The prototypical entropy-driven instability is of course 
convection, which could occur in the vertical direction if dissipation (associated with the 
physical process behind angular momentum transport) sets up an unstable entropy profile. This 
is evidently only conceivable in the region where viscous dissipation dominates, as 
irradiation prefers a nearly isothermal vertical structure. Even there, convective turbulence in 
disks is less efficient at transporting angular momentum than it is in transporting heat \cite{lesur10}, 
and this disparity creates a formidable barrier to creating consistent models in which convection 
is the primary source of disk turbulence. Convection may still be present in some regions of disks, 
perhaps especially at high accretion rates, but as a byproduct of independent angular momentum 
transport processes (for an example in dwarf novae, see \cite{hirose14}).

A disk that has a radial temperature gradient necessarily has vertical shear (equation~\ref{eq_vertical_shear}). 
The free energy associated with the vertical shear can be accessed via the vertical shear instability (VSI) 
analyzed by Nelson, Gressel \& Umurhan \cite{nelson13}. The VSI is a disk application of the 
Goldreich-Schubert-Fricke instability \cite{goldreich67,fricke68} of rotating stars, and was proposed as 
a source of protoplanetary disk transport by Urpin \& Brandenburg \cite{urpin98}\footnote{As we 
shall see, a general rule is that all disk instabilities have long histories and pre-histories.}. The VSI is a linear 
instability with a maximum growth rate that is of the order of $h \Omega_{\rm K}$ \cite{lin15}, but 
which is strongly dependent on the radiative properties of the disk. The reason is that to access the 
free energy in the vertical shear requires vertical fluid displacements, which are easy in the limit that 
the disk is strictly vertically isothermal but strongly suppressed if it is stably stratified. The local 
{\em cooling time} of the fluid is thus a critical parameter, and the VSI will only operate in regions 
of the disk where radiative cooling and heating processes result in a cooling time that is the same or 
shorter than the dynamical time $\Omega_{\rm K}^{-1}$. This, in practice, restricts the application 
of the VSI to specific radii that depend upon the disk structure 
(Lin \& Youdin suggest 5-50~AU \cite{lin15}, Malygin et al. 15-180~AU \cite{malygin17}), and limits its 
effectiveness if the dust opacity is reduced (due to coagulation into large particles). Under the 
right conditions, however, numerical simulations suggest that the VSI can generate relatively 
small but possibly significant levels of transport, with both Nelson et al. \cite{nelson13} and Stoll \& Kley \cite{stoll14} 
finding $\alpha$ of a few $\times 10^{-4}$.

The radial entropy gradient may itself be unstable. The simplest instability would be radial 
convection (a linear instability). For a disk with pressure profile $P(r)$, density profile $\rho(r)$, and 
adiabatic index $\gamma$, we define the Brunt-V\"ais\"al\"a frequency, 
\begin{equation}
 N_r^2 = - \frac{1}{\gamma \rho} \frac{{\rm d}P}{{\rm d}r} 
 \frac{\rm d}{{\rm d}r} \ln \left( \frac{P}{\rho^\gamma} \right). 
\end{equation}
The Solberg-Ho\"iland criterion indicates that a Keplerian disk is convectively unstable if, 
\begin{equation}
 N_r^2 + \Omega_{\rm K}^2 < 0.
\end{equation} 
Protoplanetary disks never (or at least almost never) have a steep enough profile of entropy 
to meet this condition, so radial convection will not set in. A different instability (the subcritical 
baroclinic instability, SBI) is possible, however, if the weaker condition $N_r^2 < 0$ (which 
is just the Schwarzschild condition for non-rotating convection) is satisfied 
\cite{petersen07a,petersen07b,lesur10b}. The SBI, which is likely related to observations 
of vortex formation in earlier numerical simulations \cite{klahr03}, is a non-linear 
instability that can be excited by finite amplitude perturbations. (Confusingly, it is 
unrelated to the linear ``baroclinic instability" studied in planetary atmospheres.) The SBI relies on radial 
buoyancy forces to sustain vortical motion via baroclinic driving. This type of effect 
is possible in disks in which surfaces of constant density are not parallel to surface 
of constant pressure. Mathematically, for a fluid with vorticity ${\bf \omega} = \nabla \times {\bf v}$, 
we can take the curl of the momentum equation to get an equation for the vortensity 
${\bf \omega} / \rho$,
\begin{equation}
 \frac{\rm D}{{\rm D}t} \left( \frac{\bf \omega}{\rho} \right) = 
 \left( \frac{\bf \omega}{\rho} \right) \cdot \nabla {\bf v} 
 - \frac{1}{\rho} \nabla \left( \frac{1}{\rho} \right) \times \nabla P.
\label{eq_vortensity}
\end{equation}
The baroclinic term, which for the SBI is responsible for generating and maintaining 
vorticity in the presence of dissipation, is the second term on the right hand side. The SBI, 
as with the VSI, is sensitive to the cooling time \cite{lesur10b,raettig13}, in this case because the baroclinic 
driving depends on the disk neither cooling too fast (which would eliminate the  
buoyancy effect) nor too slow (which would lead to constant temperature around the 
vortex). In compressible simulations, Lesur \& Papaloizou \cite{lesur10b} 
found that under favorable disk conditions the SBI could lead to outward transport 
of angular momentum with $\alpha \sim 10^{-3}$.

\subsection{Self-gravity}
\label{sec_self_gravity}
A disk is described as {\em self-gravitating} if it is unstable to the growth of 
surface density perturbations when the gravitational force between different fluid 
elements in the disk is included along with the force from the central star. For a disk 
with sound speed $c_s$, surface density $\Sigma$ and angular velocity $\Omega$ 
(assumed to be {\em close} to Keplerian) a linear analysis (for textbook 
treatments, see e.g. \cite{armitage07,pringle07}) shows that a disk becomes 
self-gravitating when the Toomre $Q$ \cite{toomre64},
\begin{equation}
 Q \equiv \frac{c_s \Omega}{\pi G \Sigma} < Q_{\rm crit},
\label{eq_toomre} 
\end{equation}
where $Q_{\rm crit} \sim 1$. We can deduce this result informally using an 
extension of the time scale argument that gives the thermal Jeans mass. 
We first note that pressure will prevent the gravitational collapse of a clump 
of gas, on scale $\Delta r$, if the sound-crossing time $\Delta r / c_s$ is 
shorter than the free-fall time $\sqrt{\Delta r^3 / G \Delta r^2 \Sigma}$. (We're 
ignoring factors of 2, $\pi$ and so on.) Equating these time scales gives 
the minimum scale that might be vulnerable to collapse as $\Delta r \sim c_s^2 / G \Sigma$. 
On larger scales, collapse can be averted if the free-fall time is longer than the 
time scale on which radial shear will separate initially neighboring fluid elements. 
For a Keplerian disk this time scale is $\sim \Omega^{-1}$. If the disk is just 
on the edge of instability the minimum collapse scale set by pressure support must 
equal the maximum collapse scale set by shear. Imposing this condition for 
marginal stability we obtain $c_s \Omega / G \Sigma \sim 1$, in accord with the 
formal result quoted above.

To glean some qualitative insight into where a disk might be self-gravitating, consider a 
steady-state disk that is described by an $\alpha$ model in which the transport arises 
from some process {\em other} than self-gravity. Collecting some previous results, 
the steady-state condition implies $\nu \Sigma = \dot{M} / 3 \pi$, the $\alpha$ 
prescription is $\nu = \alpha c_s h$, and hydrostatic equilibrium gives $h = c_s / \Omega$. 
Substituting into equation~(\ref{eq_toomre}) we find,
\begin{equation}
 Q = \frac{3 \alpha c_s^3}{G \dot{M}}.
\end{equation}  
Protoplanetary disks generically get colder (and hence have lower $c_s$) at larger 
distances from the star, and this is where self-gravity is most likely to be important.

The disk mass required for self-gravity to become important can be estimated. 
Ignoring radial gradients of all quantities, we write the disk mass $M_{\rm disk} \sim \pi r^2 \Sigma$, 
and again use the hydrostatic equilibrium result $h = c_s / \Omega$. Equation~(\ref{eq_toomre}) then 
gives, 
\begin{equation}
 \frac{M_{\rm disk}}{M_*} > \left( \frac{h}{r} \right),
\end{equation}
as the condition for instability. This manipulation of a local stability criterion into some sort 
of global condition is ugly, and begs the question of {\em where} in the disk $M_{\rm disk}$ and 
$h/r$ should be evaluated. We can safely conclude, nonetheless, that for a typical protoplanetary 
disk with $(h/r) \simeq 0.05$ a disk mass of $10^{-2} \ M_*$ will not be self-gravitating, whereas 
one with $0.1 \ M_*$ may well be.

There are two possibles outcomes of self-gravity in a disk,
\begin{itemize}
\item
The disk may establish a (quasi) stable state, characterized globally by trailing spiral 
overdensities. Gravitational torques between different annuli in the disk {\em transport 
angular momentum} outward, leading to accretion.
\item
The pressure and tidal forces, which by definition are unable to prevent the onset 
of gravitational collapse, may never be able to stop it once it starts. In this case the 
disk {\em fragments} into bound objects, which interact with (and possibly accrete) the 
remaining gas.
\end{itemize}
Both possibilities are of interest. Angular momentum transport due to self-gravity 
may be dominant, at least on large scales, at early times while the disk is still 
massive. Fragmentation, which was once considered a plausible mechanism for 
forming the Solar System's giant planets \cite{kuiper51}, remains of interest as 
a way to form sub-stellar objects and (perhaps) very massive planets. Kratter \& Lodato \cite{kratter16} 
review the physics of disk self-gravity in both the angular momentum transporting and 
fragmenting regimes. Here we summarize some mostly elementary arguments.

Gravity is a long-range force, and it is not at all obvious that we can deploy the 
machinery developed for viscous disks to study angular momentum transport in a 
self-gravitating disk. The transport could be largely non-local, driven for example 
by large-scale structures in the density field (such as bars) or by waves that 
transport energy and angular momentum a significant distance before dissipating \cite{balbus99}. 
There is no precise criterion for when self-gravitating transport can be described using a 
local theory, but numerical simulations indicate that this is a reasonable approximation 
for low-mass disks with $M_{\rm disk} / M_* \approx 0.1$ \cite{lodato04,cossins09,forgan11,steimancameron13}. 
Transport in more massive disks, such as might be present during the Class 0 and Class I phases 
of star formation, cannot be described locally (for multiple reasons, e.g. \cite{lodato05,tsukamoto15}). 

In cases where a local description of the transport is valid, we can use a thermal 
balance argument to relate the efficiency of angular momentum transport to the 
cooling time. Adopting a one-zone model for the vertical structure, we define the 
thermal energy of the disk, per unit surface area, as,
\begin{equation}
 U = \frac{ c_s^2 \Sigma}{\gamma (\gamma -1)},
\end{equation}
where $c_s$ is the mid-plane sound speed and $\gamma$ is the adiabatic index. 
The cooling time (analogous to the Kelvin-Helmholtz time for a star) is then,
\begin{equation}
 t_{\rm cool} = \frac{U}{2 \sigma T_{\rm disk}^4},
\end{equation}
where $T_{\rm disk}$ is the effective temperature. Equating the cooling rate, 
$Q_- = 2 \sigma T_{\rm disk}^4$, to the local viscous heating rate, 
$Q+ = (9/4) \nu \Sigma \Omega^2$ (equation~\ref{eq_qplus}), and adopting the 
$\alpha$-prescription (equation~\ref{eq_alpha_prescription}), we find,
\begin{equation}
 \alpha = \frac{4}{9 \gamma (\gamma-1)} \frac{1}{\Omega t_{\rm cool}}.
\end{equation} 
This relation, which is a general property of $\alpha$ disks quite independent of 
self-gravity, just says that a rapidly cooling disk needs efficient angular momentum 
transport if it to generate heat fast enough to remain in thermal equilibrium. 

For most sources of angular momentum transport we are no more able to determine 
$t_{\rm cool}$ from first principles than we are $\alpha$, so the above relation does 
not move us forward. Self-gravitating disks, however, have the unusual property 
that their Toomre $Q$, measured in the {\em saturated} (non-linear) state, is 
roughly constant and similar to the critical value $Q_{\rm crit}$ determined from 
{\em linear} theory. This property  
arises, roughly speaking, because the direct 
dependence of the linear stability criterion on temperature (via $c_s$) invites a 
stabilizing feedback loop --- a disk that cools so that $Q < Q_{\rm crit}$ is 
more strongly self-gravitating, and produces more heating, while one that 
heats so that $Q > Q_{\rm crit}$ shuts off the instability. It is therefore 
reasonable to assume that a self-gravitating disk that does not fragment 
maintains itself close to marginal stability, as conjectured by Paczynski \cite{paczynski78}.

If we assume that $Q=Q_0$ exactly (where $Q_0$ is some constant presumably close 
to $Q_{\rm crit}$) then we have enough constraints to explicitly determine the functional 
form of $\alpha$ for a self-gravitating disk. Since $Q$ depends on the mid-plane sound 
speed, $c_s = \sqrt{k_B T_c / \mu m_H}$, the condition of marginal stability directly gives 
us $T_c(\Sigma,\Omega)$,
\begin{equation}
 T_c = \pi^2 Q_0^2 G^2 \left( \frac{\mu m_H}{k_B} \right) \frac{\Sigma^2}{\Omega^2}.
\end{equation} 
We can use this to determine $t_{\rm cool}$, and from that $\alpha$, with the aid of the 
vertical structure relations developed in \S\ref{sec_viscous_heating}. To keep things 
simple, we adopt an opacity,
\begin{equation}
 \kappa_R = \kappa_0 T_c^2
\end{equation}
that is appropriate for ice grains, and assume the disk is optically thick. The opacity law, 
together with the relations for the optical depth, $\tau = (1/2) \Sigma \kappa_R$, and 
the mid-plane temperature, $T_c^4 / T_{\rm disk}^4 \simeq (3/4) \tau$, then leads to,
\begin{equation}
 \alpha = \frac{64 \pi^2 Q_0^2 G^2 \sigma}{27 \kappa_0} 
 \left( \frac{\mu m_H}{k_B} \right)^2 \Omega^{-3},
\label{eq_alpha_sg} 
\end{equation} 
which coincidentally (for this opacity law) is only a function of $\Omega$. It may look 
cumbersome --- and the numerical factors are {\em certainly} not to be trusted --- 
but what we have shown is that for a locally self-gravitating disk $\alpha$ is 
simply a constant times a determined function of $\Sigma$ and $\Omega$. 
This result allows for the evolution of low-mass self-gravitating disks to be 
modeled as a pseudo-viscous process \cite{levin07,clarke09,rafikov09}.

Self-gravity is typically important in protoplanetary disks at large radii, where 
irradiation is usually the dominant factor determining the disk's thermal 
state (except at high accretion rates). The generalization of the self-regulation 
argument given above is obvious; if irradiation is not so strong as to stabilize 
the disk on its own then viscous heating from the self-gravitating ``turbulence" 
has to make up the difference. The partially irradiated regime of self-gravitating 
disks has been studied using local numerical simulations \cite{rice11}, and the 
analytic generalization for the effective $\alpha$ that results can be found 
in Rafikov \cite{rafikov15}.

Ignoring irradiation again (and trusting the numerical factors that we just 
said were not to be trusted) we can examine the implications of equation~(\ref{eq_alpha_sg}) 
for protoplanetary disks. Taking $Q_0 = 1.5$ and $\kappa_0 = 2 \times 10^{-4} \ {\rm cm^2 \ g^{-1} \ K^{-2}}$ \cite{bell94} 
we find, across the region of the disk where ice grains would be the dominant opacity source, that,
\begin{equation}
 \alpha \sim 0.3 \left( \frac{r}{50 \ {\rm AU}} \right)^{9/2}.
\label{eq_sg_alpha_r} 
\end{equation} 
The steep radial dependence of the estimated self-gravitating $\alpha$ means that, 
unless all other sources of transport are {\em extremely} small, it will play no role in the 
inner disk. In the outer disk, on the other hand, we predict vigorous transport. The 
physical origin of the transport is density inhomogeneities that are caused by self-gravity, 
which become increasing large as $\alpha$ grows (explicitly, it is found \cite{cossins09} that 
the average fractional surface density perturbation $\delta \Sigma / \Sigma \propto \alpha^{1/2}$). 
Since even the linear threshold for gravitational instability implies that pressure forces can 
barely resist collapse, we expect that beyond some critical strength of turbulence a self-gravitating 
disk will be unable to maintain a steady-state. Rather, it will fragment into bound objects that are 
not (at least not immediately) subsequently sheared out or otherwise disrupted.

Our discussion up to this point might lead one to conjecture that the threshold for fragmentation 
could be written in terms of a critical dimensionless cooling time,
\begin{equation}
 \beta_{\rm crit} \equiv \Omega t_{\rm cool, crit},
\end{equation}
or via a maximum $\alpha_{\rm crit}$ that a self-gravitating disk can sustain without fragmenting (these are 
almost equivalent, but defining the threshold in terms of $\alpha$ incorporates the varying compressibility 
as expressed through $\gamma$). Gammie \cite{gammie01}, using local two dimensional 
numerical simulations, obtained $\beta_{\rm crit} \simeq 3$ for a two-dimensional adiabatic index 
$\gamma=2$. Early global simulations by Rice et al. \cite{rice03}, which were broadly consistent with Gammie's estimate, implied a 
maximum effective transport efficiency $\alpha_{\rm crit} \simeq 0.1$ \cite{rice05}. 

The idea that the fragmentation threshold is uniquely determined by a single number is too 
simplistic. Several additional physical effects matter. First, fragmentation requires 
that collapsing clumps can radiate the heat generated by adiabatic compression. 
There is therefore a dependence not just on the magnitude of the opacity, but also on 
how it scales with density and temperature \cite{johnson03,cossins10}. Second, if we 
view fragmentation as requiring a critical over-density in a random turbulent field there 
should be a time scale dependence, with statistically rarer fluctuations that 
lead to collapse becoming probable the longer we wait \cite{paardekooper12}. (This 
introduces an additional implicit dependence on $\gamma$, because the statistics of 
turbulent density fields depend upon how compressible the gas is \cite{federrath15}.) Finally, 
disks can be prompted to fragment not only if they cool too quickly, but also if they 
accrete mass faster than self-gravity can transport it away. This regime is clearly 
relevant for Class~0 and Class~I disks, where envelope accretion is ongoing. Accretion-induced fragmentation 
appears inevitable for very massive disks, where it would lead to binary formation \cite{kratter10}. 

In addition to these physical complexities, work by Meru \& Bate \cite{meru11} initiated a debate as 
to whether the critical cooling time scale derived from early simulations was robust. A number of 
simulations --- run with both Smooth Particle Hydrodynamics (SPH) and grid-based methods --- showed 
an increase in $\beta_{\rm crit}$ at higher numerical resolution, with little or no evidence for 
convergence. Recent work has revisited the problem using a Godunov-type mesh-free Lagrangian 
hydrodynamics scheme as implemented within the GIZMO code \cite{hopkins15}. The use of a Riemann 
solver allows for significantly less numerical dissipation compared to SPH (which is also 
mesh-free and Lagrangian), which appears to be at the root of the convergence problem. Using 
GIZMO, Deng et al. \cite{deng17} obtain a converged estimate $\beta_{\rm crit} \approx 3$ from 
global simulations of isolated self-gravitating disks with a disk-to-star mass ratio of 0.1. Baehr et al. 
\cite{baehr17} obtain the same threshold value from local three-dimensional simulations at 
high resolution. The fact that the most recent work largely agrees with the earliest relatively crude 
simulations, but not with the assuredly better calculations carried out in the intervening years, 
appears to be coincidental.

We can combine numerical estimates of $\alpha_{\rm crit}$ with the formula for 
$\alpha (r)$ (equation~\ref{eq_sg_alpha_r}) to determine where isolated protoplanetary disks 
ought to be vulnerable to fragmentation. For a Solar mass star, fragmentation 
is expected beyond $r \sim 10^2 \ {\rm AU}$ \cite{matzner05,clarke09,rafikov09}, with an uncertainty in that estimate of 
perhaps a factor of two. In most (but perhaps not all) cases, it is expected that the disk conditions 
that allow fragmentation would lead to objects with masses in the brown dwarf regime, or above \cite{kratter10b}.

\subsection{Magnetohydrodynamic turbulence and transport}
The Rayleigh stability criterion (equation~\ref{eq_rayleigh}) applies to a fluid disk. It does not apply to 
a disk containing even an arbitrarily weak magnetic field, if that field is perfectly coupled to the gas 
(the regime of ideal MHD). In ideal MHD a weakly magnetized disk has entirely different 
stability properties from an unmagnetized one, and is unstable provided only that the angular 
velocity decreases outward. This is the magnetorotational instability (MRI) \cite{balbus91,balbus98}, 
which is accepted as the dominant source of turbulence in well-ionized accretion disks (winds could 
still contribute to or dominate angular momentum loss). In protoplanetary disks the ideal MHD version 
of the MRI applies only in the thermally ionized region close to the star; across most of the disk 
we also need to consider both the dissipative (Ohmic diffusion, ambipolar diffusion) and 
the non-dissipative (the Hall effect) effects of non-ideal MHD. The Ohmic and ambipolar terms 
can be considered as modifying --- albeit very dramatically --- the ideal MHD 
MRI, while the Hall term introduces new effects (in part) via the Hall shear instability \cite{kunz08}, 
which is a different beast unrelated to the ideal MHD MRI. The phenomenology of disk instabilities 
in non-ideal MHD is rich, and appears to give rise to both turbulent and laminar angular momentum 
transport as well as phenomena, such as MHD disk winds, that may be observable.

\subsection{The magnetorotational instability}
The MRI \cite{balbus91} is an instability of cylindrical shear flows that contain a weak (roughly, if the 
field is vertical, sub-thermal) magnetic field\footnote{The mathematics of the MRI was worked out by 
Velikhov \cite{velikhov59} and Chandrasekhar \cite{chandrasekhar61} around 1960. 
Thirty years passed before Balbus \& Hawley \cite{balbus91} recognized the 
importance of the instability for accretion flows.}. In ideal MHD the condition for 
instability is simply that,
\begin{equation}
 \frac{{\rm d} \Omega^2}{{\rm d} r} < 0.
\label{eq_mri_criterion} 
\end{equation}
The fact that this condition is always satisfied in disks (though not in star-disk boundary layers) 
accounts for the MRI's central role in modern accretion theory.  

\begin{figure}[t]
\center
\includegraphics[width=\columnwidth]{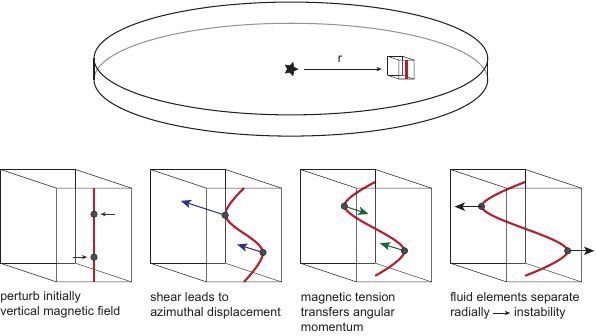}
\caption{Illustration showing why a weak vertical magnetic field destabilizes a Keplerian disk 
(the {\em magnetorotational instability} \cite{balbus98}). An initially uniform vertical field (weak 
enough that magnetic tension is not dominant) is perturbed radially. Due to the shear in the disk, 
an inner fluid element coupled to the field advances azimuthally faster than an outer one. Magnetic 
tension along the field line then acts to {\em remove} angular momentum from the inner element, 
and {\em add} angular momentum to the outer one. This causes further radial displacement, 
leading to an instability.}
\label{fig_mri}
\end{figure}

Figure~\ref{fig_mri} illustrates what is going on to destabilize a disk that contains a magnetic 
field. A basically vertical field is slightly perturbed radially, so that it links fluid elements in the 
disk at different radii. Because of the shear in the disk, the fluid closer to the star orbits faster 
than the fluid further out, creating a toroidal field component out of what was initially just 
vertical and radial field. The tension in the magnetic field linking the two elements (which can 
be thought of, even mathematically, as being analogous to a stretched spring) imparts 
azimuthal forces to both the inner fluid (in the direction {\em opposite} to its orbital motion) 
and the outer fluid ({\em along} its orbital motion). The tension force thus reduces the 
angular momentum of the inner fluid element, and increases that of the outer element. 
The inner fluid then moves further inward (and the outer fluid further outward) and we 
have an instability.

We can derive the MRI instability condition in a very similar setup as Figure~\ref{fig_mri}. 
Consider a disk with a power-law angular velocity profile, $\Omega \propto r^{-q}$, 
that is threaded by a uniform vertical magnetic field $B_0$. We 
ignore any radial or vertical variation in density (and consistent with that, ignore the vertical 
component of gravity) and adopt an isothermal equation of state, $P=\rho c_s^2$, with 
$c_s$ a constant. Our task is to determine whether infinitesimal perturbations to this 
equilibrium state are stable, or whether instead they grow exponentially with time, 
signaling a linear instability.

To proceed (largely following \cite{fromang13}) we define a locally Cartesian patch of disk 
that corotates at radius $r_0$, where the angular frequency is $\Omega_0$. The Cartesian 
co-ordinates $(x,y,z)$ are related to cylindrical co-ordinates $(r,\phi,z^\prime)$ via,
\begin{eqnarray}
 x & = & r - r_0, \nonumber \\
 y & = & r_0 \phi, \nonumber \\
 z & = & z^\prime.
\end{eqnarray} 
The local ``shearing-sheet" (or in three dimensions, ``shearing box") model is useful for 
both analytic stability studies, and for numerical simulations \cite{hawley95,latter17}. In this co-rotating frame, 
the equations of ideal MHD pick up terms representing the fictitious Coriolis and 
centrifugal forces,
\begin{eqnarray}
 \frac{\partial \rho}{\partial t} + \nabla \cdot (\rho {\bf v}) & = & 0, \nonumber \\
 \frac{\partial {\bf v}}{\partial t} + ( {\bf v} \cdot \nabla ) {\bf v} & = & 
 -\frac{1}{\rho} \nabla P + 
 \frac{1}{4 \pi \rho} ( \nabla \times {\bf B} ) \times {\bf B} - 
 2 {\Omega}_0 \times {\bf v} + 
 2q \Omega_0^2 x {\hat {\bf x}}, \nonumber \\
 \frac{\partial {\bf B}}{\partial t} & = & \nabla \times ({\bf v} \times {\bf B}).
\label{eq_sheet_mhd} 
\end{eqnarray} 
Here ${\hat{\bf x}}$ is a unit vector in the $x$-direction. As noted above, the 
initial equilibrium has uniform density, $\rho = \rho_0$, and a magnetic field 
${\bf B} = (0,0,B_0)$. There are no pressure or magnetic forces, so the velocity 
field is determined by a balance between the Coriolis and centrifugal terms,
\begin{equation}
 2 \Omega_0 \times {\bf v} = 2q \Omega_0^2 x {\hat{\bf x}}.
\end{equation}
The equilibrium velocity field that completes the definition of the initial state is,
\begin{equation}
 {\bf v} = \left( 0, -q \Omega_0 x, 0 \right),
\end{equation}
which has a linear shear (with $q=3/2$ for a Keplerian disk) around the reference radius $r_0$.  

To assess the stability of the equilibrium, we write the density, velocity and magnetic 
field as the sum of their equilibrium values plus a perturbation. We can recover the 
MRI with a particularly simple perturbation which depends on $z$ and $t$ only\footnote{An 
analysis that retains the $x$-dependence can be found in the original Balbus \& Hawley (1991) 
paper \cite{balbus91}, and follows an essentially identical approach. Studying the stability 
of non-axisymmetric perturbations (in $y$), however, requires a different and more involved 
analysis \cite{curry95,ogilvie96,terquem96,pessah05}.}. For the velocity components, for example, 
we write,
\begin{eqnarray}
 v_x & = & v_x^\prime (z,t), \nonumber \\
 v_y & = & -q \Omega_0 x + v_y^\prime (z,t), \nonumber \\
 v_z & = & v_z^\prime (z,t),
\end{eqnarray}
and do likewise for the density and magnetic field. We substitute these expressions into the 
continuity, momentum and induction equations, and discard any terms that are quadratic in 
the primed variables, assuming them to be {\em small} perturbations. This would give us 
seven equations in total (one from the continuity equation, and three each from the other 
equations), but the $x$ and $y$ components of the momentum and induction equations are 
all we need to derive the MRI. The relevant linearized equations are,
\begin{eqnarray}
 \frac{\partial v_x^\prime}{\partial t} & = & \frac{B_0}{4 \pi \rho_0} \frac{\partial B_x^\prime}{\partial z} + 
 2 \Omega_0 v_y^\prime, \nonumber \\
 \frac{\partial v_y^\prime}{\partial t} - q \Omega_0 v_x^\prime & = & \frac{B_0}{4 \pi \rho_0} \frac{\partial B_y^\prime}{\partial z} -
 2 \Omega_0 v_x^\prime, \nonumber \\ 
 \frac{\partial B_x^\prime}{\partial t} & = & B_0 \frac{\partial v_x^\prime}{\partial z}, \nonumber \\
 \frac{\partial B_y^\prime}{\partial t} & = & B_0 \frac{\partial v_y^\prime}{\partial z} - q \Omega_0 B_x^\prime.
\end{eqnarray} 
We convert these linearized differential equations into algebraic equations by taking the perturbations to have the 
form, e.g.,
\begin{equation}
 B_x^\prime = \bar{B}_x^\prime e^{i ( \omega t - k z)},
\end{equation} 
where $\omega$ is the frequency of a perturbation with vertical wave-number $k$. The time derivatives then  
pull down a factor of $i \omega$, while the spatial derivatives become $i k$. Our four equations simplify to,
\begin{eqnarray}
 i \omega v_x^\prime & = & -i k \frac{B_0 B_x^\prime}{4 \pi \rho_0} + 2 \Omega_0 v_y^\prime, \nonumber \\
 i \omega v_y^\prime & = & -i k \frac{B_0 B_y^\prime}{4 \pi \rho_0} +(q-2) \Omega_0 v_x^\prime, \nonumber \\ 
 i \omega B_x^\prime & = & - i k B_0 v_x^\prime, \nonumber \\
 i \omega B_y^\prime & = & - i k B_0 v_y^\prime -q \Omega_0 B_x^\prime.
\end{eqnarray}
(We've dropped the bars on the variables for clarity.) Eliminating the perturbation variables from these 
equations, we finally obtain the {\em MRI dispersion relation},
\begin{equation}
 \omega^4 - \omega^2 \left[ 2 k^2 v_A^2 + 2(2-q) \Omega_0^2 \right] + k^2 v_A^2 \left[ k^2 v_A^2 - 2q \Omega_0^2 \right] = 0,
\label{eq_mri_dispersion} 
\end{equation}
where $v_A^2 = B_0^2 / (4 \pi \rho_0)$ is the Alfv\'en speed associated with the net field. 

If $\omega^2 > 0$ then $\omega$ itself will be real and the 
perturbation $e^{i \omega t}$ will oscillate in time. Instability 
requires $\omega^2 < 0$, since in this 
case $\omega$ is imaginary and the perturbation will grow 
exponentially. Solving the dispersion relation we find the instability criterion is,
\begin{equation}
 \left( k v_A \right)^2 - 2 q \Omega_0^2 < 0.
\end{equation}
Letting the field strength go to zero ($B_z \rightarrow 0$, 
$v_A \rightarrow 0$) we find that the condition for instability is simply that 
$q > 0$, i.e. that the angular velocity decrease outward. Even for an 
arbitrarily weak field, the result is completely different from Rayleigh's 
for a strictly hydrodynamic disk. 

\begin{figure}[t]
\includegraphics[width=\columnwidth]{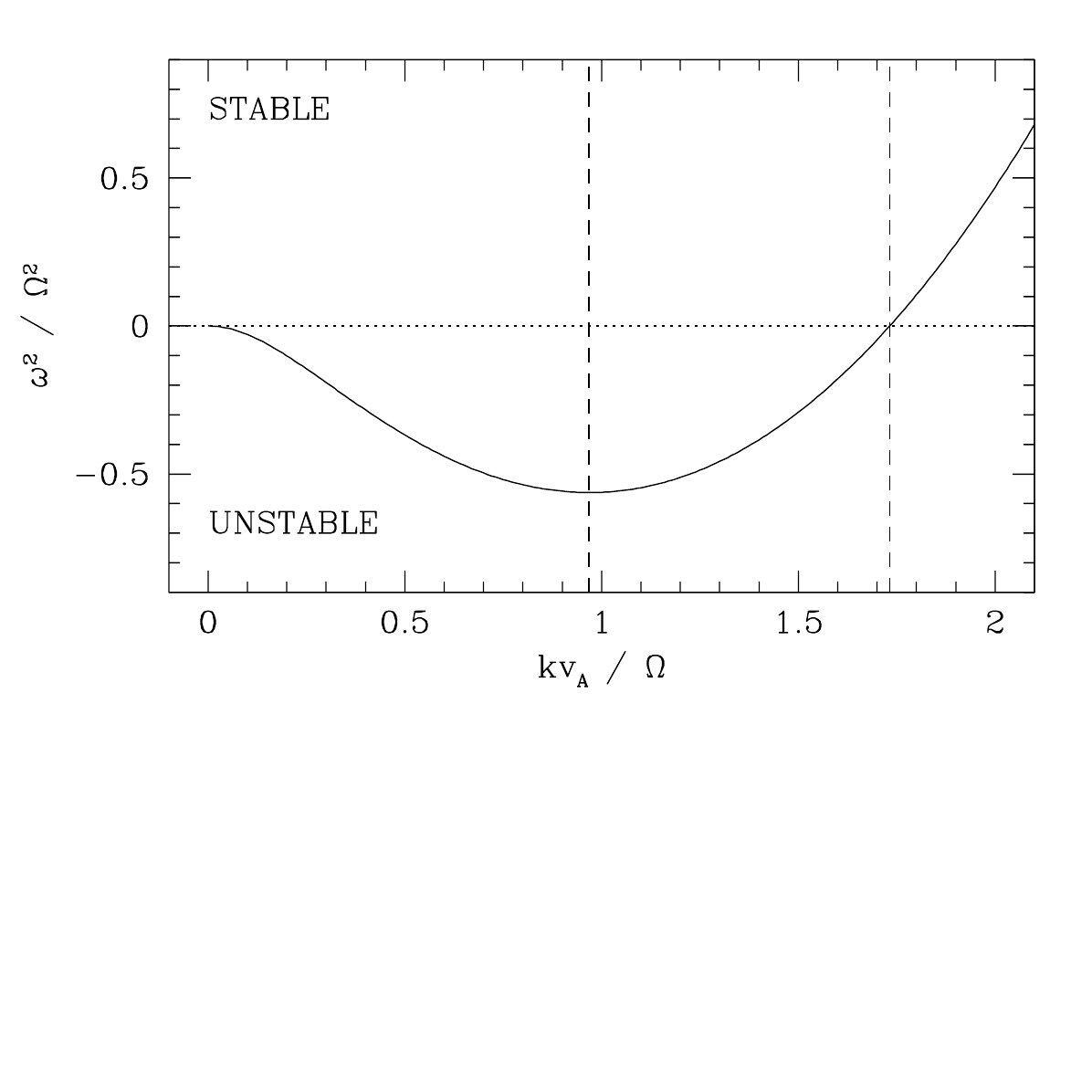}
\vspace{-4.5cm}
\caption{The unstable branch of the MRI dispersion relation is plotted for a Keplerian 
rotation law. The flow is unstable ($\omega^2 < 0$) 
for all spatial scales smaller than $kv_A < \sqrt{3} \Omega$ (rightmost 
dashed vertical line). The most unstable scale (shown as the dashed vertical line 
at the center of the plot) is close to $kv_A \simeq \Omega$.}
\label{fig_mri_dispersion}
\end{figure}

The growth rate of the instability and 
what it means for the magnetic field to be ``weak" can also be derived  
from equation~(\ref{eq_mri_dispersion}). Specializing to a Keplerian 
rotation law with $q=3/2$ the 
dispersion relation takes the form shown in Figure~\ref{fig_mri_dispersion}. 
For a fixed magnetic field strength (and hence a fixed Alfv\'en speed 
$v_A$) the flow is unstable for wavenumbers $k < k_{\rm crit}$ (i.e. on  
large enough spatial scales), where,
\begin{equation}
 k_{\rm crit} v_A = \sqrt{3} \Omega_0.
\end{equation} 
As the magnetic field becomes stronger, the {\em smallest} scale 
$\lambda = 2 \pi / k_{\rm crit}$ which is unstable grows, until eventually it 
exceeds the disk's vertical extent $\approx 2 h$. For stronger vertical fields no 
unstable MRI modes fit within the disk, and the 
instability is suppressed. Using $h = c_s / \Omega$, 
the condition that the vertical magnetic field is 
weak enough to admit the MRI (i.e. that $\lambda < 2h$) becomes
\begin{equation}
 B_0^2 < \frac{12}{\pi} \rho c_s^2
\end{equation}
If we define the plasma 
$\beta$ parameter as the ratio of gas to magnetic pressure,
\begin{equation}
 \beta \equiv \frac{8 \pi P}{B_0^2}
\label{eq_beta} 
\end{equation}
this condition can be expressed alternatively as,
\begin{equation}
 \beta > \frac{2 \pi^2}{3}.
\end{equation}  
A magnetic field whose {\em vertical} component approaches equipartition 
with the thermal pressure ($\beta \sim 1$) will be too strong 
to admit the existence of linear MRI modes, but a wide range of weaker 
fields are acceptable.

The maximum growth rate is determined by setting 
${\rm d} \omega^2 / {\rm d} (kv_A) = 0$ for the unstable branch 
of the dispersion relation plotted in Figure~\ref{fig_mri_dispersion}. 
The most unstable scale for a Keplerian disk is,
\begin{equation}
 \left( kv_A \right)_{\rm max}  = \frac{\sqrt{15}}{4} \Omega_0,
\end{equation}
where the growth rate is,
\begin{equation}
 \vert \omega_{\rm max} \vert = \frac{3}{4} \Omega_0.
\end{equation}
This result implies an {\em extremely} vigorous growth of the instability, 
with an exponential growth time scale that is a fraction of an orbital period. This 
means that if a disk is unstable to the MRI its growth rate 
will invariably be faster than that of hydrodynamic instabilities that may 
additionally be present. One cannot strictly be sure that, when several 
instabilities co-exist, the total angular momentum transport is dominated by 
the one with the fastest linear growth rate. Simulations, however, show that 
the MRI in the ideal MHD limit saturates to yield a turbulent state with a 
moderately high efficiency of angular momentum transport $\alpha \approx 0.02$ \cite{davis10,simon12}. 
It is likely to overwhelm plausible hydrodynamic sources of transport.

\subsubsection{Non-ideal MHD}
The MRI in its ideal MHD guise is relevant to protoplanetary disks only in the thermally 
ionized region close to the star (\S\ref{sec_thermal_ionization}), where $T > 10^3 \ {\rm K}$. 
The very weakly ionized gas further out is imperfectly coupled to the magnetic field, and this 
both modifies the properties of the MRI and leads to new MHD instabilities. We will begin 
by sketching the derivation of the non-ideal MHD equations (following Balbus \cite{balbus11}, 
who justifies several of the approximations that we will make), and then estimate the magnitude of 
the extra terms that arise in protoplanetary disks.

The physics of how magnetic fields affect weakly-ionized fluids is easy to visualize. We consider a 
gas that is almost entirely neutral, with only a small admixture of ions and electrons (analogous 
considerations apply if the charge carriers are dust particles, but we will not go there). Magnetic 
fields exert Lorentz forces on the charged species, but not on the neutrals. Collisions between 
the neutrals and either the ions or the electrons lead to momentum exchange whenever the 
neutral fluid has a velocity differential with respect to the charged fluids.   

We begin by considering the momentum equation. For the neutrals we have,
\begin{equation}
 \rho \frac{\partial {\bf v}}{\partial t} + \rho ( {\bf v} \cdot \nabla ) {\bf v} = 
 - \nabla P - \rho \nabla \Phi - p_{nI} - p_{ne}.
\end{equation} 
Here $\rho$, ${\bf v}$ and $P$ (without subscripts) refer to the neutral fluid, and 
$p_{nI}$ and $p_{ne}$ are the rate of momentum exchange due to collisions 
between the neutrals and the ions / electrons respectively. Identical equations 
apply to the charged species, but for the addition of Lorentz forces,
\begin{eqnarray}
 \rho_e \frac{\partial {\bf v}_e}{\partial t} + \rho_e ( {\bf v}_e \cdot \nabla ) {\bf v}_e & = & 
 - \nabla P_e - \rho_e \nabla \Phi 
 -e n_e \left( {\bf E} + \frac{ {\bf v}_e \times {\bf B} }{c} \right) - p_{en}, \nonumber \\
 \rho_I \frac{\partial {\bf v}_I}{\partial t} + \rho_I ( {\bf v}_I \cdot \nabla ) {\bf v}_I & = &
 - \nabla P_I - \rho_I \nabla \Phi 
 + Z e n_I \left( {\bf E} + \frac{ {\bf v}_I \times {\bf B} }{c} \right) - p_{In}. 
\end{eqnarray}
In these equations ${\bf E}$ and ${\bf B}$ are the electric and magnetic fields, 
the ions have charge $Ze$, where $-e$ is the charge on an electron, and of 
course $p_{ne} = -p_{en}$ and $p_{nI} = -p_{In}$. Having three momentum 
equations looks complicated, but we can make a large simplification to the system 
by noting that the time scale for macroscopic evolution of the fluid is generally 
much longer than the time scale for collisional or magnetic forces to alter a 
charged particle's momentum. We can then ignore {\em everything} in the 
charged species' momentum equations, except for the Lorentz and collisional terms. 
For the ions we have,
\begin{equation} 
 Z e n_I \left( {\bf E} + \frac{ {\bf v}_I \times {\bf B} }{c} \right) - p_{In} = 0,
\end{equation}
with a similar equation for the electrons. Imposing charge neutrality, $n_e = Z n_I$, 
we eliminate the electric field between the ion and electron equations to find an 
expression for the sum of the momentum transfer terms,
\begin{equation}
 p_{In} + p_{en} = \frac{e n_e}{c} \left( {\bf v}_I - {\bf v}_e \right) \times {\bf B}.
\end{equation} 
The current density ${\bf J} = e n_e ( {\bf v}_I - {\bf v}_e )$, so we can write this as,
\begin{equation}
 p_{In} + p_{en} = \frac{ {\bf J} \times {\bf B} }{c}.
\label{eq_collision_terms} 
\end{equation}
Finally, we go to Maxwell's equations, and note that the current can be written as,
\begin{equation}
 \frac{4 \pi}{c} {\bf J} = \nabla \times {\bf B} + \frac{1}{c} \frac{\partial {\bf E}}{\partial t}.
\end{equation}
The second term in Maxwell's equation is the displacement current, which is 
${\mathcal O} (v^2 / c^2)$ and consistently ignorable in non-relativistic MHD. Doing so, 
we substitute equation~(\ref{eq_collision_terms}) in the neutral equation of motion to obtain,
\begin{equation}
  \rho \frac{\partial {\bf v}}{\partial t} + \rho ( {\bf v} \cdot \nabla ) {\bf v} = 
 - \nabla P - \rho \nabla \Phi + \frac{1}{4 \pi} ( \nabla \times {\bf B} ) \times {\bf B}.
\end{equation}   
This is identical to the {\em ideal} MHD momentum equation (stated without 
derivation as equation~\ref{eq_sheet_mhd}) and pleasingly simple; we have reduced 
the three momentum equations to an equation for a single (neutral) fluid with a 
magnetic force term whose dependence on ${\bf B}$ is independent of the 
make-up of the gas. As one might guess, 
the consistent simplification of non-ideal MHD to a momentum equation for a single fluid 
is not always possible. Roughly speaking it works provided that the plasma's inertia is negligible 
compared to that of the neutral fluid, the coupling between charged and neutral species is strong, 
and the recombination time is short. Zweibel \cite{zweibel15} gives an accessible account of the 
conditions necessary for a valid single-fluid description. Although some of the early analytic and numerical 
work on the MRI in weakly-ionized disks utilized a two-fluid approach \cite{blaes94,hawley98}, in many 
protoplanetary disk situations a single fluid model is both justified \cite{bai11} and substantially simpler.
In the single fluid limit all of the complexities of non-ideal MHD enter only via the induction equation.

Deducing the non-ideal induction equation requires us to specify the form of the momentum coupling 
terms. Writing these in standard notation (which is different for the two terms, somewhat obscuring the 
symmetry),
\begin{eqnarray}
 p_{ne} & = & n_e \nu_{ne} m_e ( {\bf v} - {\bf v}_e ), \nonumber \\
 p_{nI} & = & \rho \rho_I \gamma ( {\bf v} - {\bf v}_I ),
\end{eqnarray} 
where $\nu_{ne}$ is the collision frequency of an electron with the neutrals, and $\gamma$ is called 
the drag-coefficient. The ion-neutral coupling involves longer-range interactions than the electron-neutral 
coupling, and is accordingly stronger \cite{balbus11}.

We now go back to the force balance deduced from the electron momentum equation,
\begin{equation}
 - e n_e \left( {\bf E} + \frac{ {\bf v}_e \times {\bf B} }{c} \right) - p_{en} = 0,
\end{equation}
and attempt to write the terms involving ${\bf v}_e$ and $p_{en}$ entirely in terms of the 
current. We start with the exactly equivalent expression,
\begin{equation}
 {\bf E} + \frac{1}{c} \left[ {\bf v} + ( {\bf v}_e - {\bf v}_I ) + ( {\bf v}_I - {\bf v} ) \right] \times {\bf B} + 
 \frac{\nu_{ne} m_e}{e} \left[ ( {\bf v}_e - {\bf v}_I ) + ( {\bf v}_I - {\bf v} ) \right] = 0,
\end{equation}
and deal with the terms in turn. We have two 
terms that involve $( {\bf v}_e - {\bf v}_I )$, which can be replaced immediately with the current, 
\begin{equation}
( {\bf v}_e - {\bf v}_I ) = - \frac{\bf J}{e n_e}.
\end{equation}
The first term with $( {\bf v}_I - {\bf v} )$ is exactly equal to $p_{In} / (\rho \rho_I \gamma)$. If, however, $| p_{In} | \gg | p_{en} |$, then equation~(\ref{eq_collision_terms}) implies that, approximately,
\begin{equation}
( {\bf v}_I - {\bf v} ) \simeq \frac{ {\bf J} \times {\bf B}}{c \rho \rho_I \gamma}.
\end{equation}
Finally it can be shown (see Balbus \cite{balbus11} for details) that the final term with $( {\bf v}_I - {\bf v} )$ can 
be consistently dropped. The version of Ohm's Law that we end up with is,
\begin{equation}
 {\bf E} + \frac{ {\bf v} \times {\bf B} }{c} 
 - \frac{ {\bf J} \times {\bf B}}{e n_e c} + 
 \frac{ ({\bf J} \times {\bf B}) \times {\bf B} }{c^2 \rho \rho_I \gamma} - 
 \frac{\nu_{ne} m_e}{e^2 n_e} {\bf J}= 0.
\end{equation} 
The non-ideal induction equation is then obtained by applying Faraday's law,
\begin{equation}
 \nabla \times {\bf E} = - \frac{1}{c} \frac{\partial {\bf B}}{\partial t},
\end{equation}
to eliminate any explicit reference to the electric field. In its usual form,
\begin{equation}
 \frac{\partial {\bf B}}{\partial t} = \nabla \times 
 \left[ {\bf v} \times {\bf B} 
 - \eta \nabla \times {\bf B} 
 - \frac{ {\bf J} \times {\bf B} }{e n_e} 
 + \frac{ ( {\bf J} \times {\bf B} ) \times {\bf B} }{c \gamma \rho \rho_I} \right].
\label{eq_induction} 
\end{equation}
We have defined the magnetic resistivity,
\begin{equation}
 \eta = \frac{c^2}{4 \pi \sigma}
\end{equation}
where $\sigma$ is (here) the electrical conductivity,
\begin{equation} 
 \sigma = \frac{e^2 n_e}{m_e \nu_{en}}.
\end{equation} 
As before we can replace the current with the magnetic field via,
\begin{equation}
 {\bf J} = \frac{c}{4 \pi} \nabla \times {\bf B},
\end{equation}
so that the induction equation is solely a function of ${\bf B}$. The terms on the 
right-hand-side are referred to as the inductive, Ohmic, Hall and ambipolar terms 
respectively. 

The non-ideal terms in the induction equation depend upon the ionization state 
of the gas (through $n_e$ and $\rho_I$) and upon the collision rates between the 
neutral and charged species (via $\eta$ and $\gamma$). Standard values 
for these quantities are \cite{blaes94,draine83}\footnote{Several assumptions are 
hardwired into these numbers. For the resistivity, it is assumed that currents are 
carried by the electrons, and that the conductivity is limited by electron-neutral collisions. 
For the drag coefficient we assume that the neutral gas is predominantly molecular 
hydrogen, and that the ions are moderately massive $m_i \simeq 30-40 m_{\rm H}$. It 
would be prudent to consult Blaes \& Balbus (1994) \cite{blaes94}, and references therein, 
should one encounter situations where these assumptions might fail.},
\begin{eqnarray}
 \eta & = & 234 \left( \frac{n}{n_e} \right) T^{1/2} \ {\rm cm^2 \ s^{-1}}, \nonumber \\
 \gamma & = & 3 \times 10^{13} \ {\rm cm^3 \ s^{-1} \ g^{-1}}.
\label{eq_eta} 
\end{eqnarray} 
We are now ready to estimate the importance of the non-ideal terms in the environment 
of protoplanetary disks, and to ask what effect they have both on the MRI, and on the 
more general question of whether there is MHD turbulence or transport in disks.

\subsubsection{Ohmic, ambipolar and Hall physics in protoplanetary disks}
The non-ideal terms in equation~(\ref{eq_induction}) all depend inversely on the electron 
or ion density, so the strength of all non-ideal MHD effects relative to the inductive term increases 
with smaller ionization fraction. The three terms also have different dependencies on density, 
magnetic field strength and temperature, so the {\em relative} ordering of the non-ideal MHD 
effects varies with these parameters.

\begin{figure}[t]
\includegraphics[width=\columnwidth]{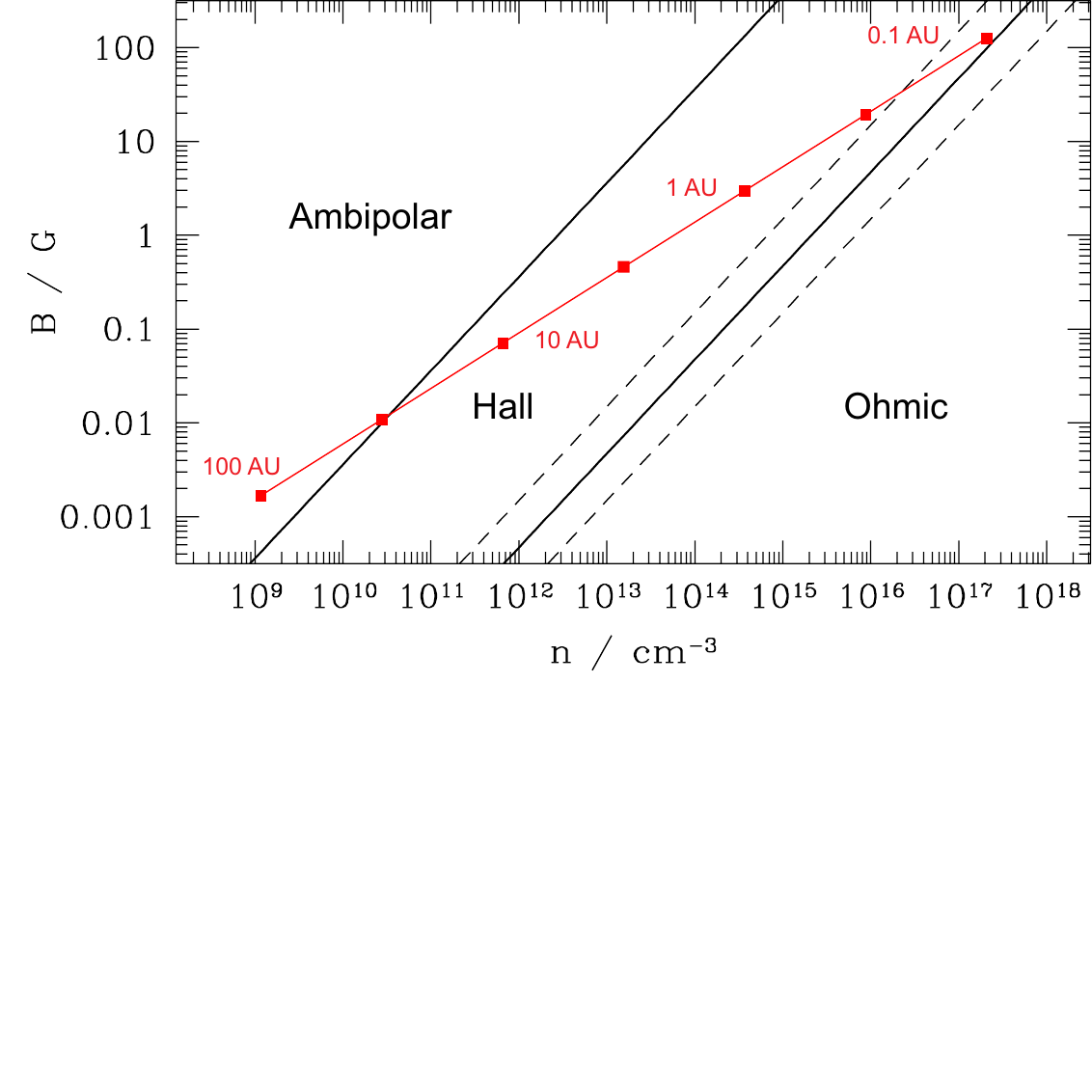}
\vspace{-4.5cm}
\caption{Regions of the $(n,B)$ parameter space in which different non-ideal terms are dominant. The 
boundary between the Ohmic and Hall regimes is plotted for $T = 100 \ {\rm K}$ (solid line) and also 
for temperatures of $10^3 \ {\rm K}$ (upper dashed line) and $10 \ {\rm K}$ (lower dashed line). The red 
line shows a very rough estimate of how the magnetic field in the disk might vary with density between 
the inner disk at 0.1~AU and the outer disk at 100~AU.}
\label{fig_non_ideal}
\end{figure}

The Ohmic, Hall and ambipolar terms have different dependencies on the magnetic field 
geometry, and in a disk setting they influence the MRI in distinct ways (most importantly, the 
Hall effect differs from the others in being non-dissipative). There is therefore no 
model-independent way to precisely demarcate when each term will affect disk evolution. 
As a first guess, however, we can treat the magnetic field as a scalar and simply take the 
ratio of the Hall to the Ohmic term and the ambipolar to the Hall term,
\begin{eqnarray}
 \frac{\rm H}{\rm O} & = & \frac{cB}{4 \pi e \eta n_e}, \nonumber \\
 \frac{\rm A}{\rm H} & = & \frac{e n_e B}{c \gamma \rho \rho_I}.
\end{eqnarray} 
Since $\eta \propto (n/n_e)$ and $n_e \propto \rho_I$ both of these ratios depend on 
$(B/n)$. Substituting for $\eta$ and $\gamma$, and taking the ion mass that enters into the ambipolar term 
as $30 m_H$, we can estimate the magnetic field strength for which the Ohmic and Hall terms 
have equal magnitude, and similarly for the Hall and ambipolar terms,
\begin{eqnarray}
 B_{ {\rm O}={\rm H} } & \approx & 0.5 \left( \frac{n}{10^{15} \ {\rm cm^{-3}}} \right) 
 \left( \frac{T}{100 \ {\rm K}} \right)^{1/2} \ {\rm G}, \nonumber \\
 B_{ {\rm H}={\rm A} } & \approx & 4 \times 10^{-3} \left( \frac{n}{10^{10} \ {\rm cm^{-3}}} \right) \ {\rm G}.
\end{eqnarray}  
Figure~\ref{fig_non_ideal} shows these dividing lines in the $(n,B)$ plane. Ohmic diffusion 
is dominant at high densities / low magnetic field strengths. Ambipolar diffusion dominates for low 
densities / high field strengths. The Hall effect is strongest for a fairly broad range of intermediate 
densities.
 
Estimating where protoplanetary disks fall in the $(n,B)$ plane can be done in various ways. 
For a Solar System-motivated estimate we can start with the disk field inferred from laboratory 
measurements of chondrules in the Semarkona meteorite \cite{fu14}, which suggest that near the 
snow line ($r \simeq 3 \ {\rm AU}$) the disk field was $B \simeq 0.5 \ {\rm G}$. (There are 
caveats and a large systematic uncertainty associated with this measurement, all of which 
we ignore for now.) Let us assume that the surface density and temperature profiles are 
$\Sigma \propto r^{-3/2}$ and $T \propto r^{-1/2}$ respectively, and that the magnetic field 
pressure is the same fraction of the thermal pressure at all radii in the disk. Taking $\Sigma \simeq 300 \ 
{\rm g \ cm^{-2}}$ and $T = 150 \ {\rm K}$ at 3~AU, the inferred scalings of mid-plane density and 
magnetic field strength are then,
\begin{eqnarray}
 n & \approx & 2 \times 10^{13} \left( \frac{r}{3 \ {\rm AU}} \right)^{-11/4} \ {\rm cm^{-3}}, \nonumber \\
 B & \approx & 0.5 \left( \frac{r}{3 \ {\rm AU}} \right)^{-13/8} \ {\rm G}.
\end{eqnarray} 
The track defined by these relations is plotted in Figure~\ref{fig_non_ideal} for radii between 
0.1~AU and 100~AU.

As we have emphasized, neither our approach to ranking the strength of the non-ideal terms, nor 
our estimate of the radial scaling of disk conditions, are anything more than crude guesses. Other 
approximations are equally valid (for example, one can order the terms in the $(n,T)$ plane 
instead \cite{balbus01,kunz04,armitage11}). Nevertheless, because $n$ and $B$ vary by 
so many orders of magnitude across Figure~\ref{fig_non_ideal} the critical inferences we 
can draw are quite robust. We predict that the Hall effect is the dominant non-ideal 
MHD process at the disk mid-plane between (conservatively) 1~AU and 10~AU. Ohmic diffusion can 
become important as we approach the thermally ionized region interior to 1~AU. Ambipolar 
diffusion dominates at sufficiently large radii, of the order of 100~AU, and in the lower density 
gas away from the mid-plane.

\subsubsection{The dead zone}
\label{sec_mhd_transport}

The linear stability of Keplerian disk flow in non-ideal MHD has been extensively investigated 
(see, e.g. \cite{blaes94,jin96,wardle99,balbus01,kunz04,desch04}), and the reader interested 
in the non-ideal analogs to the MRI dispersion relation derived as equation~(\ref{eq_mri_dispersion})   
should start there. Proceeding less formally, we follow Gammie \cite{gammie96} to 
estimate the conditions under which Ohmic dissipation (ignoring for now the Hall term) 
would damp the MRI. The basic idea is to compare the time scale on which the ideal MRI would 
generate tangled magnetic fields to that on which Ohmic diffusion would smooth them out.
We first note that diffusion erases small-scale structure 
in the field more efficiently than large scale features, so that the appropriate 
comparison is between growth and damping of the largest scale MRI models. 
Starting from the MRI dispersion relation (equation~\ref{eq_mri_dispersion}), 
for a Keplerian disk we consider the weak-field / 
long wavelength limit ($kv_A / \Omega \ll 1$). The growth rate of the MRI is,
\begin{equation}
 \vert \omega \vert \simeq \sqrt{3} kv_A.
\end{equation} 
Writing this as a function of the spatial scale $\lambda = 2 \pi / k$, we have,
\begin{equation}
 \vert \omega \vert \simeq 2 \pi \sqrt{3} \frac{v_A}{\lambda}.
\end{equation} 
Up to numerical factors the MRI on a given scale then grows on the Alfv\'en crossing 
time. Equating this growth 
rate to the Ohmic damping rate,
\begin{equation}
 \vert \omega_\eta \vert \sim \frac{\eta}{\lambda^2}
\end{equation}
we conclude that Ohmic dissipation will suppress the MRI on the largest available scale $\lambda \approx h$ 
provided that,
\begin{equation}
 \eta > 2 \pi \sqrt{3} v_A h.
\end{equation}  
We can express this result in a different form. Analogous to the fluid Reynolds number (equation~\ref{eq_Re})  
the magnetic Reynolds number ${\rm Re}_M$ is defined as,
\begin{equation}
 {\rm Re}_M \equiv \frac{UL}{\eta}
\end{equation}
where $U$ is a characteristic velocity and $L$ a characteristic 
scale. Taking $U = v_A$ and $L = h$ for a disk, the condition for Ohmic dissipation 
to suppress the MRI becomes
\begin{equation}
 {\rm Re}_M < 1
\end{equation}  
where order unity numerical factors have been omitted.

We now convert the condition for the suppression of the MRI 
into a limit on the ionization fraction $x \equiv n_e / n$. We make 
use of the formula for the magnetic resistivity (equation~\ref{eq_eta}) 
and assume that Maxwell stresses transport angular momentum, 
so that $\alpha \sim {v_A^2}/{c_s^2}$ (this follows approximately 
from equation~\ref{eq_def_alpha}). The magnetic Reynolds number can 
then be estimated to be, 
\begin{equation}
 {\rm Re}_M = \frac{v_A h}{\eta} = \frac{\alpha^{1/2} c_s^2}{\eta \Omega}.
\label{eq_ReM} 
\end{equation}
Substituting for $\eta$ and $c_s^2$, the 
magnetic Reynolds number in a protoplanetary 
disk scales as, 
\begin{equation}
 {\rm Re}_M \approx 1.4 \times 10^{12} x 
 \left( \frac{\alpha}{10^{-2}} \right)^{1/2}   
 \left( \frac{r}{1 \ {\rm AU}} \right)^{3/2} 
 \left( \frac{T}{300 \ {\rm K}} \right)^{1/2}
 \left( \frac{M_*}{M_\odot} \right)^{-1/2}.
\end{equation} 
For the given parameters, the critical ionization fraction below which Ohmic 
diffusion will quench the MRI is
\begin{equation}
 x_{\rm crit} \sim 10^{-12}.
\end{equation}
Clearly a very small ionization fraction suffices to couple the magnetic field to the gas 
and allows the MRI to operate, but there are 
large regions of the disk where even these ionization levels are not obtained and 
non-ideal effects are important.

Based on this analysis Gammie \cite{gammie96} noted, first, that the criterion for the 
MRI to operate under near-ideal MHD conditions in the inner disk coincides with the 
requirement that the alkali metals are thermally ionized (Figure~\ref{fig_saha}). The 
development of magnetized turbulence in the disk at radii where $T > 10^3 \ {\rm K}$ 
can therefore be modeled in ideal MHD. Second, he proposed that Ohmic diffusion 
would damp MHD turbulence in the low ionization environment near the disk 
mid-plane on scales of the order of 1~AU, creating a {\em dead zone} of sharply 
reduced turbulence and transport. Gammie's original model is incomplete, as it 
did not include either ambipolar diffusion or the Hall effect, but the basic idea 
motivates much of the current work on 
MHD instabilities in disks.

\subsubsection{Turbulence and transport in non-ideal MHD}
\label{sec_transport}

Once we consider the full set of non-ideal terms, the first question is to assess the level 
of turbulence and transport that is expected as a function of their strengths.  This is 
a well-defined but already difficult theoretical question, given that interesting values of 
the Ohmic, ambipolar and Hall terms span a broad range (depending physically on the temperature, 
density, ionization fraction and magnetic field strength). To define the problem in its most 
idealized form, we rewrite the non-ideal induction equation~(\ref{eq_induction}) as,
\begin{equation}
 \frac{\partial {\bf B}}{\partial t} = \nabla \times 
 \left[ {\bf v} \times {\bf B} 
 - \eta_O \nabla \times {\bf B} 
 - \eta_H \frac{ {\bf J} \times {\bf B} }{B} 
 - \eta_A \frac{ ( {\bf J} \times {\bf B} ) \times {\bf B} }{B^2} \right], 
\end{equation}
where {\em dimensionally} $\eta_O$, $\eta_H$ and $\eta_A$ are all diffusivities. There are 
different ways to construct dimensionless numbers from the diffusivities, but one useful set is,
\begin{eqnarray}
 \Lambda_O & \equiv & \frac{v_A^2}{\Omega \eta_O} \,\,\, {\rm (Ohmic \ Elsasser \ number)}, \nonumber \\
 {\rm Ha} & \equiv & \frac{v_A^2}{\Omega \eta_H} \,\,\, {\rm (Hall \ Elsasser \ number)}, \nonumber \\
 {\rm Am} & \equiv & \frac{v_A^2}{\Omega \eta_A} \,\,\, {\rm (ambipolar \ Elsasser \ number)}.
\end{eqnarray} 
(Note that the ambipolar Elsasser number can also be written as ${\rm Am} \equiv \gamma \rho_I / \Omega$, 
which is the number of ion-neutral collisions per dynamical time $\Omega^{-1}$.) 
We further specify the net vertical magnetic field 
(if any) via the ratio of the mid-plane gas and magnetic field pressures (equation~\ref{eq_beta}),
\begin{equation}
 \beta_z = \frac{8 \pi P}{B_z^2}.
\end{equation} 
Our question can then be rephrased; what is the level of angular momentum transport and 
turbulence in an MHD disk at radii where the non-ideal terms are characterized by the 
dimensionless parameters ($\Lambda_O$, Ha, Am, $\beta_z$).

Ohmic diffusion acts as a strictly dissipative process that stabilizes disks to magnetic 
field instabilities on scales below some critical value. Ambipolar diffusion is in principle 
more complex, because it does not dissipate currents that are parallel to the magnetic 
field. This distinction substantially impacts the linear stability of ambipolar-dominated 
disks \cite{kunz04}, but appears to matter less for the non-linear evolution, whose 
properties are analogous to Ohmic diffusion. For both dissipative processes, simulations 
show that MRI-driven turbulence is strongly damped when the relevant dimensionless parameter 
(either $\Lambda_0$ or Am) drops below a critical value that depends upon the initial 
field geometry but is $\sim 1-10^2$ \cite{sano01,turner07,simon09,bai11b,simon13}. Consistent with 
the dead zone idea \cite{gammie96}, we therefore expect substantial modification of the 
properties of MHD turbulence both in the mid-plane around 1~AU (where Ohmic diffusion is the dominant 
dissipative process) and in the disk atmosphere and at large radii $\sim 10^2 \ {\rm AU}$ (where 
ambipolar diffusion dominates).

\begin{figure}[t]
\center
\includegraphics[width=\columnwidth]{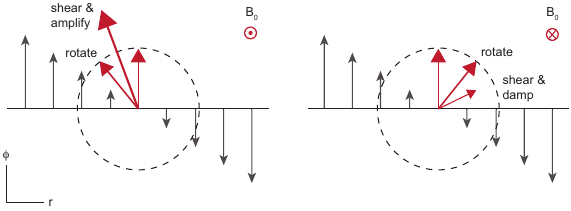}
\caption{An illustration (adapted from Kunz \cite{kunz08}) of the Hall shear instability. In the presence of a vertical field 
threading the disk, the Hall effect acts to rotate an initially toroidal field component either clockwise or 
anti-clockwise, depending upon the {\em sign} of the vertical field. The rotated field vector is then 
either amplified or damped by the shear. This instability differs from the MRI both technically (in that 
there is no reference to orbital motion, any shear flow suffices) and physically (via the dependence 
on the direction of the vertical magnetic field as well as its strength).}
\label{fig_hall_shear}
\end{figure}

The MRI dispersion relation is also modified by the Hall effect \cite{wardle99,balbus01}, which differs from the 
other non-ideal terms in that it modifies the field structure without any attendant dissipation. In this respect the 
Hall term most closely resembles the inductive term $\nabla \times ( {\bf v} \times {\bf B} )$, and its strength 
can usefully be characterized by the ratio of the Hall to the inductive term. Non-linear simulations of the 
Hall effect in disks, which were pioneered by Sano \& Stone \cite{sano02,sano02b}, have only recently been 
able to access the strongly Hall-dominated regime relevant to protoplanetary disks \cite{kunz13,lesur14,bai14,bai15,simon15}\footnote{The 
numerical implementation of the Hall effect in simulation codes remains challenging, and the presence of large-scale 
fields in the saturated state suggests that local simulations may not be adequate to describe the outcome. Observationally 
important issues such as the level of fluid turbulence that accompanies the predominantly large-scale transport by Maxwell stresses 
remain to be fully understood.}. 
In vertically stratified disks with a net vertical field, Lesur et al. \cite{lesur14} find that the Hall effect has a 
controlling influence on disk dynamics on scales between 1~AU and 10~AU. For $\beta_z = 10^5$ a strong 
but {\em laminar} Maxwell stress (i.e. one dominated by large-scale radial and toroidal fields in equation~\ref{eq_def_alpha}) 
results when the net field is aligned with the rotation axis of the disk, whereas 
anti-alignment leads to extremely weak turbulence and transport.

The results from Hall-MHD simulations of protoplanetary disks are best interpreted not as a modification 
of the MRI, but rather as the signature of a distinct {\em Hall shear instability} \cite{kunz08}. In the presence 
of a net vertical magnetic field the Hall effect acts to {\em rotate} magnetic field vectors lying in the orbital 
plane (Figure~\ref{fig_hall_shear}), with the sense of the rotation determined by whether the new field is 
aligned or anti-aligned to the rotation axis. In the aligned field case, the Hall-induced rotation allows the 
magnetic field to be amplified by the shear, while damping occurs in the anti-aligned limit. Unlike the MRI, 
the Hall shear instability does not depend on the Coriolis force, and is indifferent to the sign of the angular 
velocity gradient. By generating a radial field directly from an azimuthal one, the Hall effect (given a net field) 
supports a mean-field disk dynamo cycle \cite{tout92,lesur14} that is qualitatively different from anything that is possible in 
ideal MHD.

Going beyond the idealized question of the effect of the non-ideal terms on turbulence and transport, 
our goal is to use the results described above to predict the structure and evolution 
of protoplanetary disks. This introduces new layers of 
uncertainty. To predict disk properties from first principles, we need at a minimum to know the 
strength of the different sources of ionization (\S\ref{sec_non_thermal}), the rates of gas-phase and 
dust-induced recombination, and the global evolution of any net magnetic field (\S\ref{sec_field_transport}). 
We also need to model (or have good reasons to ignore) various non-MHD effects, including the 
hydrodynamic angular momentum transport processes already discussed and mass loss by 
photoevaporation \cite{alexander14} (\S\ref{sec_dispersal}). Given these uncertainties, the best 
that we can currently do is to highlight a number of qualitative predictions that receive 
support from numerical simulations:
\begin{itemize}
\item
Net vertical magnetic fields are important for disk evolution. A vertical magnetic field enhances 
angular momentum transport by the MRI even in ideal MHD (roughly as $\alpha \propto \beta_z^{-1/2}$ \cite{hawley95}). 
In non-ideal MHD, local simulations by Simon et al. \cite{simon13,simon13b} suggest that ambipolar 
damping of the MRI in the outer disk prevents resupply of the inner disk with gas {\em unless} a 
net field is present\footnote{It is not obvious that the inner disk {\em is} resupplied by gas, or, 
to put it more formally, that the disk attains a steady-state. Out to $\sim 10 \ {\rm AU}$ the viscous 
time scale is short enough that the disk will plausibly adjust to a steady state (provided only that a steady state is 
possible, see \S\ref{sec_episodic}), but no such argument works out to 100~AU. Ultimately the 
question of whether gas at 100~AU ever reaches the star will need to be settled by observations as 
well as by theory.}. A net field with $\beta_z \simeq 10^4 - 10^5$ suffices, comparable to the fields 
expected in global models of flux evolution \cite{guilet14} but much weaker than the likely initial 
field left over from star formation. Accretion on scales of 30-100~AU occurs predominantly through 
a thin surface layer that is ionized by FUV photons \cite{perezbecker11,simon13,simon13b}, and is 
largely independent of the Hall effect \cite{bai15}.
\item
MHD winds and viscous transport can co-exist in disks. Local numerical simulations in ideal MHD by 
Suzuki \& Inutsuka \cite{suzuki09} showed that in the presence of net vertical field, the MRI was 
accompanied by mass and angular momentum loss in a disk wind. Winds are likewise seen in 
Bai \& Stone's net field protoplanetary disk simulations at 1~AU that include Ohmic and ambipolar diffusion \cite{bai13}, 
in simulations at 30-100~AU where ambipolar diffusion dominates \cite{simon13b}, and in simulations 
that include all non-ideal terms \cite{lesur14}. Caution is required before interpreting these local simulation 
results as quantitative predictions, because although the effective potential for wind launching is 
correctly represented (equation~\ref{eq_eff_potential}) there is a known and unphysical dependence of the mass 
loss rate on the vertical size of the simulation domain \cite{fromang13b}. Outflows are also seen in recent 
{\em global} net field simulations \cite{gressel15,bethune17,bai17}, however, supporting the view that outflows 
driven by a combination of MHD and thermal processes could be a generic feature of protoplanetary disk accretion. 
\item
Turbulence and angular momentum transport are not synonymous. In classical disk theory, the value of 
$\alpha$ determines not only the rate at which the disk evolves, but also the strength of turbulence and its 
effect on small solid particles. This link is doubly broken in more complete disk models. First, as already 
noted, angular momentum loss via winds (which need not be accompanied by turbulence) may be stronger 
than viscous transport at some radii. Second, even the internal component of transport may be primarily 
a large-scale ``laminar" Maxwell stress, rather than small-scale turbulence \cite{simon13b,lesur14}.
\item
The sign of the net field could lead to bimodality in disk properties. The Hall effect is the strongest 
non-ideal term interior to about 10~AU, and simulations \cite{lesur14,bai14,bai15} confirm the 
expectation from linear theory \cite{wardle99,balbus01,kunz08} that a disk with a weak field that 
is aligned to the rotation axis behaves quite differently from one with an anti-aligned field. Although 
there are possible confounding factors --- for example the long-term evolution of the net field may 
itself differ with the sign of the field --- it appears likely that the striking asymmetry seen in 
simulations introduces {\em some} observable bimodality in disk structure \cite{simon15}. Hall MHD 
can also affect the properties of disks at an earlier time, during the collapse of molecular clouds and the formation of
protostellar disks \cite{krasnopolsky11,tsukamoto15b}, influencing for example their 
initial sizes. 
\end{itemize}

\begin{figure}[t]
\center
\includegraphics[width=\columnwidth]{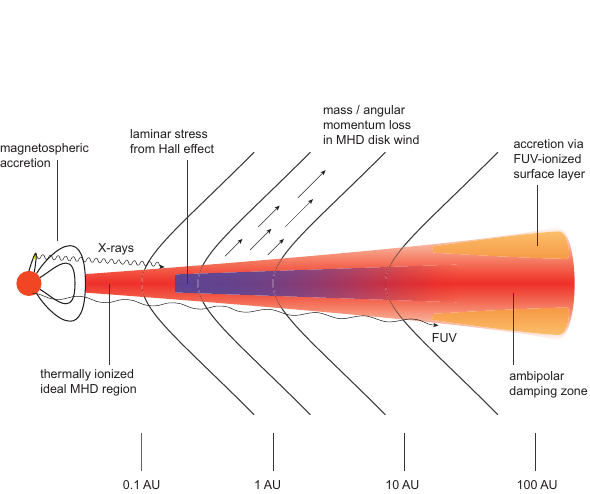}
\caption{A suggested structure for protoplanetary disks {\em if} MHD processes dominate over 
other sources of transport. The different regions are defined by the strength of non-ideal MHD 
terms (Ohmic diffusion, ambipolar diffusion and the Hall effect), and by mass and angular momentum 
loss in MHD disk winds. The Hall effect is predicted to behave differently if the net field threading the 
disk is anti-aligned to the rotation axis (here, alignment is assumed). 
Ionization by stellar X-rays and by FUV photons couples the stellar properties to those of the disk.}
\label{fig_accretion_schematic}
\end{figure}

Figure~\ref{fig_accretion_schematic} illustrates a possible disk structure implied by the above results. 
The figure should be regarded as a work in progress; there is plenty of work remaining before we fully 
understand either the physics of potential angular momentum transport and loss processes, or how to tie 
that knowledge together into a consistent scenario for disk structure and evolution.

\subsection{Transport in the boundary layer}
The nature of transport in the boundary layer deserves a brief discussion. As discussed in 
\S\ref{sec_zero_torque} boundary layers are expected when the accretion rate is 
high enough to overwhelm the disruptive influence of the stellar magnetosphere (see 
equation~\ref{eq_rm} for a semi-quantitative statement of this condition). For most 
stars this requires high accretion rates, so the boundary layer and adjacent disk 
are hot enough to put us into the regime of thermal ionization and 
ideal MHD. In the disk we then expect angular momentum transport via the MRI. 
In the boundary layer, however, we face a problem. By definition, ${\rm d \Omega / {\rm d}r} > 0$ 
in the boundary layer (see Figure~\ref{figure_boundary}), and this angular velocity profile is 
{\em stable} against the MRI (equation~\ref{eq_mri_criterion}). Something else much be 
responsible for transport in this region.

The angular velocity profile in the boundary layer is stable against the generation of turbulence 
by either the Rayleigh criterion or the MRI. It turns out to be {\em unstable}, however, to the 
generation of waves via a mechanism analogous to the hydrodynamic Papaloizou-Pringle 
instability of narrow tori \cite{papaloizou84}. Belyaev et al. \cite{belyaev13,belyaev13b}, using 
both analytic and numerical arguments, have shown that waves generated from the 
supersonic shear provide weak non-local transport of angular momentum 
(and energy) across the boundary layer. Magnetic fields are amplified by the shear 
\cite{armitage02} but do not play an essential role in boundary layer transport \cite{belyaev13b}. 
In protostellar systems boundary layers are present during eruptive accretion phases (see \S\ref{sec_episodic}) 
when strong radiation fields are present \cite{kley96}. Future work will need to combine 
the recent appreciation of the importance of wave angular momentum transport with 
radiation hydrodynamics for a full description of the boundary layer.

\begin{svgraybox}
\begin{itemize}
\item 
Purely gas-phase turbulence in protoplanetary disks can be driven by MHD instabilities of Keplerian shear flow (the MRI and 
Hall-shear instability), by hydrodynamic instabilities that occur for specific vertical and / or radial entropy 
profiles, and by self-gravity. 
\item
The magnetorotational instability probably dominates whenever the disk is thermally ionized, at temperatures 
exceeding about $10^3 \ {\rm K}$. Elsewhere, non-ideal MHD effects are important. Ohmic diffusion damps 
turbulence in high density regions, while ambipolar diffusion fulfills a similar role at large radii and in the disk 
atmosphere where the density is low. The Hall effect can enhance or inhibit angular momentum transport on $\sim 1-10 \ {\rm AU}$ scales, 
depending upon the sign of any net magnetic field relative to the disk angular momentum vector. The long-term 
evolution of the disk is thus coupled to the strength and evolution of net magnetic fields inherited from the 
star formation process.
\item
Self-gravity driven by ongoing envelope infall contributes to transport in young, massive disks. 
Fragmentation is expected to occur on sufficiently large scales (of the order of $10^2 \ {\rm AU}$), 
probably giving rise to objects with masses above the planetary mass regime.
\item
Entropy driven hydrodynamic instabilities could yield a significant level of transport in 
regions where MHD transport is suppressed. The operation of these instabilities is closely tied to the 
thermal structure and radiative properties of the disk.
\end{itemize}
\end{svgraybox}

\newpage

\section{Episodic accretion}
\label{sec_episodic}

Young stellar objects (YSOs) are observed to be variable. The short time scale (lasting hours to weeks) component of that 
variability is complex \cite{cody14}, but can probably be attributed to a combination of turbulent inhomogeneities in the inner disk, 
stellar rotation \cite{bouvier93}, and the complex dynamics of magnetospheric accretion \cite{alencar10}. There is 
also longer time scale variability --- lasting from years to (at least) many decades, that in some cases takes the 
form of well-defined outbursts in which the YSO brightens dramatically. The traditional classification of outbursting 
sources divides them into FU Orionis events \cite{herbig77,hartmann96}, 
characterized by a brightening of typically 5 magnitudes followed by a decay over decades, and EXors \cite{herbig08}, which display 
repeated brightenings of several magnitudes over shorter time scales. The statistics on these uncommon long-duration 
outbursts (especially FU Orionis events \cite{hillenbrand15}) are limited, and it is not even clear --- either observationally or theoretically --- 
whether FUOrs and EXors are variations on a theme or genuinely different phenomena \cite{audard14}. Nonetheless, it is 
established that episodic accretion is common enough to matter for both stellar accretion and for planet formation 
processes occurring in the inner disk \cite{audard14}. Our focus here is on the origin of these accretion outbursts.

Observations show that FU Orionis outbursts involve a large increase in the mass accretion rate through the inner 
disk on to the star \cite{hartmann96}. During the outburst the inner disk will be relatively thick ($h/r \approx 0.1$), 
and hot enough to be thermally ionized. We therefore expect efficient angular momentum transport from the MRI, 
with $\alpha \approx 0.02$. Writing the viscous time scale (equation~\ref{eq_viscous_time}) in terms of these 
parameters,
\begin{equation}
 t_\nu = \frac{1}{\alpha \Omega} \left( \frac{h}{r} \right)^{-2},
\end{equation}
we can estimate the disk radius associated with a (viscously driven) outburst of duration $t_{\rm burst}$, 
\begin{equation}
 r \simeq \left( G M_* \right)^{1/3} \alpha^{2/3} \left( \frac{h}{r} \right)^{4/3} t_{\rm burst}^{2/3}.
\end{equation} 
For a Solar mass star we find,
\begin{equation}
 r \simeq 0.25 \left( \frac{\alpha}{0.02} \right)^{2/3} 
 \left( \frac{h/r}{0.1} \right)^{4/3} 
 \left( \frac{t_{\rm burst}}{100 \ {\rm yr}} \right)^{2/3} \ {\rm AU}.
\end{equation} 
Disk-driven outbursts of broadly the right duration are thus likely to involve events on sub-AU scales, 
and could be associated physically with the magnetosphere, with the thermally ionized inner disk, 
or with the inner edge of the dead zone.

\begin{figure}[t]
\center
\includegraphics[width=\columnwidth]{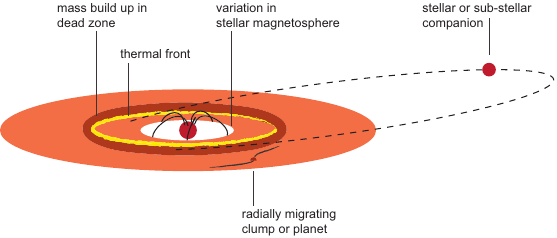}
\caption{An illustration of some of the processes suggested as the origin of episodic YSO accretion.}
\label{fig_episodic}
\end{figure}

The physical origin of episodic accretion in YSOs has not been securely identified. Mooted ideas, 
illustrated in Figure~\ref{fig_episodic}, fall 
into two categories. The first category invokes secular instabilities of protoplanetary disk structure 
that may occur on AU and sub-AU scales. The idea is that the inner disk may be intrinsically unable 
to accrete at a steady rate, and instead alternates between periods of high accretion rate when 
gas is draining on to the star and periods of low accretion rate when gas is accumulating in the 
inner disk. The instability could be a classical thermal instability \cite{bell94}, of the type accepted 
as causing dwarf nova outbursts \cite{lasota01}, or a related instability of dead zone structure \cite{gammie99,armitage01}. 
The second category appeal to triggers independent of the inner disk to initiate the outburst. Ideas 
in this class are various and include perturbations from binary companions \cite{bonnell92}, 
the tidal disruption of radially migrating gas clumps / giant planets \cite{vorobyov05,nayakshin12}, 
and disk variability linked to a stellar magnetic cycle \cite{armitage95,dangelo12,armitage16}. Neither category of ideas 
is fully compelling (in the sense of being both fully worked out and consistent with currently 
accepted disk physics), so our discussion here will focus on a few key concepts that are 
useful for understanding current and future models of episodic accretion.

\subsection{Secular disk instabilities}
The classical instabilities that may afflict thin accretion disks are the thermal and viscous 
instabilities \cite{pringle81}. These are quite distinct from the basic fluid dynamical instabilities 
(the MRI, the VSI etc) that we discussed in \S\ref{sec_turbulence}, in that they address the 
stability of derived disk models rather than the fluid per se. Thus the thermal instability 
is an instability of the equilibrium vertical structure of the disk, while viscous instability 
in an instability of an (assumed) smooth radial structure under viscous evolution.

Before discussing how thermal or viscous instability might arise, we first define what 
these terms mean. Consider an annulus of the disk that is initially in hydrostatic 
and thermal equilibrium, such that the heating rate $Q_+$ matches the cooling 
rate $Q_-$. The heating rate per unit area depends upon the central temperature 
(equation~\ref{eq_qplus}), and can be written assuming the $\alpha$-prescription as,
\begin{equation}
 Q_+ = \frac{9}{4} \nu \Sigma \Omega^2 = \frac{9}{4} \alpha \frac{k_B T_c}{\mu m_H} \Sigma \Omega.
\end{equation} 
The cooling rate directly depends upon the effective temperature, $T_{\rm disk}$, but this can 
always be rewritten in terms of $T_c$ using a calculation of the vertical thermal structure (\S\ref{sec_viscous_heating}).
In the simple case when the disk is optically thick, for example, we have from equation~(\ref{eq_C3_Tc}) that 
$T_c^4 / T_{\rm disk}^4 \simeq (3 \tau / 4)$, and hence,
\begin{equation}
 Q_- = \frac{8 \sigma}{3 \tau} T_c^4.
\end{equation} 
Both $\alpha$ and $\tau$ may be functions of $T_c$. Now consider perturbing the central temperature 
on a time scale that is long compared to the dynamical time scale (so that hydrostatic equilibrium holds) 
but short compared to the viscous time scale (so that $\Sigma$ remains fixed)\footnote{This may 
seem to require fine tuning, but in fact the ordering of time scales in a geometrically thin disk 
always allows for such a choice \cite{pringle81}.}. The disk will be unstable to runaway heating 
if an upward perturbation to the temperature increases the heating rate more than it increases the 
cooling rate, i.e. if,
\begin{equation}
 \frac{ {\rm d} \log Q_+}{{\rm d} \log T_c} > \frac{ {\rm d} \log Q_-}{{\rm d} \log T_c}.
\label{eq_thermal_instability}
\end{equation} 
The same criterion predicts runaway cooling in the event of a downward perturbation. A disk that 
is unstable in this sense is described as {\em thermally unstable}. It would heat up (or cool down) 
until it finds a new structure in which heating and cooling again balance.

To determine the condition for viscous stability, we start by considering a steady-state solution 
$\Sigma(r)$ to the diffusive disk evolution equation~(\ref{eq_disk_evolve}). Following Pringle \cite{pringle81} 
we write $\mu \equiv \nu \Sigma$ and consider perturbations $\mu \rightarrow \mu + \delta \mu$ on a 
time scale long enough that {\em both} hydrostatic and thermal equilibrium hold (in this limit $T_c$ is 
uniquely determined and $\nu = \nu(\Sigma)$). Substituting in equation~(\ref{eq_disk_evolve}) the 
perturbation $\delta \mu$ obeys,
\begin{equation}
 \frac{\partial}{\partial t} \left( \delta \mu \right) = 
 \frac{ \partial \mu}{\partial \Sigma} \frac{3}{r} 
 \frac{\partial}{\partial r} \left[ r^{1/2} 
 \frac{\partial}{\partial r} \left( r^{1/2} \delta \mu \right) \right].
\end{equation} 
The perturbation $\delta \mu$ will grow if the diffusion coefficient, which is proportional to 
$\partial \mu / \partial \Sigma$, is negative. This is {\em viscous instability}, and it occurs 
if,
\begin{equation}
 \frac{\partial}{\partial \Sigma} \left( \nu \Sigma \right) < 0.
\end{equation}
A disk that is viscously unstable would tend to break up into rings, whose amplitude would 
presumably be limited by the onset of fluid instabilities that could be thought of as 
modifying the $\nu(\Sigma)$ relation. 

\subsubsection{The S-curve: a toy model}
The instabilities of interest for YSO episodic accretion can broadly be considered to be 
thermal-type instabilities. Noting that $Q_+ \propto \alpha T_c$, and $Q_- \propto T_c^4 / \tau \propto T_c^4 / \kappa$ 
(where $\kappa$ is the opacity at temperature $T_c$), we see that instability may occur according to 
equation~(\ref{eq_thermal_instability}) if,
\begin{itemize}
\item
${\rm d} \log Q_+ / {\rm d} \log T_c$ is large, i.e. if $\alpha$ is strongly increasing with temperature.
\item
${\rm d} \log Q_- / {\rm d} \log T_c$ is small, i.e. if $\kappa$ is strongly increasing with temperature.
\end{itemize}
We expect $\alpha$ to increase rapidly with $T_c$ at temperatures around $10^3 \ {\rm K}$, as we 
transition between damped non-ideal MHD turbulence at low temperature and the more vigorous 
ideal MHD MRI at higher temperature. We expect $\kappa$ to increase most strongly at temperatures 
around $10^4 \ {\rm K}$, as hydrogen is becoming ionized and there is a strong contribution to the 
opacity from $H^-$ scattering (in this regime $\kappa$ can vary as something like $T^{10}$ \cite{bell94}). 
Either of these changes can result in instability.

Before detailing the specifics of possible thermal and dead zone instabilities in protoplanetary disks, 
it is useful to analyze a toy model that displays their essential features. We consider an optically 
thick disk described by the usual classical equations \cite{frank02} whose angular momentum 
transport efficiency $\alpha$ and opacity $\kappa$ are both piece-wise constant functions of 
central temperature $T_c$. Specifically,
\begin{eqnarray}
 T_c & < & T_{\rm crit}: \alpha = \alpha_{\rm low}, \kappa = \kappa_{\rm low}, \nonumber \\
 T_c & > & T_{\rm crit}: \alpha = \alpha_{\rm high}, \kappa = \kappa_{\rm high}, 
\end{eqnarray} 
with $\alpha_{\rm low} \leq \alpha_{\rm high}$ and $\kappa_{\rm low} \leq \kappa_{\rm high}$. 
Our goal is to calculate the explicit form of the $\dot{M} (\Sigma)$ relation in the ``low" and 
``high" states below and above the critical temperature $T_{\rm crit}$. For a steady-state disk, 
at $r \gg R_*$, heated entirely by viscous dissipation, the equations we need (mostly from \S\ref{sec_evolution})  
read,
\begin{eqnarray}
 T_{\rm disk}^4 & = & \frac{3 \Omega^2}{8 \pi \sigma} \dot{M}, \nonumber \\
 \frac{T_c^4}{T_{\rm disk}^4} & = & \frac{3}{4} \tau, \nonumber \\
 \tau & = & \frac{1}{2} \kappa \Sigma, \nonumber \\
 \nu \Sigma & = & \frac{\dot{M}}{3 \pi}, \nonumber \\
 \nu & = & \alpha \frac{c_s^2}{\Omega} = \frac{\alpha}{\Omega} \frac{k_B T_c}{\mu m_H}.
\end{eqnarray} 
Note that the stellar mass $M_*$ and radius in the disk $r$ enter these formulae only via 
their combination in the Keplerian angular velocity $\Omega$. Eliminating $T_c$, $T_{\rm disk}$, 
$\tau$ and $\nu$ between these equations, we obtain a solution for $\dot{M} (\Sigma)$,
\begin{equation}
 \dot{M} = \frac{9 \pi}{4} \left( \frac{k_B^4}{\mu^4 m_H^4 \sigma} \right)^{1/3} 
 \kappa^{1/3} \alpha^{4/3} \Omega^{-2/3} \Sigma^{5/3},
\end{equation}
valid on either the low or the high branch when the appropriate values for $\alpha$ and $\kappa$ are 
inserted. A solution on the low branch is possible provided that $\Sigma \leq \Sigma_{\rm max}$, where 
$\Sigma_{\rm max}$ is defined by the condition that $T_c = T_{\rm crit}$. Similarly, a high branch solution 
requires $\Sigma \geq \Sigma_{\rm min}$ with $T_c = T_{\rm crit}$ at $\Sigma_{\rm min}$. The limiting 
surface densities are given by,
\begin{eqnarray}
 \Sigma_{\rm max} & = & \frac{8}{3^{3/2}} \left( \frac{\mu m_H \sigma}{k_B} \right)^{1/2} 
 \Omega^{-1/2} T_{\rm crit}^{3/2} \kappa_{\rm low}^{-1/2} \alpha_{\rm low}^{-1/2}, \nonumber \\
 \Sigma_{\rm min} & = & \frac{8}{3^{3/2}} \left( \frac{\mu m_H \sigma}{k_B} \right)^{1/2} 
 \Omega^{-1/2} T_{\rm crit}^{3/2} \kappa_{\rm high}^{-1/2} \alpha_{\rm high}^{-1/2}.
\end{eqnarray} 
If $\kappa_{\rm high} > \kappa_{\rm low}$ and / or $\alpha_{\rm high} > \alpha_{\rm low}$, then 
$\Sigma_{\rm max} > \Sigma_{\rm min}$ and there will be a range of surface densities where 
accretion rates corresponding to either the low or the high branch are possible.

\begin{figure}[t]
\center
\includegraphics[width=0.75\columnwidth]{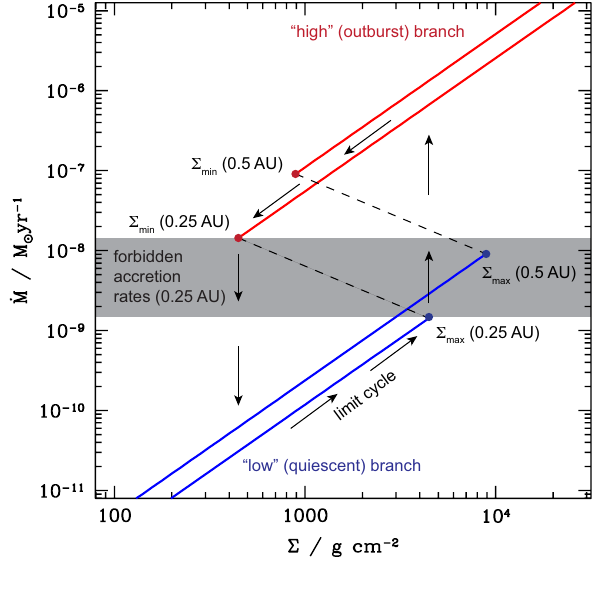}
\caption{Example S-curves in the accretion rate--surface density plane from the toy 
model described in the text. For these curves we take $\kappa_{\rm low} = \kappa_{\rm high} = 1 \ {\rm cm^2 \ g^{-1}}$, 
$\alpha_{\rm low} = 10^{-4}$, $\alpha_{\rm high} = 10^{-2}$, and $T_{\rm crit} = 10^3 \ {\rm K}$. The lower of the two 
curves is for $\Omega = 1.6 \times 10^{-6} \ {\rm s^{-1}}$ (0.25~AU for a Solar-mass star), the upper for 
$\Omega = 5.6 \times 10^{-7} \ {\rm s^{-1}}$ (0.5~AU).}
\label{fig_scurve}
\end{figure}

Figure~\ref{fig_scurve} shows, for a fairly arbitrary choice of the model parameters, the thermal equilibrium 
solutions that correspond to the low and high states of the disk annulus. One should not take the results 
of such a toy model very seriously, but it captures several features of more realistic models,
\begin{itemize}
\item
The solution has stable thermal equilibrium solutions on two branches, a low state branch where $\dot{M}$ 
for a given surface density is small, and a high state branch where it is substantially larger. In the toy model 
these branches are entirely separate, but in more complete models they are linked by an {\em unstable} 
middle branch (giving the plot the appearance of an ``S"-curve).
\item
There is a range of surface densities for which either solution is possible.
\item
There is a band of accretion rates for which no stable equilibrium solutions exist.
\item
The position of the S-curve in the $\Sigma - \dot{M}$ plane is a function of radius, with the band of 
forbidden accretion rates moving to higher $\dot{M}$ further from the star.
\end{itemize}

The S-curve is derived from a local analysis, and the existence of annuli whose 
thermal equilibrium solutions have this morphology is a necessary but not sufficient condition for a global disk 
outburst. That time dependent behavior of some sort is inevitable can be seen by supposing that the annulus at 0.25~AU in 
Figure~\ref{fig_scurve} is fed with gas from outside at a rate that falls into the forbidden band. No stable thermal 
equilibrium solution with this accretion rate is possible. If the disk is initially on the lower branch, the rate of 
gas supply exceeds the transport rate through the annulus, and the surface density increases. This continues 
until $\Sigma$ reaches and exceeds $\Sigma_{\rm max}$, at which point the only available solution lies on the 
high branch at much higher accretion rate. The annulus transitions to the high branch, where the transport rate 
is now larger than the supply rate, and the surface density starts to drop. The cycle is completed when 
$\Sigma$ falls below $\Sigma_{\rm min}$, triggering a transition back to the low state.

The evolution of a disk that is potentially unstable (i.e. one that has some annuli with S-curve thermal 
equilibria) is critically dependent upon the {\em radial} flow of mass and heat, which are the key extra ingredients needed 
if an unstable disk is to ``organize" itself and produce a long-lived outburst. To see this, imagine a disk 
where annuli outside $r_f$ are already on the high branch of the S-curve, while those inside remain 
on the low branch. The strong radial gradient of $T_c$ implies a similarly rapid change in $\nu$, which 
leads to a large mass flux from the annuli that are already in outburst toward those that remain 
quiescent (equation~\ref{eq_vr}). The resulting increase in surface density, along with the heat that 
goes with it, can push the neighboring annulus on to the high branch, initiating a propagating 
``thermal front" that triggers a large scale transition of the disk into an outburst state.

The quantitative modeling of global disk evolution including these thermal processes is well-developed 
within the classical $\alpha$ disk formalism \cite{lasota01}. For a minimal model, all that is needed is 
to supplement the disk evolution equation~(\ref{eq_disk_evolve}) with a model for the vertical 
structure (conceptually as described for the toy model above) and an equation for the evolution of the central 
temperature. This takes the form \cite{cannizzo93,lasota01},
\begin{equation}
 \frac{\partial T_c}{\partial t} = \frac{Q_+ - Q_-}{c_p \Sigma} - \frac{{\cal{R}}T_c}{\mu c_p} 
 \frac{1}{r} \frac{\partial}{\partial r} \left( r v_r \right) - v_r \frac{\partial T_c}{\partial r} + ...
\end{equation}  
Here $c_p$ is the specific heat capacity at constant pressure, and $\cal{R}$ is the gas 
constant. The first term on the right hand side describes the direct heating and cooling 
due to viscosity and radiative losses, while the second and third terms describe 
$P {\rm d}V$ work and the advective transport of heat associated with radial mass 
flows. In general there should be additional terms to represent the radial flow of heat 
due to radiative and / or turbulent diffusion (these effects are small in most thin disk 
situations, but become large when there is an abrupt change in $T_c$ at a thermal 
front). The treatment of these additional terms is somewhat inconsistent in published 
models, though they can significantly impact the character of derived disk outbursts 
\cite{owen14}. 

\subsubsection{Classical thermal instability}
In physical rather than toy models for episodic accretion $\alpha$ and $\kappa$ are smooth 
rather than discontinuous functions of the temperature. A local instability, with a resulting 
S-curve, occurs if one or both of these functions changes sufficiently rapidly with $T_c$ 
(so that equation~\ref{eq_thermal_instability} is satisfied). No simple condition specifies 
when a disk that has some locally unstable annuli will generate well-defined global 
outbursts, but loosely speaking outbursts occur provided that the branches of the 
S-curve (and the values of $\Sigma_{\rm max}$ and $\Sigma_{\rm min}$) are well 
separated.

The classical cause of disk thermal instability is the rapid increase in opacity associated 
with the ionization of hydrogen, at $T \simeq 10^4 \ {\rm K}$. Around 
this temperature $\kappa$ can rise as steeply as $T^{10}$, and the disk will invariably 
satisfy at least the condition for local thermal instability. The evolution of disks subject 
to a hydrogen ionization thermal instability was first investigated as a model for dwarf 
novae (eruptive disk systems in which a white dwarf accretes from a low-mass companion 
star) \cite{hoshi79,meyer81}, and subsequently applied to low-mass X-ray binaries. 
Thermal instability models provide a generally good match to observations of outbursts 
in these systems (which are of shorter duration that YSO outbursts, and correspondingly 
better characterized empirically), and are accepted as the probable physical cause. Good 
fits to data require models to include not only the large change in $\kappa$ that is 
the cause of classical thermal instability, but also changes in $\alpha$ between the low and 
high branches of the S-curve. Typically $\alpha_{\rm low} \leq 10^{-2}$, whereas 
$\alpha_{\rm high} \sim 0.1$ \cite{king07}. MHD simulations that include radiation transport 
have shown that the S-curve derived from $\alpha$ disk models can be approximately recovered 
as a consequence of the MRI, and that the difference in stress between the quiescent and 
outburst states may be attributable in part to the development of vertical convection within the 
hot disk \cite{hirose14,coleman16}.

By eye, the light curves of FU Orionis events look quite similar to scaled versions of 
dwarf novae outbursts, so the success of thermal instability models in the latter sphere 
makes them a strong candidate for YSO accretion outbursts. The central temperature of 
some outbursting FUOr disks, moreover, almost certainly does exceed the $10^4 \ {\rm K}$ 
needed to ionize hydrogen, making it inevitable that thermal instability physics will play 
some role in the phenomenon. Detailed thermal instability models of FU Orionis events 
were constructed by Bell \& Lin \cite{bell94}, who combined a one-dimensional (in $r$) 
treatment of the global evolution with detailed $\alpha$ model vertical structure 
calculations. They were able to find periodic solutions that describe ``self-regulated" 
disk outbursts (i.e., requiring no external perturbation or trigger), with the disk 
alternating between quiescent periods (with $\dot{M} = 10^{-8} - 10^{-7} \ M_\odot \ {\rm yr}^{-1}$) 
and outburst states (with $\dot{M} \geq 10^{-5} \ M_\odot \ {\rm yr}^{-1}$). These properties, 
and the inferred outburst duration of $\sim 10^2 \ {\rm yr}$, are in as good an agreement 
with observations as could reasonably be expected given the simplicity of the model.

The weakness of classical thermal instability models as an explanation for YSO accretion 
outbursts is that they require unnatural choices of the viscosity parameter $\alpha$ \cite{armitage98}. Thermal 
instability is immutably tied to the hydrogen ionization temperature, which exceeds even 
the mid-plane temperatures customarily attained in protoplanetary disks. The required temperatures 
can be reached (if at all) only extremely close to the star, and the radial region affected by 
instability extends to no more than 0.1-0.2~AU. The viscous time scale on these scales is 
short, so matching the century-long outbursts seen in FUOrs requires a very weak 
viscosity --- Bell \& Lin \cite{bell94} adopt $\alpha_{\rm high} = 10^{-3}$. This is at least an 
order of magnitude lower than the expected efficiency of MRI transport under ideal MHD 
conditions \cite{davis10,simon12}. Moreover, in the specific case of FU Orionis itself, 
radiative transfer models suggest that the region of high accretion rate during the outburst 
extends out to 0.5-1~AU \cite{zhu07}, substantially larger than would be expected in the 
thermal instability scenario.

\subsubsection{Instabilities of dead zones}
A dead zone in which the MRI is suppressed by Ohmic resistivity (\S\ref{sec_mhd_transport}) supports a 
related type of instability whose high and low states are distinguished primarily by different values of 
$\alpha$, rather than by the thermal instability's different values of $\kappa$. The origin of instability 
is clear within Gammie's original conception \cite{gammie96} of the dead zone, which has a simple 
two-layer structure. The surface of the disk, ionized by X-rays\footnote{Cosmic rays in the original model, though this is 
an unimportant distinction.}, is MRI-active and supports accretion with a local $\alpha \sim 10^{-2}$. Below 
a critical column density Ohmic resistivity completely damps MRI-induced turbulence (according to equation~\ref{eq_ReM}), 
and the disk is dead with $\alpha = 0$. This structure can be bistable if the surface density exceeds that of the 
ionized surface layer. The low accretion rate state corresponds to a cool, externally heated disk with a dead zone; 
the high accretion rate state to a hot thermally ionized disk at the same $\Sigma$.

Martin \& Lubow \cite{martin11b,martin14} have shown that the local physics of Ohmic dead zone instability 
can be analyzed in an manner closely analogous to thermal instability. The sole difference lies in the reason why 
the lower state ceases to exist above a critical surface density ($\Sigma_{\rm max}$ in Figure~\ref{fig_scurve}). 
For thermal instability $\Sigma_{\rm max}$ is set by the onset of ionization at the disk mid-plane, and the attendant 
rise in opacity. A simple dead zone, however, {\em does not get hotter with increasing $\Sigma$}, because the 
heating (either from irradiation, or viscous dissipation in the ionized surface layer) occurs at low optical depth 
at a rate that is independent of the surface density. It is then not obvious how even a very thick dead 
zone can be heated above $10^3 \ {\rm K}$, ``ignited", and induced to transition to the high state of the 
S-curve.

One way to trigger a jump to the high state is to postulate that some source of turbulence other than the MRI 
is present to heat the disk. Gammie \cite{gammie99} suggested that inefficient transport through the ionized 
surface layer would lead to the build of mass in the dead zone \cite{gammie96} until $Q \approx 1$ and 
self-gravity sets in (see \S\ref{sec_self_gravity}). We can readily estimate the properties we might expect for 
an instability triggered in this manner by the onset of self-gravity at small radii. Suppose that at 1~AU the 
temperature in the quiescent (externally irradiated) disk state is 150~K. Then to reach $Q = 1$ requires 
gas to build up until $\Sigma \simeq 7 \times 10^4 \ {\rm g \ cm^{-2}}$, at which point the mass interior 
is $M \sim \pi r^2 \Sigma \sim 0.025 \ M_\odot$. This is comparable to the amount of mass accreted per 
major FU Orionis-like outburst. In the outburst state the disk will be moderately thick (say $h/r = 0.3$ 
\cite{bell97}), and the viscous time scale $(1 / \alpha \Omega)(h/r)^{-2}$ works out to be about 200~yr 
for $\alpha = 0.01$, again similar to inferred FUOr time scales. At this crude level of estimate, it therefore 
seems possible that a self-gravity triggered dead zone instability could be consistent both with the 
inferred size of the outbursting region \cite{zhu07} and with theoretical best guesses as to the strength 
of angular momentum transport in ideal MHD conditions.

Time-dependent models for outbursts arising from a dead zone instability were computed by Armitage 
et al. \cite{armitage01}, and subsequently by several groups in both one-dimensional \cite{zhu10a,zhu10b,martin11b,owen14} 
and two-dimensional models \cite{zhu09,bae14}. The more recent studies show that a self-gravity triggered instability of an Ohmic 
dead zone can give rise to outbursts whose properties are broadly consistent with those of observed 
FUOrs. Instability persists even if there is a small residual viscosity within the dead zone \cite{bae13,martin14}, 
which could arise hydrodynamically in response to the ``stirring" from the overlying turbulent surface layer \cite{fleming03}. 
How well such models work when either or both of the Hall effect and MHD winds contribute to the dynamics close 
to the star remains unclear. In principle these effects could also lead to entirely different types of 
eruptive behavior.

\subsection{Triggered accretion outbursts}
Accretion variability, including (perhaps) the large scale outbursts of FUOrs, can also be triggered by 
processes largely independent of the inner disk itself. Stellar activity cycles, binaries with small periastron 
distances, and tidal disruption of gaseous clumps or planets may all contribute.

\subsubsection{Stellar activity cycles}
As discussed in \S\ref{sec_magnetospheric} the inner disk is expected \cite{konigl91} 
and observed \cite{bouvier07} to be disrupted by the stellar magnetosphere. The 
complex dynamics of the interaction between the field --- which may be misaligned to the 
stellar spin axis and have non-dipolar components \cite{johnstone14} --- and the disk \cite{kurosawa13} 
is likely the dominant cause of T Tauri variability on time scales comparable to the 
stellar rotation period (i.e. days to weeks). If the strength of the field also varies systematically 
due to the presence of activity cycles analogous to the Solar cycle, these could trigger 
longer time scale (years to decades) accretion variability. The simplest mechanism 
is modulation of the magnetospheric radius across corotation \cite{clarke95}. When the field is strong 
and $r_m > r_{\rm co}$ the linkage between the stellar field lines and the disks adds 
angular momentum to the disk, impeding accretion in the same way as gravitational 
torques from a binary (\S\ref{sec_binary_bc})\footnote{In compact object accretion, this is 
described as the ``propeller" regime of accretion \cite{illarionov75}.}. 
Gas then accumulates just outside the magnetospheric radius, and can subsequently be accreted 
in a burst when the field weakens.

The viability of such magnetically ``gated" accretion as an origin for large scale 
variability is limited by the short viscous time scale of the disk at $r \approx r_m$, 
which makes it hard to accumulate large masses of gas if the stellar fields are, 
as expected, of no more than kG strength. Models \cite{armitage95} suggest that 
significant decade-long variability could be associated with protostellar activity cycles, 
but there is no clear path to generating FU Orionis outbursts. Activity cycles are 
more promising as an explanation for lower amplitude, periodic EXor outbursts \cite{dangelo10,dangelo12}. 
The interaction between the time-variable stellar magnetic field and the non-ideal physics 
of the inner disk, just outside the thermally ionized region, could be significant \cite{armitage16}. 

\subsubsection{Binaries}
An eccentric binary with an AU-scale periastron distance could funnel gas into the inner 
disk, increasing the accretion rate and leading to an outburst if the increased surface density 
is high enough to trigger thermal or dead zone instability. This mechanism was proposed 
by Bonnell \& Bastien \cite{bonnell92}, and has subsequently been studied with higher 
resolution simulations \cite{pfalzner08,forgan10}. Although there are some differences 
between the predicted outbursts and those observed (this is almost inevitable, as the 
limited sample of FUOrs is already quite diverse), it is clear that close encounters from 
binary or cluster companions induce episodes of substantially enhanced accretion. The 
obvious prediction --- that FUOrs ought to be found with observable binary companions 
or preferentially associated with higher-density star forming regions  --- is neither 
confirmed nor ruled out given the small sample of known objects. 

\subsubsection{Clump tidal disruption}
\label{sec_roche}
A final possibility is that accretion outbursts could be triggered by the tidal disruption of a 
bound object (a planet or gas cloud) that migrates too close to the star. The necessary 
condition for this to occur is given by the usual argument for the Roche limit. If we consider 
a planet with radius $R_p$ and mass $M_p$, orbiting a star of mass $M_*$ at radius $r$, 
the differential (tidal) gravitational force between the center of the planet and its surface is,
\begin{equation}
 F_{\rm tidal} = \frac{GM_*}{r^2} - \frac{GM_*}{(r + R_p)^2} \simeq \frac{2 GM_*}{r^3} R_p.
\end{equation} 
Equating the tidal force to the planet's own self-gravity, $F_{\rm self} = GM_p / R_p^2$, we 
find that tidal forces will disrupt the planet at a radius $r_{\rm tidal}$ given approximately by, 
\begin{equation}
 r_{\rm tidal} = \left( \frac{M_*}{M_p} \right)^{1/3} R_p.
\end{equation}
An equivalent condition is that tidal disruption occurs when the mean density of the planet 
${\bar{\rho}} < M_* / r^3$. 

It is difficult to tidally disrupt a mature giant planet. A Jupiter mass planet has a radius of 
$R_p \simeq 1.5 \ R_J$ at an age of 1~Myr \cite{marley07}, and will not be disrupted 
outside the photospheric radius of a typical young star. (Though such planets, if 
present in the inner disk, could alter the course of thermal or dead zone instability \cite{clarke96,lodato04b}.) 
If tidal disruption is to be relevant to episodic accretion we require, first, that the outer disk 
is commonly gravitationally unstable to fragmentation, and, second, that the clumps that 
form migrate rapidly inward (in the Type~1 regime discussed by Kley in this volume) without 
contracting too rapidly. Numerical evidence supports the idea that clump migration can be 
rapid \cite{vorobyov05,baruteau11,cha11,michael11}, though it is at best unclear whether contraction 
can be deferred sufficiently to deliver clumps that would be tidally disrupted on sub-AU 
scales \cite{galvagni12}. Assuming that these pre-conditions are satisfied, however, 
Nayakshin \& Lodato \cite{nayakshin12} studied the tidal mass loss from the close-in 
planets and its impact on the disk. They found that the tidal disruption of $\sim 20 \ R_J$ 
clumps, interior to 0.1~AU, led to accretion outbursts consistent with the basic properties 
of FUOrs.

The primary theoretical doubts about tidal disruption as a source of outbursts concern 
the relative rates of inward migration and clump contraction, which are both hard to 
calculate at substantially better than order of magnitude level. Observationally, this 
process would produce outbursts in systems whose disks were young, massive, and 
probably still being fed by envelope infall.

\begin{svgraybox}
\begin{itemize}
\item
Episodic accretion on to Young Stellar Objects occurs in several different flavors, 
including high-amplitude outbursts such as FU~Orionis events that would certainly 
have an important influence on disk chemistry and planet formation close to the star.
\item
A possible explanation for episodic accretion is an instability in the equilibrium disk 
structure on AU or sub-AU scales, related to the 
classical thermal instability that explains outbursts in dwarf nova systems. A 
potential instability of dead zone structure has been demonstrated in one- and 
two-dimensional models, but may need to be modified to take account of additional 
non-ideal MHD physics and disk winds.
\item
Outbursts could also be {\em internally} triggered, for example by changes in the 
stellar magnetic field, or {\em externally} triggered, if 
dense clumps of gas migrate quickly into the inner disk and are tidally 
disrupted.
\end{itemize}
\end{svgraybox}

\newpage

\section{Single and collective particle evolution}
\label{sec_single}

The evolution of solid particles within disks differs from that of gas 
because solid bodies are unaffected by pressure gradients but do experience 
aerodynamic forces. We discuss here how these differences affect the motion of 
single particles orbiting within the gas disk, and how we can describe the 
evolution of a ``fluid" made up of small solid particles interacting aerodynamically 
with the gas. Issues such as the rate and outcome of particle collisions, that are 
central to early stage planet formation, are treated in the accompanying part by Kley, 
and elsewhere \cite{armitage10}.

The key parameter describing the aerodynamic coupling between solid particles and 
gas is the {\em stopping time}. For a particle of mass $m$ that is moving with velocity 
$\Delta v$ relative to the local gas, the stopping time is defined as,
\begin{equation}
 t_s \equiv \frac{m \Delta v}{| F_{\rm drag} |},
\end{equation}
where $F_{\rm drag}$ is the magnitude of the aerodynamic drag force that acts 
in the opposite direction to $\Delta v$. Very frequently, what matters most is how 
the stopping time compares to the orbital time at the location of the particle. We 
therefore define a dimensionless stopping time by multiplying $t_s$ by the orbital 
frequency $\Omega_{\rm K}$,
\begin{equation}
 \tau_s \equiv t_s \Omega_{\rm K}.
\end{equation} 
The dimensionless stopping time is also called the {\em Stokes number}.

For our immediate purposes it largely suffices to describe aerodynamic 
effects in terms of the stopping time, but eventually you will want to translate 
the results into concrete predictions for how particles of various sizes and 
material properties behave. This requires specifying $F_{\rm drag}$, whose 
form depends on the size of the particle relative to the mean free path of 
gas molecules, and (in the fluid regime) on the Reynolds number of the 
flow around the particle \cite{whipple72}. If the particle radius $s$ is 
small compared to the mean free path $\lambda$ ($s < 9 \lambda / 4$) 
the particle experiences {\em Epstein drag}, with a drag force,
\begin{equation}
 {\bf F}_{\rm drag} = - \frac{4 \pi}{3} \rho s^2 v_{\rm th} \Delta {\bf v}.
\label{eq_epstein} 
\end{equation}
Here $\rho$ is the density of the surrounding gas, and the thermal 
speed of the molecules,
\begin{equation}
 v_{\rm th} = \sqrt{ \frac{8 k_B T}{\pi \mu m_H} },
\end{equation}
is roughly the same as the sound speed. Because Epstein drag is 
proportional to the velocity difference $\Delta v$ (rather than the more 
familiar square of the velocity difference), the stopping time is a function 
of the particle properties that is independent of the velocity difference. For a 
spherical particle of material density $\rho_m$, 
\begin{equation}
 t_s = \frac{\rho_m}{\rho} \frac{s}{v_{\rm th}}.
\label{eq_stopping_epstein} 
\end{equation} 
The mean free path in protoplanetary disks is of the order of cm (larger in the 
outer disk), so the Epstein regime is relevant for particles that range from dust 
to those of small macroscopic dimensions. Drag laws appropriate for larger bodies, 
which fall into the Stokes regime of drag, are given by Whipple \cite{whipple72}. 

\subsection{Radial drift}
\label{subsec_drift}
The most important consequence of aerodynamic forces is the phenomenon of 
{\em radial drift}. In \S\ref{sec_radial_structure} we showed that radial pressure 
gradients result in a gas orbital velocity that differs from the Keplerian value by 
50-100~$\rm m \ s^{-1}$ (using typical disk parameters at 1~AU). Most commonly, 
the gas is partially supported against gravity by the pressure gradient, and so 
rotates more slowly than the Keplerian value. We will consider how this velocity 
differential affects the evolution of solids in various limits. We will start by ignoring both turbulence 
and the feedback of aerodynamic forces {\em on the gas}, before revisiting the problem 
with these processes included.

\subsubsection{Particle drift without feedback}
\label{sec_particle_v}
Large bodies ($\tau_s \gg 1$) orbit at close to the Keplerian speed, and the effect 
of gas on their evolution can be considered as a simple ``headwind" if the disk 
is sub-Keplerian. Suppose that the gas orbits at a speed $v_K - \Delta v$, with 
$\Delta v \ll v_K$. The drag force $| F_{\rm drag} | = m \Delta v \Omega_{\rm K} / \tau_s$ 
does work at a rate,
\begin{equation}
 \dot{E} \simeq - |F_{\rm drag}| v_K,
\end{equation} 
that leads to a change in the orbital energy $E = - GM_* m / 2 a$, where $a$ is the 
radius of the orbit. Noting that,
\begin{equation}
 \dot{E} = \frac{GM_*m}{2 a^2} \frac{{\rm d}a}{{\rm d}t},
\end{equation}
and equating the two expressions for $\dot{E}$, we find that the orbit decays at 
a speed $v_r = {\rm d}a / {\rm d}t$ that is given by,
\begin{equation}
 v_r = -\frac{2}{\tau_s} \Delta v.
\end{equation}  
The radial drift of large bodies is inversely proportional to their Stokes number.

The simple headwind argument fails for small particles with $\tau_s \ll 1$, which 
instead are forced to orbit {\em at the gas speed} by the strong aerodynamic 
coupling. The particles do not feel the pressure gradient, so their non-Keplerian 
orbital motion results in a net radial force,
\begin{equation}
 \frac{F_r}{m} = \frac{(v_K - \Delta v)^2}{a} - \frac{GM_*}{a^2} \simeq -\frac{2 v_K \Delta v}{a}.
\end{equation} 
Equating this to the drag force for {\em radial} motion at speed $v_r$, 
$|F_{\rm drag}| / m = v_r \Omega_{\rm K} / \tau_s$, we find that radial drift 
for small particles occurs at the terminal drift speed,
\begin{equation}
 v_r = -2 \tau_s \Delta v.
\end{equation}
This is the speed relative to the gas, so for a disk that is accreting there is an additional 
component given by the gas' radial velocity. 

Intermediate-sized particles orbit at some speed between that of the gas and that 
given by the Keplerian velocity. To derive the general rate of radial drift 
\cite{weidenschilling77,takeuchi02}, we consider a gas disk whose orbital 
velocity is,
\begin{equation}
 v_{\phi,{\rm gas}} = v_K \left( 1 - \eta \right)^{1/2}.
\label{eq_vphi_gas} 
\end{equation}
The parameter $\eta \propto (h/r)^2$. For example, if the disk has $\Sigma \propto r^{-1}$ and 
central temperature $T_c \propto r^{-1/2}$, we showed in \S\ref{sec_radial_structure}  
that $\eta = (11/4)(h/r)^2$. Defining the particle radial and 
azimuthal velocities to be $v_r$ and $v_\phi$ respectively, 
the equations of motion are,
\begin{eqnarray}
 \frac{{\rm d}v_r}{{\rm d}t} & = & \frac{v_\phi^2}{r} 
 - \Omega_{\rm K}^2 r - \frac{1}{t_{\rm s}} \left( v_r - v_{r,{\rm gas}} \right) \\
 \frac{\rm d}{{\rm d}t} \left( r v_\phi \right) & = & 
 - \frac{r}{t_{\rm s}} \left( v_\phi - v_{\phi,{\rm gas}} \right).
\end{eqnarray}
The azimuthal equation can be simplified by noting that the specific angular 
momentum remains close to Keplerian,
\begin{equation} 
 \frac{\rm d}{{\rm d}t} \left( r v_\phi \right) \simeq 
 v_r \frac{\rm d}{{\rm d}r} \left( r v_K \right) = 
 \frac{1}{2} v_r v_K.
\end{equation}
This yields,
\begin{equation}
 v_\phi - v_{\phi,{\rm gas}} \simeq - \frac{1}{2} \frac{t_{\rm s}v_r v_K}{r}.
\label{eq_drift1} 
\end{equation}   
We now substitute for $\Omega_K$ in the radial equation using 
equation~(\ref{eq_vphi_gas}). Discarding higher order terms we obtain,
\begin{equation}
 \frac{{\rm d}v_r}{{\rm d}t} = - \eta \frac{v_K^2}{r} + 
 \frac{2 v_K}{r} \left( v_\phi - v_{\phi,{\rm gas}} \right) 
 - \frac{1}{t_{\rm s}} \left( v_r - v_{r,{\rm gas}} \right).
\label{eq_drift2}
\end{equation}
The ${{\rm d}v_r}/{{\rm d}t}$ term is negligible. Dropping that, we eliminate $(v_\phi - v_{\phi,{\rm gas}})$ 
between equations~(\ref{eq_drift1}) and (\ref{eq_drift2}) to obtain,
\begin{equation}
 v_r = \frac{(r/v_K) t_{\rm s}^{-1} v_{r,{\rm gas}} -\eta v_K}{(v_K/r) t_{\rm s} + 
 (r/v_K) t_{\rm s}^{-1}}.
\end{equation} 
In terms of the Stokes number the final result for the particle radial velocity is,
\begin{equation}
 v_r = \frac{\tau_{\rm s}^{-1} v_{r,{\rm gas}}-\eta v_K}{\tau_{\rm s} + 
 \tau_{\rm s}^{-1}}.
\label{eq_radial_drift} 
\end{equation}
The previously derived results for very small and very large particles are recovered 
by taking the appropriate limits.

\begin{figure}[t]
\includegraphics[width=\columnwidth]{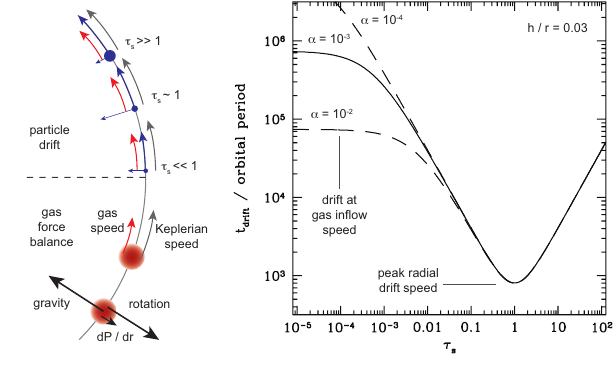}
\caption{Particles drift inwards in a disk wherever ${\rm d}P / {\rm d}r$, due to the  
combination of aerodynamic forces and sub-Keplerian gas rotation. 
The radial drift time scale $t_{\rm drift} = r / |v_r|$, in units of the local orbital period, 
is plotted as a function 
of the dimensionless stopping time $\tau_s$. The fastest 
radial drift occurs for $\tau_s = 1$. The specific numbers shown in the plot are 
appropriate for a disk with $h/r = 0.03$ and $\alpha$ values of $10^{-2}$ (lowest 
curve on the left-hand side of the figure), $10^{-3}$ and $10^{-4}$.}
\label{fig_radial_drift}
\end{figure}

The speed of the radial drift implied by equation~({\ref{eq_radial_drift}) is shown 
in Figure~\ref{fig_radial_drift}. Very rapid drift is predicted for particles with 
$\tau_s \sim 1$. For our fiducial disk model with $\Sigma \propto r^{-1}$, 
$T_c \propto r^{-1/2}$ and $h/r = 0.03$, the radial drift time scale 
$t_{\rm drift} = r / |v_r|$ is just $10^3$ orbital periods --- one thousand years 
at 1~AU! Particles with this stopping time are very roughly meter-sized, and 
their rapid drift is the origin of the ``meter-sized barrier" that severely 
constrains models of planetesimal formation \cite{chiang10,johansen14}.

\subsubsection{Drift with diffusion}
\label{sec_particle_diffuse}
If the disk is turbulent, small dust particles that are aerodynamically well-coupled to 
the gas will diffuse radially and vertically. (The same physics will apply to trace gas 
species, such as water or CO molecules.) Diffusion will tend to equalize the 
{\em concentration} of the dust or trace gas species relative to the dominant 
gaseous component in the disk. To derive an equation for the evolution of the 
trace species \cite{clarke88}, we initially ignore any radial drift due to aerodynamic effects and 
write the concentration of the trace gas or dust (generically the ``contaminant") as, 
\begin{equation}
 f = \frac{\Sigma_d}{\Sigma},
\end{equation} 
where $\Sigma_d$ is the surface density of the contaminant. 
If the contaminant is neither created nor destroyed within the region of the 
disk under consideration, continuity demands that,
\begin{equation}
 \frac{\partial \Sigma_d}{\partial t} + 
 \nabla \cdot {\bf F}_d = 0,
\label{eq_Ccontinuity} 
\end{equation}
where ${\bf F}_d$, the flux, can be decomposed into two parts: an advective 
piece describing transport of the dust or gas with the mean disk flow, and 
a diffusive piece describing the tendency of turbulence to equalize the 
concentration of the contaminant across the disk. For $f \ll 1$ we can 
assume that the diffusive properties of the disk depend only 
on the {\em gas} surface density, in which case the flux can be written as,
\begin{equation}   
 {\bf F}_d = \Sigma_d {\bf v} - D \Sigma \nabla \left( \frac{\Sigma_d}{\Sigma} 
 \right).
\label{eq_Cflux}
\end{equation}
Here ${\bf v}$ is the mean velocity of gas in the disk and $D$ is the  
turbulent diffusion co-efficient. The diffusive term vanishes if $f$ is constant. 
Combining this equation 
with the continuity equation for the gaseous component, we obtain an 
evolution equation for $f$ in an axisymmetric disk. In cylindrical 
polar co-ordinates,
\begin{equation}
 \frac{\partial f}{\partial t} = 
 \frac{1}{r \Sigma} \frac{\partial}{\partial r}
 \left( D r \Sigma \frac{\partial f}{\partial r} \right) 
 - v_r \frac{\partial f}{\partial r}.
\label{eq_concentration} 
\end{equation}   
This result has the form of an advection-diffusion 
equation, with the advective component being due to the radial flow 
of the disk gas. It is easy to generalize 
this equation to account for the radial drift of larger particles that 
are imperfectly coupled to the gas, by adding an additional flux 
representing the radial drift speed \cite{alexander07}. 

Determining what is the appropriate value for the turbulent diffusion co-efficient $D$ 
involves many of the same uncertainties that afflict the determination of the turbulent 
viscosity. For trace gas species and very small dust particles the zeroth-order 
expectation is that $D \approx \nu$ \cite{morfill83}, though simulations of 
non-ideal MHD disk turbulence show both significant deviations from $\nu$ and anisotropy 
between the vertical and radial directions \cite{zhu15}. For larger bodies 
there is a well-determined analytic scaling with the Stokes number of the 
particles \cite{youdin07}.

\subsubsection{Particle pile-up}
\label{sec_pileup}
Solids that experience significant radial drift tend to become concentrated (``pile-up") 
in the inner disk \cite{youdin02,youdin04}. The basic effect is present in the simplest case 
where diffusion and inward drag by the mean flow are small, and feedback of the 
particles on the gas can be neglected. The radial drift speed is then, approximately,
\begin{equation}
 v_r \simeq - \tau_s \eta v_K,
\end{equation}
with $\eta \propto (h/r)^2$. For particles that are in the Epstein regime of drag, the 
stopping time in the mid-plane is (from equations~\ref{eq_stopping_epstein} and 
\ref{eq_rho0}),
\begin{equation}
 t_s = \frac{\rho_m}{\rho_0} \frac{s}{v_{th}} = \sqrt{2 \pi} \frac{\rho_m h}{\Sigma} \frac{s}{v_{th}}.
\end{equation}  
Since $h = c_s / \Omega$, and $c_s$ and $v_{th}$ differ only by a numerical factor, we 
obtain,
\begin{equation}
 \tau_s = \frac{\pi}{2} \frac{\rho_m}{\Sigma} s.
\end{equation} 
Suppose now (perhaps not very realistically) that the surface density profile of solids 
has attained a steady-state, such that the mass flux is constant with radius. Then,
\begin{equation}
 \dot{M}_d = - 2 \pi r \Sigma_d v_r = {\rm constant},
\end{equation} 
and substituting for $v_r$ we find,
\begin{equation}
 \frac{\Sigma_d}{\Sigma} \propto \left( \frac{h}{r} \right)^{-2} r^{-1/2}.
\end{equation}
For a disk with constant $(h/r)$ the steady-state concentration of solids increases 
closer to the star as $r^{-1/2}$. In the more realistic case of a flaring disk with 
mid-plane temperature profile, say, $T_c \propto r^{-1/2}$, the effect is stronger. 
A constant $\alpha$ model of such a disk has a steady-state gas surface density 
profile $\Sigma \propto r^{-1}$, with the solids following $\Sigma_d \propto r^{-2}$.

\subsubsection{The Nakagawa-Sekiya-Hayashi equilibrium}
Up till now we have implicitly assumed that the evolution of the gas is 
unaffected by the evolution of the solids within it. Obviously this can never be 
strictly correct. If a population of solid particles are losing angular momentum 
to the gas through aerodynamic forces and spiraling inward, the gas must 
gain a corresponding amount of angular momentum. If the solid to gas ratio 
is only of the order of 1\%, however, one might suppose that the effect of the 
angular momentum exchange on the gas would be small and, perhaps, ignorable. 
This is only partially true. First, a number of processes, including vertical particle 
settling \cite{dubrulle95}, gas loss via photo-evaporation \cite{throop05,alexander07} 
or MHD winds, and radial drift itself \cite{youdin04}, can boost the solid to gas ratio, 
at least locally. Second, the equilibrium solution for radial drift in the presence of 
back reaction on to the gas can be unstable to the {\em streaming instability} \cite{youdin05}, 
which can result in strong localized clumping of the solids.

The generalization of the radial drift formula (equation~\ref{eq_radial_drift}) to account 
for the back reaction of the drift on the gas is known as the Nakagawa-Sekiya-Hayashi (NSH) 
equilibrium \cite{nakagawa86}. The NSH solution is derived by considering the interaction 
between solids and gas in a simple disk model that ignores the effects of turbulence and 
vertical gravity. Both the gas, with density $\rho_g$, pressure $P$ and velocity ${\bf v}_g$, 
and the solid particles, with density density $\rho_p$ and velocity ${\bf v}_p$, are treated as 
fluids that interact with each other via aerodynamic drag. They obey continuity and 
momentum equations of the form \cite{youdin05},
\begin{eqnarray}
 \frac{\partial \rho_p}{\partial t} + \nabla \cdot (\rho_p {\bf v}_p) & = & 0, \\
 \nabla \cdot {\bf v}_g & = & 0, \\
 \frac{\partial {\bf v}_p}{\partial t} + {\bf v}_p \cdot \nabla {\bf v}_p & = & 
 - \Omega_{\rm K}^2 {\bf r} - \frac{1}{t_s} \left( {\bf v}_p - {\bf v}_g \right), \\
 \frac{\partial {\bf v}_g}{\partial t} + {\bf v}_g \cdot \nabla {\bf v}_g & = & 
 - \Omega_{\rm K}^2 {\bf r} + \frac{1}{t_s}\frac{\rho_p}{\rho_g} \left( {\bf v}_p - {\bf v}_g \right) 
 - \frac{\nabla P}{\rho_g}. 
\end{eqnarray} 
We have replaced the continuity equation for the gas with the condition for incompressibility, 
which is valid provided that velocities remain highly subsonic. The only other differences between 
the equations for the two species are the presence of a pressure gradient 
term in the gas momentum equation, and the pre-factor in the aerodynamic drag term 
expressing the differing inertia of the two fluids.
 
It is straightforward to derive the steady-state axisymmetric solution for the drift of 
solids and gas from the above equations \cite{nakagawa86,youdin05}. To illustrate 
the difference between the NSH and no-feedback solutions, we quote here just the 
result for the relative radial velocity of the solids and the gas. With our definition of 
$\eta$ (equation~\ref{eq_vphi_gas}) this takes the form,
\begin{equation}
 v_{r,{\rm rel}} = - \frac{\rho_g}{\rho} \frac{\eta \tau_s v_K}{1 + (\tau_s \rho_g / \rho)^2},
\end{equation}  
where $\rho = \rho_g + \rho_p$. Equation~(\ref{eq_radial_drift}) is recovered in the 
case where $\rho_p \ll \rho_g$ and feedback can be neglected (note that we have ignored  
any radial gas motions due to processes {\em other} than particle-gas coupling in this version  
of the NSH solution).

\begin{figure}[t]
\center
\includegraphics[width=\columnwidth]{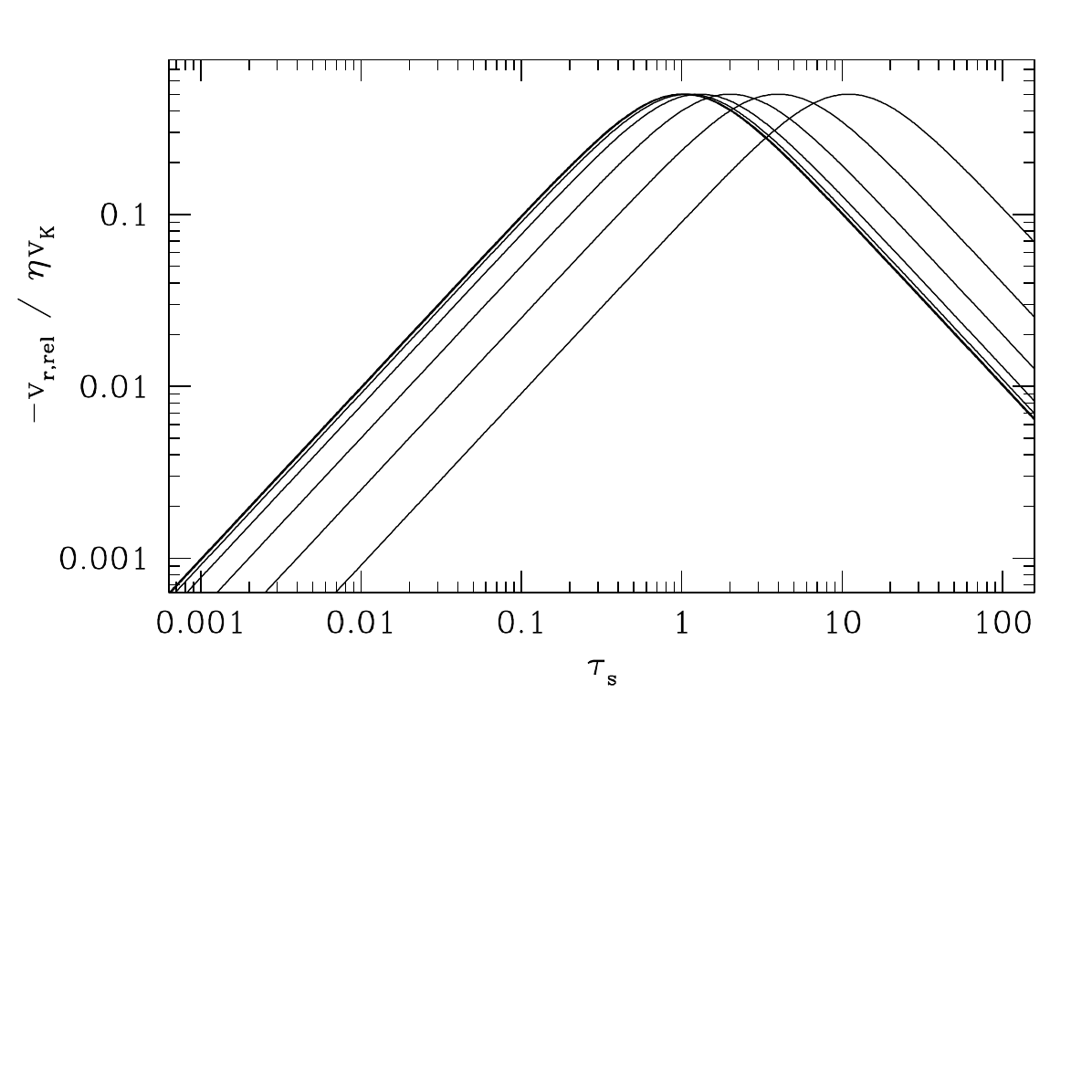}
\vspace{-4.5cm}
\caption{The relative velocity of solids compared to gas (normalized by $\eta v_K$, the parameter 
specifying departure of the gas disk from Keplerian rotation) is plotted as a function of the Stokes number $\tau_s$. 
From left to right the curves show the NSH equilibrium solutions for $\rho_p / \rho_g = 10^{-2}$ 
(effectively identical to the no-feedback solution), $3 \times 10^{-2}$, $0.1$, $0.3$, $1$, $3$ and 
$10$.}
\label{fig_nsh}
\end{figure}

The predicted relative radial velocity between the solid and gas components is plotted in 
Figure~\ref{fig_nsh} for various values of the solid to gas ratio. For $\rho_p / \rho_g = 10^{-2}-10^{-1}$ 
the rate of drift at a given value of the stopping time is very similar to the no-feedback solution 
given by equation~(\ref{eq_radial_drift}). For large values of $\rho_p / \rho_g$, however, the 
peak of the relative velocity curve (which in this limit is predominantly outward motion of the 
{\em gas}) shifts to somewhat higher values of $\tau_s$.

\subsection{Vertical settling}
\label{sec_settling}

Aerodynamic drag also modifies the vertical distribution of solids relative 
to gas. We first consider the forces acting on a small particle of mass $m$ 
at height $z$ above the mid-plane of a laminar disk. The vertical component of stellar 
gravity exerts a force,
\begin{equation}
 | F_{\rm grav} | = m \Omega_{\rm K}^2 z.
\end{equation}
The gas in the disk is supported against this force by the vertical pressure 
gradient, but solid particles are not. If started at rest a particle will accelerate 
until the gravitational force is balanced by drag. In the Epstein 
regime (given by equation~\ref{eq_epstein}) the resulting terminal 
velocity is,
\begin{equation}
 v_{\rm settle} = \frac{\rho_m}{\rho} \frac{s}{v_{\rm th}} \Omega_{\rm K}^2 z.
\label{eq_settle} 
\end{equation}
Using numerical values appropriate for a 1~$\mu$m particle 
at $z \sim h$ at 1~AU ($\rho = 6 \times 10^{-10} 
\ {\rm g} \ {\rm cm}^{-3}$, $z = 3 \times 10^{11} \ {\rm cm}$, 
$v_{\rm th} = 10^5 \ {\rm cm} \ {\rm s}^{-1}$) gives a settling speed 
$v_{\rm settle} \approx 0.06 \ {\rm cm} \ {\rm s}^{-1}$. The 
settling time, defined as,
\begin{equation}
 t_{\rm settle} = \frac{z}{| v_{\rm settle} |},
\end{equation}
is about $1.5 \times 10^5 \ {\rm yr}$. In the absence of turbulence 
micron-sized particles ought to settle out of the upper layers of the 
disk on a time scale that is shorter than the disk lifetime.

Turbulent diffusion acts to counteract the effects of settling. If the particle fluid with 
density $\rho_p$ behaves as a trace species (i.e. $\rho_p / \rho \ll 1$) then it 
obeys an advection-diffusion equation \cite{dubrulle95,fromang06},
\begin{equation}
 \frac{\partial \rho_p}{\partial t} = 
 D \frac{\partial}{\partial z} \left[ \rho 
 \frac{\partial}{\partial z} \left( \frac{\rho_p}{\rho} \right) \right] 
 + \frac{\partial}{\partial z} \left( \Omega_{\rm K}^2 t_s \rho_p z \right).
\end{equation}
Steady-state solutions to this equation can be found in the 
limit where the particle layer is thin enough that the {\em gas} density 
is approximately constant across the particle scale height. The 
dimensionless friction time $\tau_s$ is then independent of 
$z$ and we find,
\begin{equation}
 \frac{\rho_p}{\rho} = \left( \frac{\rho_p}{\rho} \right)_{z=0} 
 \exp \left[-\frac{z^2}{2 h_p^2} \right],
\end{equation}
where $h_p$, the scale height of the particle concentration $\rho_p / \rho$, is,
\begin{equation}
 h_p = \sqrt{ \frac{D}{\Omega_{\rm K}^2 t_s} }.
\end{equation}
If the turbulent diffusivity is comparable to the turbulent viscosity, i.e. $D \sim \nu$, the ratio of the concentration scale height to the gas scale height is just,
\begin{equation}
 \frac{h_p}{h} \simeq \sqrt{ \frac{\alpha}{\tau_s} }.
\end{equation}   
Solid particles become strongly concentrated 
toward the disk mid-plane whenever their dimensionless friction 
time substantially exceeds $\alpha$. For reasonable 
values of $\alpha$ this requires substantial particle growth.

\subsection{Streaming instability}
\label{sec_streaming}
Youdin \& Goodman \cite{youdin05} demonstrated that the aerodynamically coupled system 
of gas and solids, described by the NSH equilibrium, is linearly unstable to the growth of  
perturbations. The instability, known as the {\em streaming instability} by (very rough) 
analogy with the two-stream instability of plasmas, taps the free energy present in the 
relative motion between the solid and gaseous fluids, which is ultimately sustained by 
the background gradient in the pressure. It provides a physically plausible route to forming 
planetesimals (km or larger bodies that are largely decoupled from the gas) rapidly from smaller 
solids with $\tau_s \sim 1$ or less.

The pre-requisites for the existence of the streaming instability are two-way aerodynamic 
coupling between gas and dust within a rotating system (with shear and Coriolis force). 
A minimal mathematical analysis can be performed in the ``terminal velocity approximation", 
in which the relative velocity between gas and dust is,
\begin{equation}
 \Delta {\bf v} = -\frac{\nabla P}{\rho} t_s.
\end{equation}
This approximation yields a third-order dispersion relation \cite{youdin05}, while a 
more complete analysis (still neglecting vertical stratification) results in a sixth-order system. 
The linear growth rates, plotted in Youdin \& Goodman \cite{youdin05}, are functions of 
$\tau_s$ and the ratio of solid to gas density, $\rho_d / \rho_g$. Growth is typically 
substantially slower than dynamical, with the most unstable modes having scales $\ll h$. 
For $\tau_s \sim 10^{-2}$ and $\rho_p / \rho_g \sim 0.1$, for example, the linear growth 
time scale is a few hundred orbits.
 
A simple physical (as opposed to mathematical) explanation of the streaming instability 
is frustratingly elusive. (Analogies to ``traffic jams", or to the drag reducing properties 
of pelotons in bicycle races, are more relevant to the strong clustering that {\em results} 
from the streaming instability than to its existence as a linear instability.) The reader 
distressed by this state of affairs may seek solace in papers by Chiang \& 
Youdin \cite{chiang10}, and by Jacquet et al. \cite{jacquet11}, who discuss the 
origin of instabilities in simplified or related physical systems.

The relationship between the saturated state of the streaming instability (which is of 
greatest interest when the fluctuations in particle density are very strongly non-linear) 
and the linear growth phase is non-obvious, and requires numerical simulations. 
Starting with the work of Johansen \& Youdin \cite{johansen07}, several authors have 
quantified the outcome of the instability in protoplanetary disks 
(for recent examples, see e.g. \cite{bai10,johansen11,yang14}). Simplifying greatly, 
the streaming instability depends on the particle stopping time $\tau_s$, the local solid to gas ratio (or metallicity, with 
super-Solar metallicities $Z > 10^{-2}$ being favored \cite{johansen09,bai10}) and 
on the magnitude of the deviation from Keplerian velocity $\eta v_K / c_s$ (with 
{\em small} values of this parameter promoting clumping \cite{bai10b}). Using 
two-dimensional simulations Carrera et al. \cite{carrera15} and Yang et al. \cite{yang17} 
find that strong clumping occurs for $Z > Z_{\rm crit} (\tau_s)$, where the critical 
metallicity is fit by a piece-wise function,
\begin{eqnarray}
\log Z_{\rm crit} & = & 0.10 ( \log \tau_s )^2 + 0.20 \log \tau_s - 1.76 \,\,\, (\tau_s < 0.1), \nonumber \\
\log Z_{\rm crit} & = & 0.30 ( \log \tau_s )^2 + 0.59 \log \tau_s - 1.57 \,\,\, (\tau_s > 0.1).
\end{eqnarray}
These results suggest that the sweet spot where the lowest metallicity is required for strong 
clustering occurs for $\tau_s \approx 0.1$ at $Z \approx 0.015$.

\begin{figure}[t]
\center
\includegraphics[width=\columnwidth]{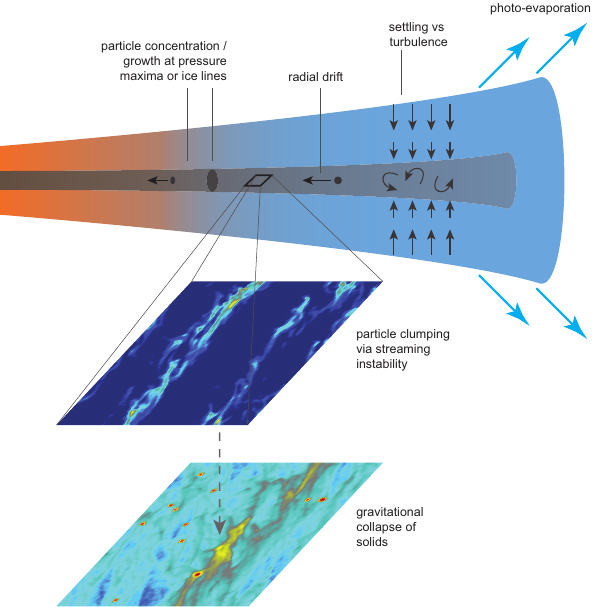}
\caption{Illustration of some of the processes that can lead to streaming instability in the 
aerodynamically coupled particle-gas system. (Simulations of the streaming instability 
and gravitational collapse by Jake Simon \cite{simon16})}
\label{fig_streaming}
\end{figure}

The existence of a strongly inhomogenous distribution of solid particles has important 
implications for particle growth and planetesimal formation, irrespective of whether the 
over-densities are strong or very strong. There is particular interest, however, in 
determining whether the streaming instability can yield over-densities that 
exceed the Roche density (\S\ref{sec_roche}), given approximately by,
\begin{equation}
 \rho_p \sim \frac{M_*}{r^3}.
\end{equation}
Particle clumps whose density exceeds the  Roche density can collapse gravitationally 
into planetesimals, whose properties (such as size and binarity \cite{nesvorny10}) will 
depend upon the statistics of the particle density field generated by the streaming instability. 
Collapse is a likely outcome in regions of the disk where the streaming instability is 
strong \cite{johansen07b,johansen11}. Current simulation results suggest that 
collapse forms a population of planetesimals with an initial mass function that 
can be fit with a power-law \cite{johansen15,simon16,schafer17},
\begin{equation}
 \frac{{\rm d}N}{{\rm d} M_p} \propto M_p^{-1.6},
\end{equation} 
whose exponent displays at most a weak dependence on the stopping time of the 
particles that participate in the instability \cite{simon17}. This predicted mass 
function is top-heavy, in the sense that most of the mass resides in a handful 
of the largest objects (in contrast to the stellar IMF, which has the opposite 
property \cite{bastian10}). The role of intrinsic gas-phase disk turbulence (which 
is not included in most of the recent calculations) in modifying the predicted 
mass function has not, however, been fully established.

The circumstances that lead to the formation of streaming-unstable regions within 
protoplanetary disks, along with the outcome of the instability when it occurs, are 
by no means definitively established. Figure~\ref{fig_streaming} illustrates the 
flavor of theoretical models now under investigation \cite{johansen14}, which 
invoke the single particle processes discussed in this section 
as essential elements. Vertical settling and radial drift, operating on particles that 
have grown through collisions \cite{blum08} to be imperfectly coupled to the gas, act to enhance 
the local metallicity toward the values where the streaming instability would operate. 
Settling and pile-up, however, may not always be sufficient, and the next section 
is devoted to processes that can generate structure and additional enhancement 
in the local metallicity within the disk.

\begin{svgraybox}
\begin{itemize}
\item
Particles with sizes from microns up to centimeters are coupled to the disk gas by 
aerodynamic forces, operating in the Epstein regime (where the particle size is 
smaller than the molecular mean-free-path, leading to a linear dependence of the 
drag on relative velocity).
\item
Aerodynamic forces acting on particles lead to radial drift, because the gas in the 
disk feels radial pressure forces that are not experienced by the particles. Drift 
is directed toward pressure maxima. In a disk with a monotonically declining 
pressure profile drift is inward, extremely rapid, and can lead to a transient pile-up of 
solids in the inner disk regions.
\item
Vertical settling also occurs on a time scale that is short compared to the disk 
lifetime. It can be inhibited by even modest levels of fluid turbulence, and is 
usually accompanied by coagulation and fragmentation processes.
\item
When the feedback of a radially drifting particle fluid on the gas is accounted for, 
the resulting coupled system of gas and particles is often linearly unstable to 
the streaming instability. Given appropriate conditions this can lead to very 
strong clustering of the particles, followed by gravitational collapse.
\end{itemize}
\end{svgraybox}

\newpage

\section{Structure formation in protoplanetary disks}
\label{sec_structure_form}

Up until now we have largely assumed that the gas and dust in protoplanetary disks 
follow axisymmetric distributions, with smooth (but very probably different \cite{andrews12}) radial 
profiles. This is an approximation, which is known to fail spectacularly in some 
observed systems. As discussed in \S\ref{sec_observations} disks show a variety of 
structures:
\begin{itemize}
\item
Classification of a significant fraction of protoplanetary disks as {\em transitional disks} \cite{espaillat14}, 
based on evidence of inner cavities in (at least) the dust distribution.
\item
Radial structure in molecular emission linked to the presence of ice lines, for example of CO \cite{qi13}.
\item
Multiple rings of emission seen in high resolution mm / sub-mm observations of HL Tau \cite{alma15} 
and other systems.
\item
Pronounced non-axisymmetric (``horseshoe"-shaped) sub-mm emission in systems including 
IRS~48 \cite{vdm13}.
\item
Spiral arms and other non-axisymmetric structures seen in scattered light images of 
disks \cite{grady13}.
\end{itemize}
An open and important question is whether these structures are a consequence of --- 
or a precursor to --- planet formation. That question cannot yet be answered, but keeping 
it in mind we discuss here a number of 
processes that can lead to the formation of directly observable (and hence necessarily 
large-scale) structure in one or both of the gas and dust distributions within disks. 
Independent of the topical observational interest, any process that can generate 
inhomogeneity in the solid distribution is potentially important theoretically. In particular 
planetesimal formation could be made easier if there are processes that enhance the ratio of 
solids relative to the gas (in rings, vortices etc), creating conditions more favorable for 
both direct collisions and for the streaming instability.

\subsection{Ice lines}
The water snow line, together with the silicate sublimation front and various ice lines in the 
outer disk, are potentially critical locations for planet formation. Most often, this importance is 
quantified by noting that the equilibrium chemical composition of a Solar abundance gas has a 
substantially larger mass of condensible solids outside the snow line than inside (by about a 
factor of 4 in the classical Minimum Mass Solar Nebula \cite{hayashi81}, rather less than that 
using more modern calculations of the chemical equilibrium \cite{lodders03}). The likelihood that 
this leads to a jump in solid surface density at the snow line is then invoked as the reason why 
the Solar System has only terrestrial planets at smaller radii, and giants beyond.

These arguments are valid but incomplete. First, the equilibrium chemical composition is only 
linked directly to the solid surface density in the limit where the solid particles remain small and 
well-coupled to the gas. Particles that grow to be large enough that radial drift 
becomes significant will instead develop a surface density profile that is both different from that 
of the gas (\S\ref{sec_pileup}, \cite{youdin02}), and dependent on the size distribution. If icy 
particles are typically substantially larger than silicates (as is frequently suggested) their more rapid 
radial drift could lead to an instantaneously {\em lower} surface density of solids outside the snow 
line than inside. Second, although it is {\em possible} to construct models in which an assumed 
jump in planetesimal surface density at the snow line contributes to efficient core formation, 
the compositional effect is not the whole story. The greater area of planetary feeding zones at 
larger radii, along with more complex effects such as Type~I migration and pebble accretion, 
affect the outcome of planet formation at different radii to a similar extent. 
The most important role of ice lines may instead be as a preferential site for planetesimal 
formation, or perhaps as a location where Type~I migration stalls.

The pressure in the protoplanetary disk is substantially below that of the triple point of water, 
and hence the snow line marks a radial transition between ice and water vapor. Under 
mid-plane conditions, the corresponding temperature is typically $T = 150 - 180 \ {\rm K}$. 
Where this isotherm lies in the disk is a function of the stellar luminosity and 
accretion rate --- neither of which are constant over time --- and of the disk opacity 
which may change due to coagulation. Theoretical models \cite{lecar06,garaud07,min11} 
suggest that when $\dot{M} \approx 10^{-7} \ M_\odot \ {\rm yr}^{-1}$ $r_{\rm snow}$ is at $\approx 3 \ {\rm AU}$, 
before moving inward to within 1~AU as the accretion 
drops to $\dot{M} \approx 10^{-9} \ M_\odot \ {\rm yr}^{-1}$. At still lower 
$\dot{M}$ the inner disk becomes optically thin, and the resultant rise in temperature 
pushes the snow line back out to 2-3~AU.

The above estimates, calculated within the framework 
of relatively simple disk models that include viscous heating and irradiation, are significantly 
modified  \cite{martin12} if the true disk structure instead resembles Gammie's \cite{gammie96} layered model 
with a mid-plane dead zone. The calculation of snow line evolution may require further revision 
if winds, which potentially change both the surface density profile and the fraction of potential 
energy that goes into disk heating, are important on AU-scales.

\begin{figure}[t]
\center
\includegraphics[width=\columnwidth]{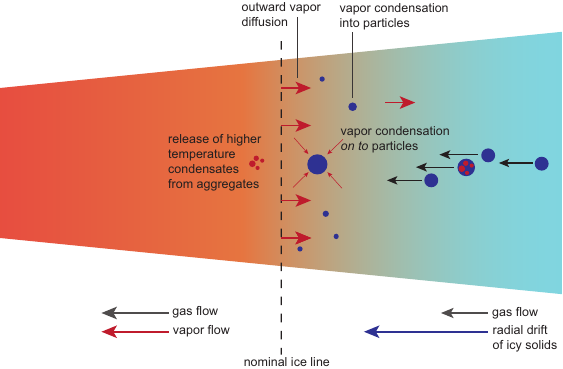}
\caption{Illustration of some of the physical processes occurring near ice lines in the 
protoplanetary disk. Icy materials drifting radially inward sublimate when they reach the 
ice line, releasing any higher temperature materials that were embedded into aggregates \cite{ciesla06}. 
The resulting vapor flows toward the star at the same speed as the rest of the gas, but 
also  diffuses outward down the steep gradient in concentration \cite{stevenson88}. It may 
then recondense, either into new particles or on to pre-existing particles \cite{ros13}. Some 
combination of these effects may feedback upon the gas physics via changes to either the 
opacity or, in models where MHD processes dominate angular momentum transport, 
the ionization state \cite{kretke07}.}
\label{fig_iceline_cartoon}
\end{figure}

Figure~\ref{fig_iceline_cartoon} illustrates some of the key physical processes occurring 
near ice lines. {\em Where} ice lines occur can be calculated by use of the 
Clausius-Clapeyron relation, which gives the saturated vapor pressure $P_{\rm eq}$ 
at temperature $T$ in terms of the latent heat $L$ of the phase transition,
\begin{equation}
 P_{\rm eq} = C_L e^{-L / {\cal R} T}.
\end{equation}
Here ${\cal R}$ is the gas constant, and $C_L$ is a constant that depends upon the 
species involved. For water, $L / {\cal R} = 6062 \ {\rm K}$ and 
$C_L  = 1.14 \times 10^{13} \ {\rm g \ cm^{-1} \ s^{-2}}$ \cite{ciesla06}. We can 
compare this pressure to the actual partial pressure of vapor in the disk. Using 
water (molecular weight $\mu_{\rm H_2 O} = 18$) as an example, if the surface 
density of vapor is $\Sigma_v$, the mid-plane pressure is,
\begin{equation}
 P_v = \frac{1}{\sqrt{2 \pi}} \frac{\Sigma_v}{h} \frac{k_B T}{\mu_{{\rm H_2 O}}m_H}.
\end{equation}  
If $P_v < P_{\rm eq}$ water ice will sublimate, whereas if $P_v > P_{\rm eq}$ vapor 
will condense into solid form. For small particles whose sizes are measured in mm 
or cm sublimation is rapid \cite{ciesla06}, and hence to a reasonable approximation 
sublimation and condensation processes act to maintain the vapor pressure close to the 
equilibrium value.

The large value of $L / {\cal R}$ implies that the snow line is a sharply defined transition 
within the disk. At 150~K, the characteristic temperature interval over which the 
equilibrium vapor pressure varies, $P_{\rm eq} / ( {\rm d}P_{\rm eq} / {\rm d}T$), is 
just a few Kelvin. If sublimation of icy particles that drift through the snow line is fast, 
this has the consequence of imposing a 
sharp radial gradient in water vapor concentration at the snow line which, in a 
turbulent disk, will in turn result in a diffusive outward flux of vapor (equation~\ref{eq_Cflux}). 
Re-condensation of the vapor can then lead to an enhancement of the solid surface density 
immediately outside the snow line.

\begin{figure}[t]
\center
\includegraphics[width=\columnwidth]{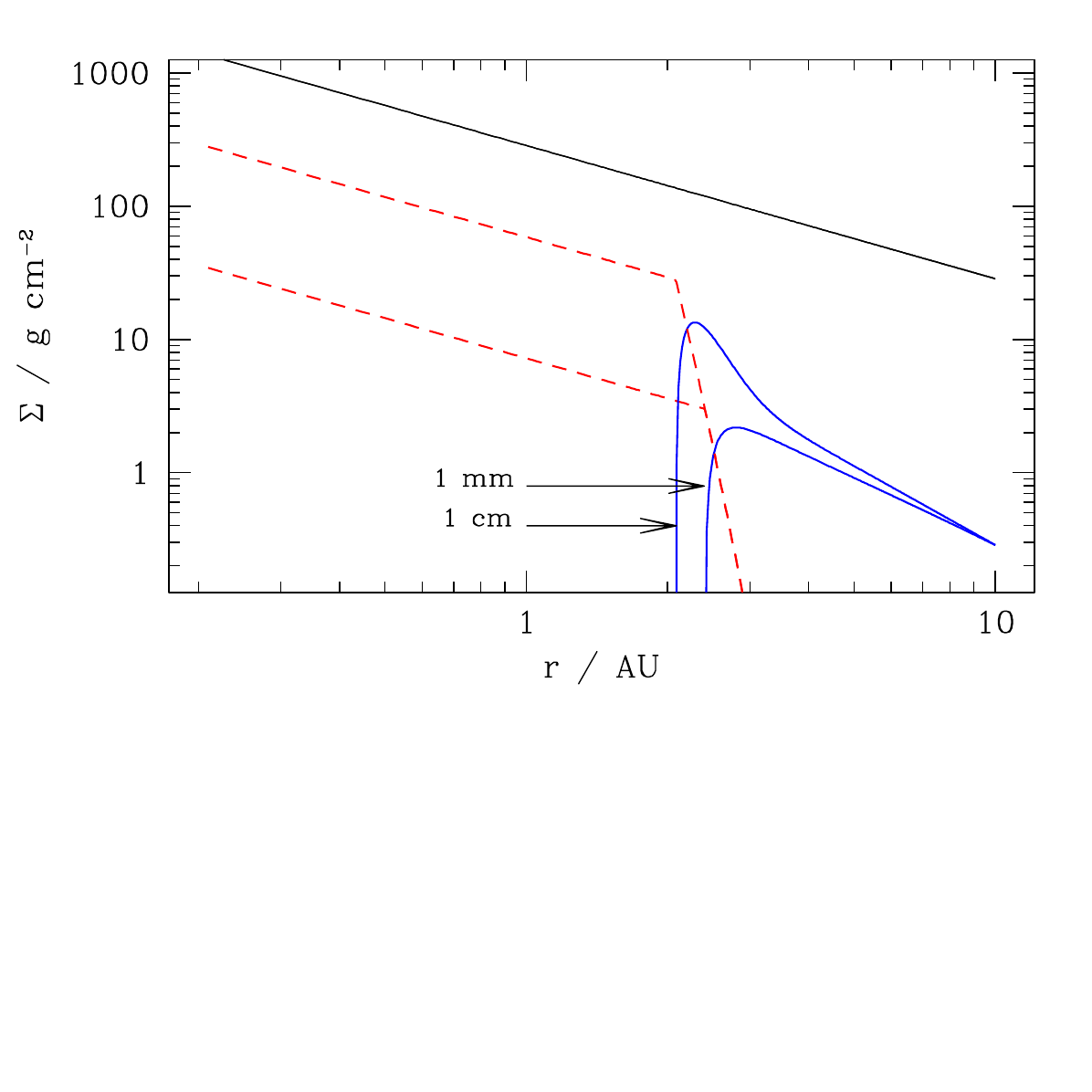}
\vspace{-4.5cm}
\caption{Example steady-state profiles of gas (upper solid line), radially drifting 
icy particles (solid blue lines) and water vapor (red dashed lines) in a turbulent 
protoplanetary disk. The assumed disk model has an accretion rate of 
$10^{-8} \ M_\odot \ {\rm yr}^{-1}$, a temperature $T = 150 (r/{3 \ {\rm AU}})^{-1/2}$, 
and an $\alpha$ parameter of $5 \times 10^{-3}$. The ratio of the turbulent diffusivity 
to the turbulent viscosity is taken to be unity, and the concentration of icy solids is 
set (arbitrarily) to be $10^{-2}$ at 10~AU. The rapid radial drift of cm-sized particles 
leads to a high concentration of water vapor in the inner disk. Outward diffusion and 
re-condensation of this vapor --- assumed here to form particles of a single fixed size --- 
leads to an enhancement of solids just outside the snow line \cite{stevenson88}. Note also 
the more elementary conclusion that the vapor concentration in the inner disk is a direct 
probe of the mass flux of radially drifting solids encountering the snow line.}
\label{fig_iceline_calc}
\end{figure}

Stevenson \& Lunine \cite{stevenson88} proposed that a vapor diffusion / condensation 
cycle of this type could enhance the surface density of ice enough to promote the rapid 
formation of Jupiter's core. The strength of the effect depends upon the size of the icy 
solids drifting in toward the snow line, and upon what is assumed about processes 
including disk turbulence, condensation and coagulation / fragmentation. Figure~\ref{fig_iceline_calc} 
shows the results of a particularly simple calculation (\cite{armitage16b} after \cite{clarke88,ciesla06}), which 
assumes that vapor condenses (or, condenses and rapidly coagulates) into solid particles 
of a single size that {\em matches} the size of icy solids drifting in from larger radii. For the 
adopted disk model, mm-sized icy solids have relatively low drift velocities, and these particles 
sublimate into vapor without generating any local enhancement in the surface density of solids. 
Larger cm-sized particles, conversely, are enhanced by a factor of several outside the snow line 
as a consequence of the diffusive transport of vapor followed by condensation. Surface density 
enhancements of this magnitude could be important, particularly in models where planetesimal 
formation depends upon the disk locally exceeding a threshold value of  
metallicity (as is the case for the streaming instability, see \S\ref{sec_streaming}, \cite{johansen09}).

Ros \& Johansen \cite{ros13} investigated a related possibility. Instead of assuming that vapor 
condenses into new particles (or, on to very small grains released when aggregates break 
up at the snow line), they modeled the growth of pre-existing solids as vapor condenses on to 
their surfaces. A simple collisional argument gives the growth rate due to 
vapor condensation / sublimation as,
\begin{equation}
 \frac{{\rm d}m}{{\rm d}t} = 4 \pi s^2 v_{\rm th} \rho_v \left( 1 - \frac{P_{\rm eq}}{P_v} \right),
\end{equation}
for a particle of mass $m$ and radius $s$, surrounded by vapor of density $\rho_v$ and 
thermal speed $v_{\rm th}$. (As noted by Supulver \& Lin \cite{supulver00}, this is not 
an exact expression \cite{haynes92}.) Using a Monte Carlo approach, Ros \& Johansen \cite{ros13} 
found that condensation on to particle surfaces could provide an efficient growth mechanism 
up to sizes of the order of 10~cm. This could aid planetesimal formation by boosting the 
stopping time of particles into the range preferred by the streaming instability \cite{carrera15}. 
Moreover, by removing mass from the directly observable mm-size regime, condensation-driven 
growth could suppress the mm and sub-mm flux from disks in the vicinity of ice lines \cite{zhang15}. 

The aforementioned physics affects only trace components of the disk --- the icy solids and 
the resulting vapor. It is easy, however, to contemplate feedback processes that couple the 
evolution of solids at ice lines to the bulk of the gas disk. At a minimum the opacity will 
vary depending upon the radial distribution of solids. Beyond that, we have already noted 
that the efficiency of angular momentum transport (in MHD models) is expected to be a 
function of the local ionization state, and that the ionization balance is affected by the 
abundance of small grains. Kretke \& Lin \cite{kretke07} suggested that the enhanced 
abundance of solids near the snow line would act to suppress the rate of angular momentum 
transport, and that this could lead to the formation of a local pressure maximum that 
would act to trap particles (producing, in principle, a positive feedback loop). This argument 
(which has been invoked in some models of collisional growth \cite{brauer08}) is highly plausible, 
though the breadth and complexity of the physics involved makes quantitative 
investigation challenging. 

Although we have focused here on the snow line, analogous considerations carry over to other 
ice lines. The CO ice line corresponds to a temperature of $T = 17-19 \ {\rm K}$ \cite{bisschop06}, 
and is somewhat more complicated to model because the CO is typically mixed with N$_2$ and 
water ices. The silicate ``ice line" (or sublimation front) could also be important, since it lies 
at radii where a high fraction of stars are observed to host short-period planetary systems.

\subsection{Particle traps}
In a disk with a monotonically declining pressure profile radial drift is always inward. More 
generally, however, aerodynamic drift is directed toward pressure maxima, and can be outward 
if there is a local pressure maximum within the disk. This possibility was recognized in a 
prescient paper by Whipple \cite{whipple72}, who appealed to it as part of a model for the 
formation of comets\footnote{Quoting from his paper, ``should it be possible for a toroid of higher 
density to occur in the Solar nebula, the growing planetesimals would be drawn toward it 
from the inside as well as from the outside...".}. The tendency for solids to be aerodynamically 
enhanced in the vicinity of pressure maxima is often described as particle ``trapping", though 
this term is somewhat misleading; in a turbulent disk small particles are at most temporarily 
detained by pressure maxima rather than being permanently trapped.

The effect of local pressure maxima on the radial distribution of solids can be derived, in 
the limit where the solids remain a trace contaminant, following the methods described in 
\S\ref{sec_particle_v} and \S\ref{sec_particle_diffuse}. For an axisymmetric disk with an 
arbitrary mid-plane pressure profile, the radial velocity $v_r$ of particles under the action 
of aerodynamic forces remains as given by equation~(\ref{eq_radial_drift}), with the 
parameter $\eta$ describing the deviation from Keplerian velocity becoming \cite{takeuchi02},
\begin{equation}
 \eta = - \left( \frac{h}{r} \right)^2 
 \left[ \frac{{\rm d}\ln \Sigma}{{\rm d} \ln r} + (q-3) \right].
\end{equation} 
In this formula $q$ is defined as the local power-law index describing the flaring of the disk,
\begin{equation}
 \frac{h}{r} \propto r^{q-1},
\end{equation}
such that a non-flaring disk has $q=1$. In axisymmetry and steady-state, equations~(\ref{eq_Ccontinuity}) 
and (\ref{eq_Cflux}) can be immediately integrated to give,
\begin{equation}
 r ( F_{\rm diff} + F_{\rm adv} ) = k,
\end{equation}
where the diffusive and advective fluxes are,
\begin{eqnarray}
 F_{\rm diff} & = & - D \Sigma \frac{{\rm d}}{{\rm d}r} \left( \frac{\Sigma_d}{\Sigma} \right), \nonumber \\
 F_{\rm adv} & = & \Sigma_d v_r,
\end{eqnarray}
and the constant $k$ is just the radial flux of solid material. Written out explicitly, the concentration 
of particles $f \equiv \Sigma_d / \Sigma$ obeys a first-order differential equation,
\begin{equation}
 \frac{{\rm d}f}{{\rm d}r} - \frac{v_r}{D} f = \frac{k}{D\Sigma} \frac{1}{r}.
\label{eq_ODE_C} 
\end{equation} 
Analytic solutions to this equation are possible for simple choices of $v_r$, $D$ and $\Sigma$. 
A straightforward quadrature gives the solution for the concentration profile given more realistic 
choices of these functions.

\begin{figure}[t]
\center
\includegraphics[width=0.9\columnwidth]{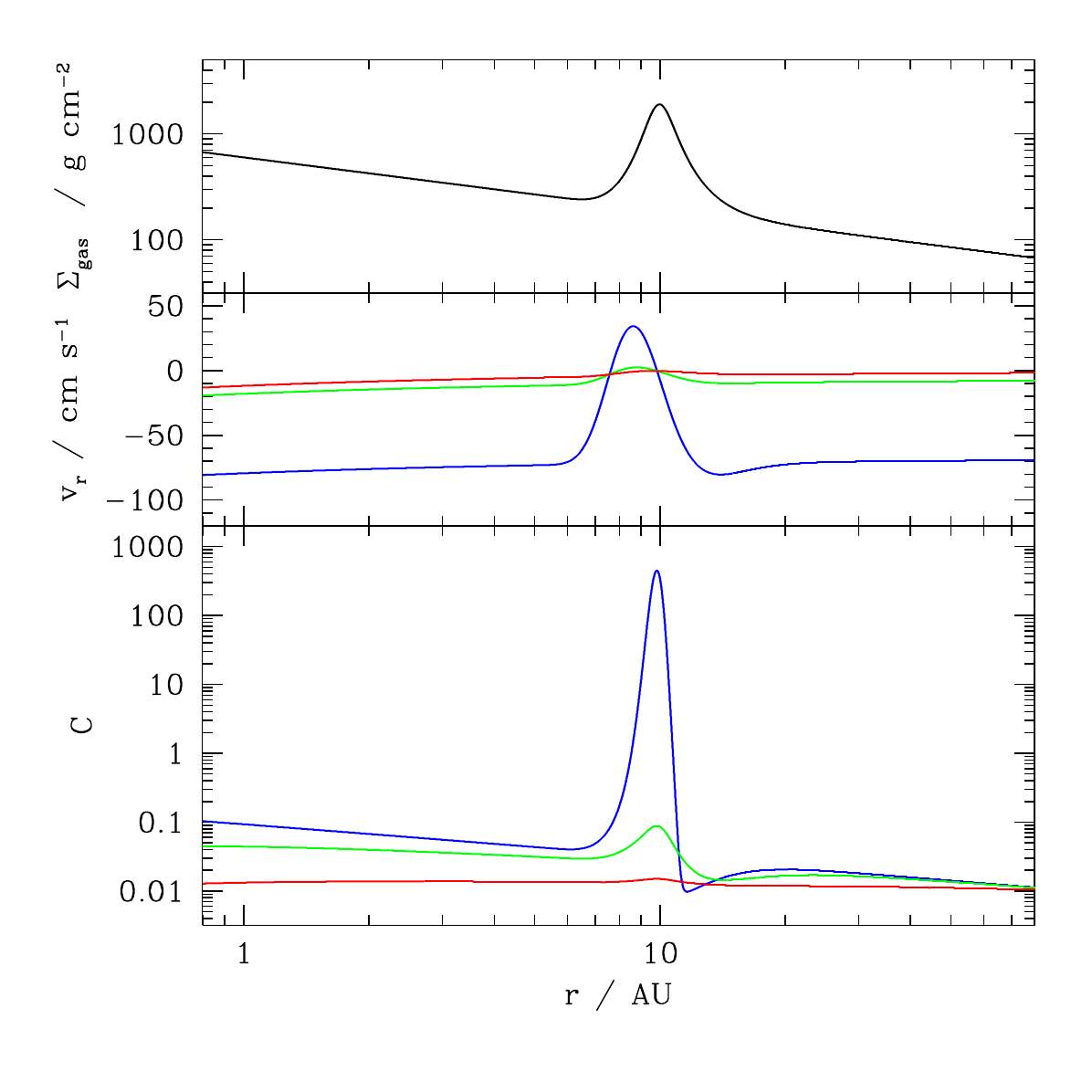}
\vspace{-1.0cm}
\caption{The steady-state radial distribution of solids in a turbulent disk with an axisymmetric 
local pressure maximum (a particle ``trap"). Upper panel: the surface density of the gas. 
Middle panel: the radial velocity of 0.1~mm (red), 1~mm (green) and 1~cm (blue) particles. 
The smallest particles have a radial velocity that is almost indistinguishable from that of the 
gas, while the larger particles experience rapid radial drift that can be outward near the 
location of the pressure maximum. Lower panel: the concentration $C = \Sigma_d / \Sigma_{\rm gas}$, 
normalized to an arbitrary value of $10^{-2}$ at 100~AU. The assumed disk model for this calculation 
has $\dot{M} = 10^{-8} \ M_\odot \ {\rm yr}^{-1}$, $M_* = M_\odot$, $h/r = 0.5$, $\alpha = 10^{-3}$ and $D = \nu$ (c.f. 
equation~\ref{eq_concentration}). The trap is modeled as a gaussian-shaped reduction in $\alpha$ 
to a minimum of $10^{-4}$, with a width of $4h$. The particles are assumed to be spherical, with a 
material density of $1 \ {\rm g \ cm}^{-3}$, and to follow the Epstein drag law.}
\label{fig_trap}
\end{figure}

Figure~\ref{fig_trap} shows an illustrative numerical solution to equation~(\ref{eq_ODE_C}). For this 
example, we have modeled a trap by assuming (arbitrarily) that the viscosity in the gas disk is 
reduced across a moderately narrow annulus. The lower viscosity leads to a higher surface density, 
producing a pressure maximum which in turn concentrates particles. The concentration effect is 
strongly size-dependent. Small particles (in this example those with radii of 0.1~mm and 1~mm) 
have a radial velocity that is similar to that of the gas, and are largely unaffected by the pressure 
maximum. Larger cm-sized particles, on the other hand, have a larger magnitude of radial 
drift, and can be strongly concentrated at the location of the pressure maximum. The 
enhancement in the local particle density can reach several orders of magnitude, depending 
both on the particle size and on the ``strength" (radial width and amplitude) of the pressure 
maximum within the gas disk. 

It should be noted that in a turbulent disk (and ignoring 
particle feedback on the gas) it is {\em always} possible to find a steady-state solution 
for the radial particle concentration in which the particle mass flux is a constant at all 
radii. Physically, this is because particles accumulate near pressure maxima until 
the radial gradient in concentration becomes large enough for turbulent diffusion to 
allow them to leak out \cite{ward09,zhu12}. Pressure maxima only act as a ``filter" \cite{rice06}, 
{\em permanently} removing large particles from the inward radial drift flow, when additional 
physical effects are included \cite{zhu12}. If the local concentration of large particles becomes 
large enough, for example, planetesimal formation \cite{haghighipour03} could cause some fraction of the solid 
material to drop out of the radial flow.

A number of physical effects have been identified that could lead to the formation of 
local pressure maxima within protoplanetary disks. These include the exterior edges 
of planet-carved gaps \cite{paardekooper04,rice06,zhu12}, the outer edges of cavities 
created by photoevaporation (\S\ref{sec_photoevaporation}, \cite{alexander07}), and a 
photoelectric heating instability that may operate in gas-poor systems (primarily debris 
disks) \cite{besla07,lyra13}. Spiral arms in self-gravitating disks \cite{rice04,gibbons14} 
and the inner edges of dead zones \cite{lyra09} can also concentrate solids via 
closely related physical processes, though these environments typically involve 
significant non-axisymmetry. With the exception of self-gravity, these possible locations for 
pressure maxima either form at specific places within the disk (e.g. at the inner dead 
zone edge, defined by a characteristic mid-plane temperature), or at a time after when 
we expect planets to have formed (during photoevaporative disk clearing, or in 
{\em response} to pre-existing massive planets). It is possible, however, for the 
turbulence within the gas disk to be generically unstable to the formation of 
pressure maxima within zonal flows. This would be interesting because it would 
imply the (possibly transient) existence of multiple particle traps within the disk, 
that could play a role in early-stage planet formation \cite{pinilla12}.

\subsection{Zonal flows}
It is evident from equation~(\ref{eq_force_balance}) that an equilibrium can be set up 
in which radial forces from a complex pressure profile (that may include local pressure maxima) 
are balanced by radial variations in azimuthal velocity. In a local description (equation~\ref{eq_sheet_mhd}, 
but here ignoring magnetic fields) the balance is between the pressure gradient term $\nabla P / \rho$ and 
that describing the Coriolis force $2 \Omega_0 \times {\bf v}$. When these terms balance, such that 
\begin{equation}
 \frac{{\rm d}P}{{\rm d}x} = 2 \Omega_0 \rho v_y,
\end{equation}
the system is said to be in {\em geostrophic balance} (here $x$ is the radial direction, and $y$ the 
azimuthal). The pressure gradient is compensated by variations in the orbital velocity, creating 
{\em zonal flows} analogous to the banded structure of winds in giant planet atmospheres.

A disk zonal flow is an equilibrium solution to the fluid equations, but that equilibrium may not be 
stable. Too pronounced a deviation from Keplerian rotation results in a shear profile that is 
unstable to Rossby wave instability \cite{lovelace99}. We will discuss this instability, which is 
similar to Kelvin-Helmholtz instability, in \S\ref{sec_rossby}. Even if the rotation profile is 
stable, the diffusive nature of a classical viscosity (equation~\ref{eq_disk_evolve_general}) 
would tend to erase any small-scale perturbations in the pressure that are sourced from 
surface density fluctuations. Persistent zonal flows are thus not expected in classical disk 
theory. They have been observed, however, in local numerical simulations of MHD 
turbulent disks \cite{johansen09b}. The key to their formation appears to be the ability 
of MHD disk turbulence to generate large-scale structure in the magnetic field \cite{simon12}, 
which could be viewed as an inverse cascade of turbulent power. The details of how and when 
zonal flows form are not entirely clear, though Johansen et al. \cite{johansen09b} 
describe a simplified dynamical model in which large-scale variations in the Maxwell stress lead 
first to azimuthal velocity perturbations and thence to axisymmetric structure in the pressure and 
density.

The lifetime and radial scale of zonal flows in protoplanetary disks depend 
upon the same factors that determine the properties of MHD disk turbulence more 
generally (\S\ref{sec_transport}), namely the strength of non-ideal terms in the 
induction equation and the presence of net vertical magnetic field. In ideal MHD, 
lifetimes of tens of orbital periods and radial scales of the order of $10 \ h$ appear 
to be typical \cite{johansen09b,simon12}, though these results require a double dose 
of caveats --- first because the inferred scales are not much smaller than the size 
of the local simulation domains used, and second because they are large enough that 
curvature terms neglected in local models may be important. Nonetheless, the 
amplitude, scale and lifetime of zonal flows under ideal MHD conditions plausibly 
lead to strong local enhancements in the dust to gas ratio for particles with 
stopping times $\tau_s \sim 0.1 - 1$ \cite{dittrich13}. The presence of net 
vertical fields substantially enhances the amplitude of zonal flows \cite{bai14b}.

Zonal flows are also found in MHD disk simulations that include ambipolar diffusion, 
and can be comparable in amplitude to the ideal MHD case if the net field is 
sufficient to stimulate a significant $\alpha \sim 10^{-2}$ \cite{simon14}. However, 
both inferences from local simulations \cite{simon14}, and explicit tracking of 
particles in global simulations \cite{zhu15}, suggest that zonal flows in the outer 
regions of protoplanetary disks, where ambipolar diffusion dominates, have 
properties close to the boundary beyond which strong particle enhancement would 
be expected. 

In summary, zonal flows are likely to be present in the inner disk, where ideal MHD 
is a good approximation, though these flows would only strongly influence the dynamics 
of relatively large solid particles. In the outer disk, where ambipolar diffusion is 
important and even mm-sized particles have significant stopping times, zonal flows 
could introduce observable large-scale axisymmetric structure and may contribute 
to particle concentration. In the Hall-dominated regime that prevails around 1~AU 
theoretical expectations are less clear. Extremely strong zonal structures were 
observed in local vertically unstratified Hall-MHD simulations \cite{kunz13}, 
whereas comparable stratified runs instead led to large-scale Maxwell stress \cite{lesur14}. 
It is therefore unclear whether there are circumstances in which zonal flows on 
AU-scales could contribute to particle concentration and planetesimal formation.

\subsection{Vortices}
Few issues in planet formation are as long-debated as the possible role of vortices. Very 
general arguments suggest that large-scale vortices could be present in protoplanetary 
disks and play an important role in planetesimal formation. We note first that disks are 
(approximately) two dimensional fluid systems, and in contrast to three dimensional systems 
they therefore support an inverse cascade of turbulent energy toward large scales \cite{kraichnan67}.
Second, for a barotropic disk (i.e. $P=P(\rho)$ only) the vortensity ${\bf \omega} / \rho$ is 
conserved (equation~\ref{eq_vortensity}). Taken together, these properties imply that 
vortices within disks have the potential to form persistent long-lived structures. Indeed, 
simulations of strictly two dimensional flows show that disks seeded with small-scale 
vorticity perturbations evolve to form a small number of large and persistent anticyclonic 
vortices \cite{godon00,johnson05}. Anticyclonic vortices are high pressure regions 
that attract marginally coupled solids \cite{barge95,tanga96}, potentially catalyzing the 
subsequent formation of planetesimals. Even absent planetesimal formation, the natural 
tendency of vortices to form large-scale non-axisymmetric dust features makes it 
tempting to identify them with observed disk asymmetries \cite{vdm13}.

The basic properties of disk vortices are well-established. What is 
much trickier is to determine (1) whether vortices form spontaneously in disks or only 
{\em after} planet formation (for the reasons already mentioned, spontaneous vortex 
formation generally requires non-barotropic processes), and (2) whether three dimensional 
instabilities and / or particle feedback are fatal impediments to their survival. Observations 
as well as theory are probably needed to resolve these issues.

\subsubsection{The Kida solution}
The magnitude of the vorticity in a strictly Keplerian disk is $\omega_K = -(3/2) \Omega_K$. 
A vortex can be modeled as a spatially localized elliptical perturbation, within which the vorticity 
$\omega = \omega_K + \omega_v$, with $\omega_v$ a constant. Other types of vortex 
are possible, but rather remarkably this type can be described by an exact non-linear 
solution \cite{kida81} that is useful for both analytic and numerical studies.

The Kida solution \cite{kida81} describes a vortex within a shearing-sheet approximation to 
disk flow. Following Lesur \& Papaloizou \cite{lesur09} we define a cartesian co-ordinate 
system $(x,y)$ that is centered at radius $r_0$ and which co-rotates with the background 
disk flow at angular velocity $\Omega_K = \Omega_K (r_0)$,
\begin{eqnarray}
 x & = & r_0 \phi, \nonumber \\
 y & = & - (r - r_0).
\end{eqnarray} 
Kida considered time-dependent vortex solutions, but here we will worry only about 
vortices that are steady\footnote{For a derivation of the steady Kida solution, see e.g. 
the appendix of Chavanis \cite{chavanis00}.}. Time-independent solutions are possible if the semi-major 
axis of the vortex is aligned with the azimuthal direction ($x$ in the shearing sheet 
model), and the vorticity perturbation satisfies,
\begin{equation}
 \frac{\omega_v}{\omega_K} = \frac{1}{\chi} \left( \frac{\chi + 1}{\chi - 1} \right).
\end{equation}
Here $\chi = a/b$ is the aspect ratio of the vortex, which forms an elliptical patch 
with semi-major axis $a$ and semi-minor axis $b$. The right-hand-side is 
evidently positive, which implies that the only steady Kida vortices in Keplerian 
disks are anticyclonic (with $\omega_v$ having the opposite sign to $\Omega_K$). 

The complete Kida solution is written in terms of a streamfunction $\psi$ in an 
elliptic co-ordinate systems $(\mu,\nu)$, where,
\begin{eqnarray} 
 x & = & f \cosh (\mu) \cos (\nu), \nonumber \\
 y & = & f \sinh (\mu) \sin (\nu), 
\end{eqnarray}
and $f = a \sqrt{ (\chi^2 -1)/\chi^2}$. The solution can be split into a core and an 
exterior part,
\begin{eqnarray}
 \psi_{\rm core} =  -\frac{3 \Omega_K f^2}{4 (\chi -1)} 
   \left[ \chi^{-1} \cosh^2 (\mu) \cos^2 (\nu) + \chi \sinh^2 (\mu) \sin^2 (\nu) \right], \nonumber \\
 \psi_{\rm ext}  =  -\frac{3 \Omega_K f^2}{8 (\chi -1)^2} 
   \left[ 1 + 2 (\mu - \mu_0) + 2 (\chi-1)^2 \sinh^2 (\mu) \sin^2 (\nu) \right. \nonumber \\
   \left.  +  \frac{\chi-1}{\chi+1} \exp[ -2 (\mu-\mu_0)] \cos(2 \nu) \right],
\end{eqnarray}   
which match at $\mu = \mu_0 = \tanh^{-1} (\chi^{-1})$. The cartesian velocity 
field is then given by $v_x = - \partial \psi / \partial y$, $v_y = \partial \psi / \partial x$. 
In general the cartesian representation of the velocity has no simple form, but 
within the core it is,
\begin{eqnarray} 
  v_{x,{\rm core}} & = & \frac{3 \Omega_K \chi}{2 (\chi-1)} y, \nonumber \\
  v_{y,{\rm core}} & = & - \frac{3 \Omega_K}{2 \chi (\chi -1)} x,
\end{eqnarray}  
describing simple elliptical motion.
Figure~\ref{fig_vortex1} shows the contours (logarithmically spaced) of the 
full streamfunction for Kida vortices of varying aspect ratio.

\begin{figure}[t]
\center
\includegraphics[width=0.9\columnwidth]{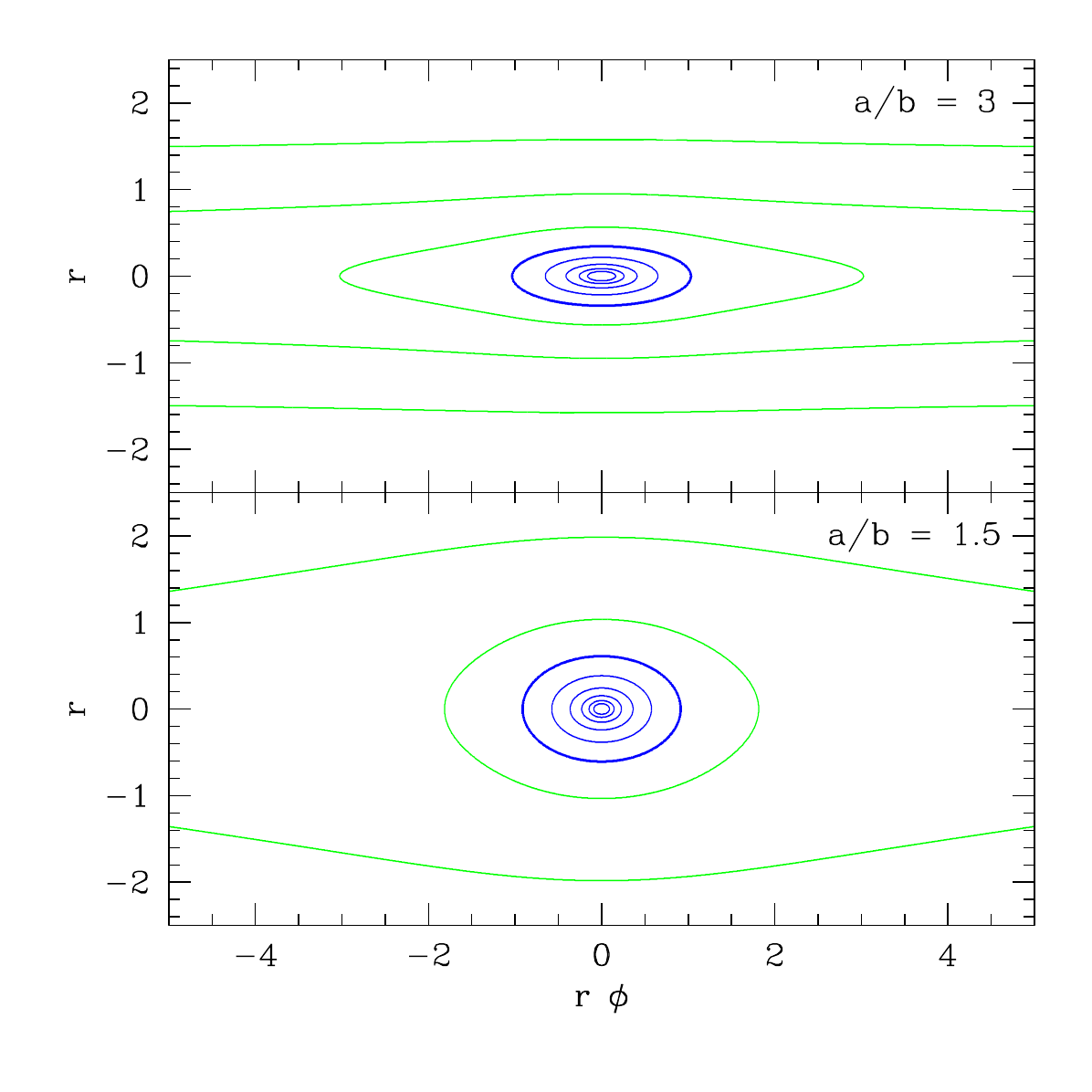}
\vspace{-0.7cm}
\caption{Contours of the Kida vortex streamfunction $\psi(x,y)$ are shown for different 
values of the vortex aspect ratio $\chi \equiv a/b$. Within the vortex core, delineated by the 
bold contour, the streamlines defined by the solution are elliptical, with fixed aspect ratio. 
Outside the core, the vortex merges smoothly into the background shear flow of the disk. }
\label{fig_vortex1}
\end{figure}

Barge \& Sommeria \cite{barge95} studied the trajectories of aerodynamically coupled 
solids that encounter vortices (using a different and approximate vortex model). The 
high pressure within anticyclonic vortex cores acts as an attractor for solids. The 
cross-section for capture is maximized for particles with dimensionless stopping 
time $\tau_s = 1$ \cite{chavanis00}. Since vortices can potentially grow to have 
radial extents $\Delta r \approx h$ (the supersonic velocity perturbations of larger 
vortices would radiate sound waves sapping their energy) a single large vortex 
can potentially trap a substantial mass of solids flowing radially toward it due 
to ordinary radial drift.

\subsubsection{Stability of vortices}
Although the geometry of protoplanetary disks is approximately two dimensional, the 
fact that disk vortices are limited in radial size to $\Delta r \approx h$ means that 
they are three dimensional objects that notice the vertical stratification. Barranco \& 
Marcus \cite{barranco05} and Shen et al. \cite{shen06}, using three dimensional 
simulations, found that mid-plane vortices that would be highly stable in 2D  
are rapidly destroyed by three dimensional instabilities. The origin of these 
observed instabilities, at least in part, appears to be the {\em elliptical instability} 
\cite{kerswell02}, which occurs whenever there is a resonance between the 
vortex rotation period and inertial waves within the disk. In a disk environment, 
Lesur \& Papaloizou \cite{lesur09} find that purely gaseous vortices are unstable 
for almost all choices of the vortex aspect ratio and degree of vertical stratification, 
though these parameters strongly affect the linear growth time of the instability. 
The instability that afflicts 3D vortices, however, is typically slow-growing and of 
small radial scale. Numerically this makes studies of vortex survival particularly 
challenging. Physically it means that the questions of vortex formation and vortex 
survival are closely linked, except perhaps at very large radii (where ``primordial 
vortices" might persist for interesting periods of time) the vortex population in a 
disk is expected to reflect an equilibrium between formation and destruction processes.

A significant loading of solids will also impact vortex longevity, generally for the worse. 
Railton \& Papaloizou \cite{railton14} studied the stability of generalized Kida vortices, 
containing both gas and dust in the limit of strongly aerodynamically coupled particles. 
They found that these configurations were vulnerable to parametric instabilities in the 
same way as gas-only vortices. This is consistent with a wide range of other analytic 
and numerical work \cite{chang10,inaba06,fu14b}, which suggests that dust to gas 
ratios in the range between 0.1 and 1 are sufficient to imperil the survival of disk 
vortices. This limitation may not, however, preclude vortices playing a role in 
planet formation. If we assume that there is a continual source of vorticity within the 
disk, then vortex particle concentration up to $\rho_p \sim \rho_g$ could be 
sufficient to initiate planetesimal formation via a variation of the streaming instability 
\cite{raettig15}.

Observations have identified a number of systems (primarily transition disks) that show a 
non-axisymmetric distribution of sub-mm emission \cite{casassus13,isella13,vdm13}, consistent with 
that expected if aerodynamically coupled solids are accumulating in a vortex \cite{zhu14}. The putative 
vortices in these examples may all be {\em caused} by planets. It would 
be interesting to ask whether useful constraints on vortices could be derived from observations of the 
{\em most-axisymmetric} disks, where there is no independent suspicion that planets already exist.

\subsection{Rossby wave instability}
\label{sec_rossby}
The Rossby wave instability (RWI) \cite{lovelace99,li00,li01} is a well-characterized mechanism for 
producing vortices within protoplanetary disks. The RWI is a linear instability that grows whenever 
there is ``sufficiently sharp" radial structure in the disk. Specifically, for a two dimensional disk model 
with angular velocity $\Omega(r)$, vertically integrated pressure $P(r)$, and adiabatic index $\gamma$, 
we define the entropy $S$ and epicyclic frequency $\kappa$ via,
\begin{eqnarray}
 S & = & \frac{P}{\Sigma^\gamma}, \nonumber \\
 \kappa^2 & = & \frac{1}{r^3} \frac{\partial}{\partial r} (r^4 \Omega^2).
\end{eqnarray}
In terms of these quantities, the stability of the disk to the RWI is determined by the radial 
profile of a generalized potential vorticity,
\begin{equation}
 {\cal{L}} (r) = \frac{\kappa^2}{2 \Omega \Sigma} \times S^{-2/\gamma}.
\label{eq_rwi} 
\end{equation} 
A necessary condition for RWI is that $\cal{L}$ have an extremum. A precise sufficient 
condition is not known, but variations in the potential vorticity of the 
order of 10\% over radial scales $\approx h$ appear to be enough to trigger instability, 
which leads to the formation of anticyclonic vortices on time scales that can be rapid --- 
of the order of $10 \Omega^{-1}$ \cite{li00}. The instability, which can be understood 
in terms of the local trapping of waves in the vicinity of the vortensity perturbation 
\cite{goldreich86,umurhan10}, has similarities to the Papaloizou-Pringle \cite{papaloizou84} instability 
of accretion tori. The RWI is essentially a two dimensional instability \cite{meheut10,lin13}, 
though the vortices that it forms are as vulnerable to unrelated three dimensional instabilities 
as any others.

We have already remarked that the edges of dead zones (and ice lines \cite{kretke07}) are 
places where local pressure maxima may form. The RWI criterion (equation~\ref{eq_rwi}) is not 
necessarily satisfied at every local pressure maximum, but it is nonetheless true that dead 
zone edges are plausible locations where the RWI may occur. Hydrodynamic models \cite{varniere06,inaba06,lyra08},  
and MHD simulations that include Ohmic diffusion as a simple model of dead 
zones \cite{lyra12,faure15,lyra15,flock15}, support this expectation, and show that the radial disk structure 
introduced by dead zones is likely to be unstable to the RWI and subsequent 
formation of vortices. The vortices in turn act as sites of particle concentration \cite{lyra08}. Depending 
upon the radii involved, a single vortex generated by the RWI may have a lifetime that is very 
short as compared to the disk lifetime (this will be particularly true at the inner edge of the dead zone). 
However, in this scenario fresh generations of vortices may be expected to form as long as the 
non-ideal disk physics that sustains the RWI-unstable dead zone structure persists.

\begin{figure}[t]
\center
\includegraphics[width=\columnwidth]{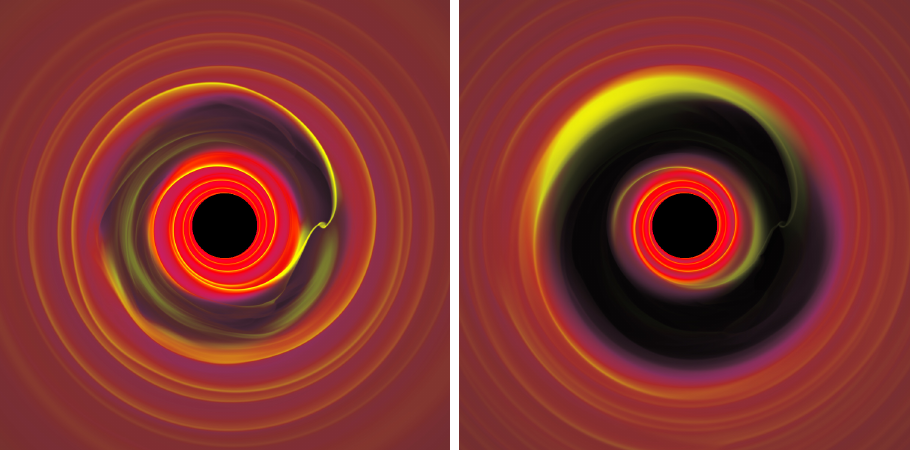}
\caption{Snapshots showing the hydrodynamic evolution of an almost inviscid disk containing a 
massive planet \cite{armitage13}. The planet rapidly clears an annular gap within the disk, whose 
edge is unstable to the generation of vortices. The system then evolves through a phase when the 
outer gap edge hosts a single large vortex, which can be an efficient trap for solid particles. The 
disk has $\alpha = 10^{-4}$ and $h/r = 0.05$ at the location of the planet, which has 
a mass ratio to the star of $5 \times 10^{-3}$.}
\label{fig_rwi}
\end{figure}

A second location where RWI may occur is at the edge of a gap created dynamically by 
a massive planet \cite{koller03,devalborro07}. Figure~\ref{fig_rwi} shows the evolution of a 
disk in a two-dimensional simulation of this scenario. The formation of a massive planet creates 
an approximately axisymmetric gap within the disk, whose edges can be unstable to the RWI. 
The vortices that are generated fairly rapidly merge, creating a single large vortex at either 
edge of the gap then can subsequently trap particles. This process works best if the disk 
in the vicinity of the planet's orbit has a low viscosity (roughly $\alpha \sim 10^{-3}$ or lower), 
and is at least partially a transient effect --- the vortices form during the phase when the 
planet accretes its gaseous envelope. For these reasons, observable structure from planet-initiated 
vortices is likely to be easiest to see in the outer disk, where ambipolar diffusion damps 
turbulence \cite{simon13} and the absolute lifetime of a single generation of vortices against 
disruptive instabilities is long. The particles trapped efficiently in the outer disk 
include those with $s \sim {\rm mm}$ that can be seen in sub-mm 
observations of protoplanetary disks \cite{zhu14b,zhu14}.

\begin{svgraybox}
\begin{itemize}
\item Condensation and sublimation processes are expected to modify the surface 
density and size distribution of particles in annuli that are adjacent to ice lines (of 
water, silicates, and possibly species such as CO with cooler condensation temperatures).
\item
Pressure maxima, whether caused by planets or by intrinsic disk processes such as 
zonal flows, lead to local axisymmetric enhancements in the particle surface density.
\item
Vortices can also trap particles. A population of vortices in a disk tend to merge to form 
a small number of large vortices, which unless replenished have a finite lifetime.
\item
Sharp features in the radial disk structure can trigger the Rossy Wave Instability, 
which generates vortices.
\end{itemize}
\end{svgraybox}

\newpage

\section{Disk dispersal}
\label{sec_dispersal}
There is no leading order mystery as to where the gas in protoplanetary disks goes to. 
If we divide the mean disk mass estimated from sub-mm studies in Taurus 
($M_d \sim 5 \times 10^{-3} \ M_\odot$ \cite{andrews05}) by the median accretion 
rate ($\dot{M} \sim 10^{-8} \ M_\odot \ {\rm yr}^{-1}$ \cite{gullbring98}) estimated in the 
same region, we obtain a characteristic evolution timescale of 0.5~Myr. The lifetime of 
detectable disks might be expected to be a small multiple of this timescale --- say a 
few Myr --- which is indeed what is observed \cite{hernandez07}.

The above exercise establishes that, from a purely observational perspective, most or 
all of the gas in protoplanetary disks {\em could} be lost via accretion on to the star. 
Beyond that, it proves nothing. Disk mass estimates and measurements of 
stellar accretion rates are subject to uncertainties, which when combined probably 
allow us to shift the inferred characteristic lifetime by an order of magnitude in 
either direction. In particular, if disk masses are systematically {\em under-estimated} 
from sub-mm continuum observations (because of particle growth to sizes too 
large to contribute to the sub-mm opacity) the characteristic lifetime could 
approach or exceed the observed one.

The next order observational diagnostic probes the time-dependence of the decay 
of disk signatures. Here a puzzle does emerge. The classical disk evolution equation~(\ref{eq_disk_evolve}) 
admits a self-similar solution \cite{lyndenbell74} that ought to approximate the evolution 
at sufficiently late times. For a disk with a viscosity that scales with radius as $\nu \propto r^\gamma$, 
the late-time evolution of the surface density close to the star is predicted to follow (e.g. \cite{armitage10}),
\begin{equation}
 \Sigma \propto t^{-(5/2 - \gamma) / (2-\gamma)}.
\end{equation} 
If $\gamma = 1$, for example, as would be the case for a disk with a steady-state surface 
density profile $\Sigma \propto r^{-1}$, then the predicted late-time decay goes as 
$t^{-3/2}$. This is relatively slow, and would probably lead to a population of stars with weak 
disk signatures (in gas and dust tracers) that are not observed (for a review of these  
observational arguments, see Alexander et al. \cite{alexander14}). One might argue, 
of course, that this is an illusory problem that we have created for ourselves by placing unwarranted faith in the 
validity of classical viscous disk theory. A resolution along these lines is possible. More 
commonly, however, the discrepancy between the simple theory and observations is 
taken to imply that some distinct process acts to rapidly disperse the disk and 
terminate accretion. Photoevaporation is almost certainly part of the story, on account both of 
robust theoretical estimates that suggest it should be important, and observations 
that are consistent with photoevaporation occurring in some relatively extreme situations.

\subsection{Photoevaporation}
\label{sec_photoevaporation}
Disk photoevaporation is a purely hydrodynamic process that occurs when molecular 
gas in the disk is dissociated or ionized by high energy (UV or X-ray) photons. If the 
gas is heated sufficiently to become unbound, it accelerates away from the disk 
under the influence of pressure gradient forces to form a thermally driven wind. 
Early models for photoevaporative flows were developed in the 1980s by Bally \& Scoville 
\cite{bally82} in the context of massive stars surrounded by neutral disks, 
and in more quantitative detail by Begelman et al. \cite{begelman83} who 
studied X-ray heated disks around compact objects. The essential physics is 
thus very well-established. We will begin by considering how photoevaporation 
works if the disk is exposed to extreme ultraviolet (EUV) photons that have sufficient energy 
($h \nu > 13.6 \ {\rm eV}$) to ionize hydrogen. This is probably {\em not} the 
dominant driver of photoevaporation from protoplanetary disks around low mass stars, but it is 
amenable to an analytic treatment that exposes the main principles \cite{hollenbach94}.

\subsubsection{Thermal winds from disks}
We consider a disk whose surfaces are illuminated by a source of high energy photons, that 
may come either from the central star or from other luminous stars within a cluster. The 
radiation heats a surface layer of the disk to a temperature $T$, with a sound speed $c_s$. 
A characteristic scale $r_g$ (the ``gravitational radius") can be defined by asking where 
the sound speed equal the local Keplerian velocity,
\begin{equation}
 r_g = \frac{GM_*}{c_s^2}.
\end{equation}
Noting that the thermal energy per particle is $\sim k_B T$, the radius $r_g$ is approximately 
equivalent to the smallest radius where the {\em total} energy of the gas in the heated surface 
layer (i.e. thermal plus gravitational) is zero. For radii $r > r_g$ the total energy is positive, 
and it is energetically possible for the surface gas to flow away in a thermal wind.

This is all quite rough, and we should really consider both the hydrodynamic structure of the 
wind (which is similar to the textbook example of a Parker wind \cite{waters12}) and the role 
of rotation \cite{liffman03}. Doing so results in an improved estimate of the critical radius 
beyond which a thermal wind is launched, which scales with but is significantly smaller than $r_g$,
\begin{equation}
 r_c \approx 0.2 \frac{GM_*}{c_s^2} \approx 1.8 \left( \frac{M_*}{M_\odot} \right) 
 \left( \frac{c_s}{10 \ {\rm km \ s^{-1}}} \right)^{-2} 
 \ {\rm AU}.
\end{equation} 
We have picked 10~${\rm km \ s}^{-1}$ as a fiducial sound speed because this is approximately 
the sound speed in EUV-ionized gas, which has a temperature near $10^4 \ {\rm K}$. X-rays or 
far ultraviolet (FUV) photons (those with $6 \ {\rm eV} < h \nu < 13.6 \ {\rm eV}$ that dissociate 
but do not ionize hydrogen) heat the surface to lower temperatures, $\sim 10^3 \ {\rm K}$, 
resulting in correspondingly larger critical radii.

Interior to $r_g$ the hot gas is bound, and unless some other process intervenes (such as a 
stellar wind or an MHD disk wind) it will form a static isothermal atmosphere with a scale 
height that varies with radius as $h \propto r^{3/2}$ (see \S\ref{sec_vertical_structure}). 
Outside $r_g$ it will flow away, at a speed of the order of the sound speed. In the case of 
EUV illumination the hot and ionized surface layer is separated from the underlying cool 
gas by a sharp ionization front, which gives a clearly defined ``base" to the wind. If the 
number density at the base is $n_0$, we expect a mass loss rate per unit area of the disk that 
is given by,
\begin{equation}
 \dot{\Sigma}_w \simeq 2 \mu m_H n_0 c_s,
\end{equation}
up to factors of order unity that depend again on the detailed hydrodynamic structure of the 
flow. In the EUV case the mass loss profile due to photoevaporation is then determined 
by the radial scaling of the base density $n_0 (r)$. Noting that the integrated mass loss 
rate is just,
\begin{equation}
 \dot{M}_w = \int_{r_c}^{r_{\rm out}} 2 \pi r \dot{\Sigma}_w {\rm d}r,
\end{equation}
we conclude that if $n_0 (r)$ declines more steeply that $r^{-2}$ the mass loss is 
predominantly from the inner disk (near $r_c$), whereas a shallower profile leads 
to most of the mass being lost from the outer disk. 

\begin{figure}[t]
\center
\includegraphics[width=0.8\columnwidth]{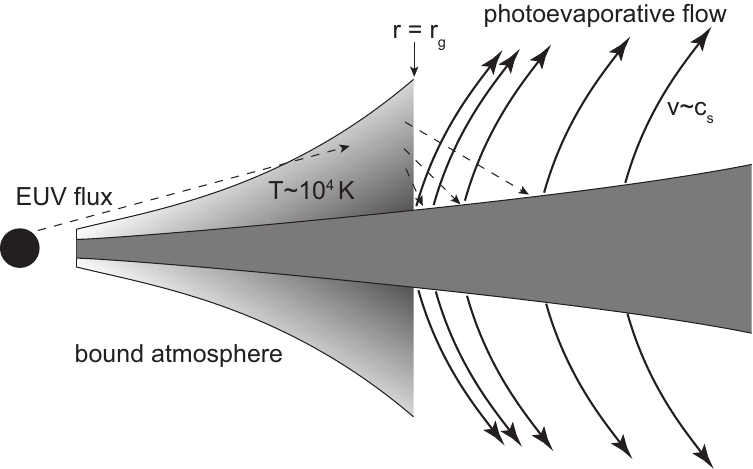}
\caption{Illustration of the simplest model of internal photoevaporation driven by extreme 
ultraviolet (EUV) radiation (based on the ``weak stellar wind" case from Hollenbach et al. \cite{hollenbach94}). Stellar 
EUV radiation ionizes and heats the surface layers of the disk. Where the thermal energy of the surface layer remains 
small compared to the binding energy, the hot gas forms a bound atmosphere. At larger radii, where the gas is more weakly bound, the hot gas flows away in a thermally driven wind. Details of the radiative transfer and heating 
processes differ depending upon the nature of the high energy radiation, but a qualitatively similar scenario 
applies also to X-ray and FUV-driven photoevaporation.}
\label{fig_photoevap}
\end{figure}

Actually finding $n_0 (r)$ for photoevaporation driven by EUV radiation from the central 
star requires the solution of a radiative transfer problem, whose geometry is illustrated in 
Figure~\ref{fig_photoevap}. The base density is determined by equating the rate of 
ionization to that of recombination, which occurs at a rate per unit volume $\alpha n_e n_p$ 
with $\alpha$, the recombination co-efficient, given by $\alpha \approx 3 \times 10^{-13}$.
The main difficulty is determining the radial scaling of the ionization rate. This in principle 
has two components, a ``direct" flux from the star and a ``diffuse" field that originates from 
the fraction (about one third) of recombinations within the bound atmosphere that go to the 
ground state and hence regenerate an ionizing photon. Hollenbach et al. \cite{hollenbach94} 
presented analytic and approximate numerical solutions to the radiative transfer problem that 
imply $n_0 (r) \propto r^{-5/2}$, and this scaling has been widely adopted in subsequent 
work. Important features of the Hollenbach et al. solution have been verified in more detailed 
radiation hydrodynamic simulations (including the $\Phi^{1/2}$ of the mass loss 
with the ionizing photon flux \cite{richling97}), though the slope of $n_0(r)$ at $r > r_g$ remains 
to be confirmed (indeed, a recent radiative transfer calculation is inconsistent 
with the canonical slope \cite{tanaka13}).

\subsubsection{Drivers of photoevaporation}
Photoevaporation driven by a photon flux of EUV radiation from the central star $\Phi$ is 
estimated to result in a mass loss rate \cite{font04,alexander14},
\begin{equation}
 \dot{M}_w \simeq 1.6 \times 10^{-10} 
 \left( \frac{M_*}{M_\odot} \right)^{1/2} 
 \left( \frac{\Phi}{10^{41} \ {\rm s}^{-1}} \right)^{1/2} \ M_\odot {\rm yr}^{-1},
\end{equation}
provided that the disk is present and optically thick at all radii. The dominant 
uncertainty in applying this estimate to specific systems comes from lack of 
knowledge of $\Phi$, which can be constrained but which remains hard to 
pin down precisely \cite{alexander05,pascucci14}. For low mass stars ($M \leq M_\odot$) 
reasonable estimates imply that EUV photoevaporation rates are negligible when 
compared to the median accretion rate of T~Tauri stars \cite{gullbring98}, but 
large enough to matter for disk dispersal if no stronger mass loss processes 
are operative. (EUV photon fluxes are of course vastly larger for massive 
stars, for which the theory was originally developed.)

Low mass pre-main-sequence stars are strong emitters of FUV and X-ray radiation 
\cite{france14,getman05,gudel07}. The FUV luminosity has a base level that is set by 
chromospheric activity, on top of which there is a potentially much larger 
component from accretion \cite{ingleby11}. The X-ray luminosity scales 
linearly with the bolometric luminosity, $\log (L_X / L_{\rm bol}) = -3.6$ \cite{getman05}, 
but with a large scatter. Qualitatively, these photons affect the disk in the same way 
as the EUV. The surfaces of the disk are heated, albeit to a somewhat lower 
temperature than the $10^4 \ {\rm K}$ that is characteristic of HII regions, and 
where this heating results in unbound gas a wind ensues. Quantitatively, the main 
difference is that X-ray and FUV heated layers are not separated from the cool 
underlying disk by any analog of a sharp ionization front, and this makes 
modeling of FUV and X-ray photoevaporation more difficult. State of the art 
calculations \cite{gorti09,owen10,owen12}, however, suggest that X-rays and 
FUV radiation drive mass loss rates that are substantially higher than the 
EUV prediction --- with values of the order of $10^{-8} \ M_\odot \ {\rm yr}^{-1}$ 
being possible --- with X-rays likely dominating in the inner region. The exact 
mass loss rates have a significant dependence on the adopted thermochemistry 
within the disk \cite{wang17}

\subsubsection{Disk evolution including photoevaporation}
Including photoevaporation in classical viscous models for disk evolution is 
particularly simple, because thermal winds exert no torque on the disk 
(equation~\ref{eq_disk_evolve_pe}). The rate and radial dependence 
of the mass loss, moreover, is primarily a function of the spectral 
energy distribution of the irradiation. This may depend upon the stellar 
accretion rate (if the FUV luminosity is an important driver of photoevaporation) 
but it is not coupled at leading order to details of the disk structure\footnote{In more 
detail, however, the grain population within the disk will affect the absorption of high energy 
photons and hence the local mass loss rate \cite{gorti15}.}. These properties mean 
that disks evolving under the joint action of viscosity and photoevaporation  
exhibit two distinct phases of evolution, an early phase in which viscosity 
dominates and a short subsequent phase in which the wind results in rapid 
disk dispersal \cite{clarke01}. 

\begin{figure}[t]
\center
\includegraphics[width=\columnwidth]{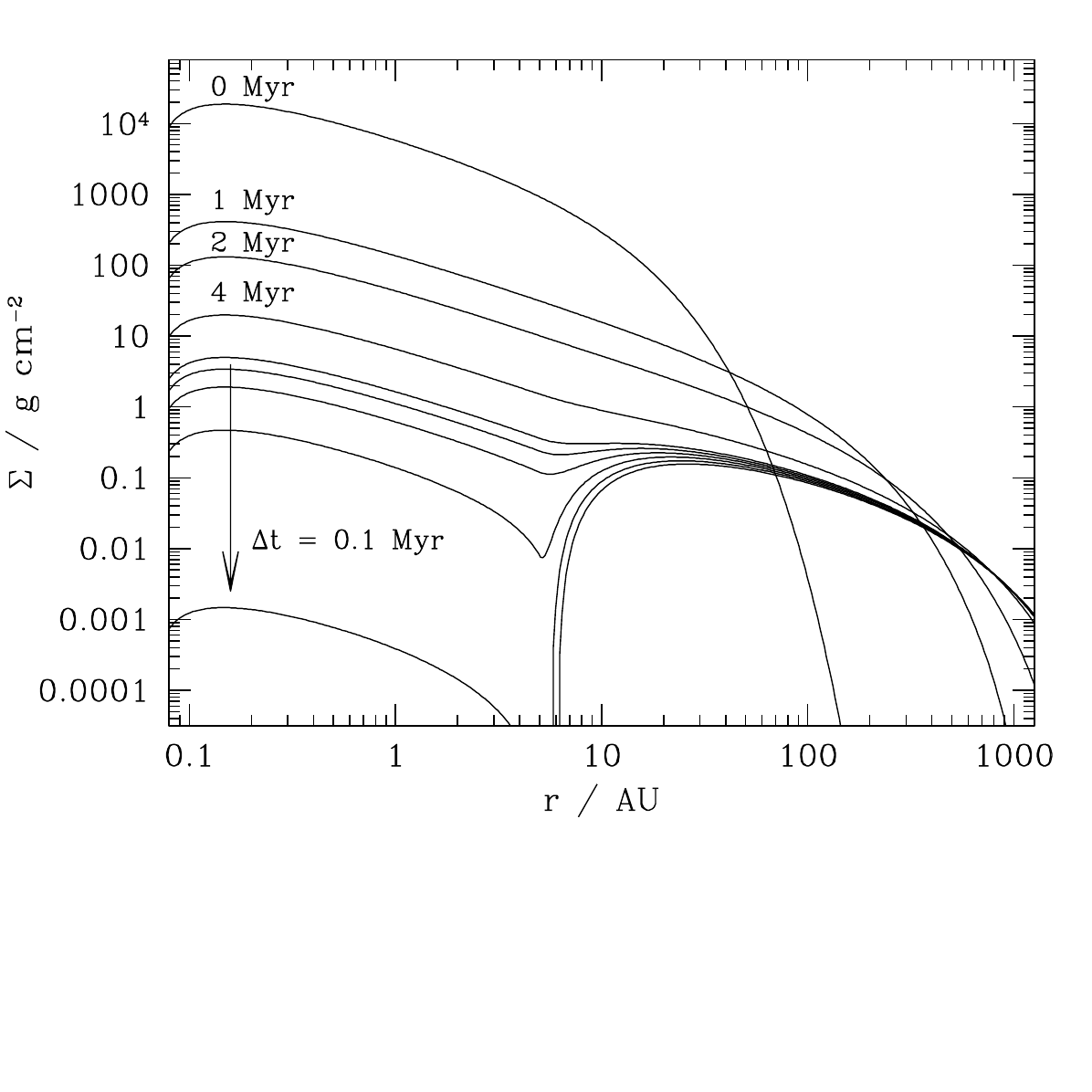}
\vspace{-3.0cm}
\caption{An illustrative calculation of disk evolution including photoevaporative mass loss. The 
model plotted is based on a disk with $\nu \propto r$, and wind mass loss that scales as 
$\dot{\Sigma}_w \propto r^{-5/2}$ outside 5~AU. The disk displays the two time scale evolutionary 
behavior that is reasonably generic to internal photoevaporation models, with a long period of 
slow ``viscous" evolution being followed by rapid inside-out dispersal.}
\label{fig_disperse}
\end{figure}

Figure~\ref{fig_disperse}, based on the original calculations by Clarke et al. \cite{clarke01}, 
shows how a disk evolves under the action of viscosity and EUV photoevaporation from the 
central star. In addition to the two time scale evolutionary behavior, the concentration of EUV mass loss 
toward the inner disk leads to a characteristic radial structure of disk dispersal. As the 
disk accretion rate drops, photoevaporation first dominates the evolution near to the 
innermost radius where mass loss is possible (in the illustrated example, this is taken 
to be 5~AU). A gap opens at this location, separating the inner disk (with a short 
viscous time scale) from the outer disk (where the viscous time scale is much longer). 
The inner disk then drains on to the star, stellar accretion ceases, and the disk is 
dispersed from the inside-out. The final dispersal is rapid, because the 
formation of a hole in the disk allows EUV \cite{alexander06} or X-ray \cite{owen13} 
radiation to directly illuminate the inner edge of the disk, which typically accelerates 
mass loss. The dust to gas ratio in whatever is left of the disk increases during 
the dispersal phase \cite{throop05,alexander07,gorti15}. Although the details depend on the radial profile 
of the photoevaporative wind, broadly similar evolution is predicted in most 
models \cite{gorti09b}. The observed lifetimes of disks are broadly consistent 
with theoretically estimates that are based on ``best-guess" values of  
photoevaporative mass loss rates \cite{bae13b}. 

There is no universal form describing how the action of photoevaporation cuts off accretion 
in the inner disk. For simple models of EUV photoevaporation, however, it can be shown 
that the inner accretion rate $\dot{M} (t)$ is related to the accretion rate $\dot{M}_{\rm SS}$ 
that would be predicted by a self-similar model \cite{lyndenbell74} without mass loss via \cite{ruden04},
\begin{equation}
 \dot{M} = \left[ 1 - \left( \frac{t}{t_0} \right)^{3/2} \right] \dot{M}_{\rm SS},
\end{equation}
where $t_0$ is the time at which accretion ceases. This formula is derived for a specific 
viscosity law ($\nu \propto r$) and photoevaporation model, but provides a qualitative 
idea of how the inner disk drains under more general circumstances.

The inside-out character of photoevaporative dispersal applies only in the limit 
where radiation from the central star is dominant. In sufficiently rich stellar 
clusters, photoevaporation due to intense FUV radiation fields from other 
(massive) stars is more important. Unlike in the case of central star photoevaporation, 
for which the observational evidence is indirect \cite{alexander14}, photoevaporative 
flows driven by external UV fields can sometimes be seen directly, most spectacularly 
in the core of the Orion Nebula \cite{bally00}. Adams et al. \cite{adams04} and 
Clarke \cite{clarke07} have modeled the evolution of viscous disks under 
external photoevaporation, and shown that it results in destruction of the disk 
from the outside-in. For the fraction of stars that form in such clusters, this 
process evidently limits the time over which gas-rich disks would be present 
on scales comparable to the Solar System's Kuiper Belt.

\subsection{MHD winds}
The theoretical and observational arguments for photoevaporation being an important 
component of disk evolution are strong, but other processes may 
also contribute to disk dispersal. The obvious alternate candidate is MHD winds, 
which are likely to be present if mature disks retain a dynamically significant net 
magnetic field (see the discussion in \S\ref{sec_transport}). MHD disk winds have 
obvious qualitative differences from their photoevaporative cousins,
\begin{itemize}
\item
Their strength depends upon the disk's net flux, rather than on the stellar radiation field.
\item
They can be accelerated from arbitrarily small radii, where even EUV-ionized gas 
would be bound, with a velocity proportional to the Keplerian velocity at the 
magnetic field footpoint.
\item
The local mass loss rate is (roughly) expected to scale with the disk surface density, 
rather than being (approximately) a constant independent of the underlying column.
\item
Predominantly neutral or molecular gas can, at least in principle, be accelerated 
(though it might subsequently be dissociated or ionized by stellar radiation).
\end{itemize}
If MHD disk winds are sufficiently strong, they can affect disk dispersal via some 
combination of mass and angular momentum loss. The resultant evolution can 
be quite similar to the photoevaporative case. In particular, if mass rather than 
angular momentum loss is dominant, Suzuki et al. \cite{suzuki10,suzuki16} showed that 
MHD winds lead to the formation of a shallow surface density profile at small 
radii and, eventually, an expanding inner hole. In the opposite limit where 
angular momentum loss is strong (and mass loss negligible) Armitage et al. 
\cite{armitage13b} suggested that dispersal could occur through the late 
onset of magnetic braking. Whether this is possible depends entirely on 
how the mass to flux ratio of the disk changes over time, and hence on the 
uncertain question of how net flux is transported and lost (\S\ref{sec_field_transport}). 
The most realistic scenario is one in which winds are driven by a combination 
of thermal (i.e. photoevaporative) and MHD processes, with the rate and 
radial profile of the mass and angular momentum loss depending jointly 
on the magnetic field structure and on the strength of impinging high 
energy photons \cite{bai16}.

\begin{svgraybox}
\begin{itemize}
\item
High-energy radiation in the form of FUV, EUV and X-ray photons heats the upper 
layers of protoplanetary disks, and drives a thermal wind from radii where the hot 
gas is unbound.
\item
Photoevaporative winds lead to distinct dispersal pathways for relatively isolated 
disks (where the high energy flux comes from the central star) and disks 
in rich clusters where FUV from massive stars dominates.
\item
Both thermal (photoevaporative) and MHD driving may contribute to disk evolution 
and dispersal, though the latter depends on the persistence of a significant net 
magnetic flux to late times.
\end{itemize}
\end{svgraybox}

\begin{acknowledgement}
My work on protoplanetary disk physics and planet formation has been supported by 
the National Science Foundation, by NASA under the Origins of Solar Systems, Exoplanet Research and 
Astrophysics Theory programs, and by the Space Telescope Science Institute. I acknowledge 
the hospitality of the IIB at the University of Liverpool, where much of this chapter was 
written, and thank Kaitlin Kratter for an informal review of the manuscript.
\end{acknowledgement}

\newpage

\bibliographystyle{spmpsci.bst}

\end{document}